\documentclass[12pt]{article}
\usepackage{amsmath}
\usepackage{mathtools} 
\usepackage[T1]{fontenc}
\usepackage{amssymb}
\usepackage{amsfonts}
\usepackage{amsthm}
\usepackage{mathrsfs}
\usepackage[all]{xy}
\usepackage{tensor}
\usepackage{comment}
\usepackage{stmaryrd}
\usepackage{bm}
\usepackage[normalem]{ulem}
\usepackage[usenames,dvipsnames]{xcolor}
\usepackage[colorlinks=true, citecolor=Blue, linkcolor=Blue]{hyperref}
\usepackage{ragged2e}
\usepackage{tikz}
\hypersetup{
	breaklinks=true,   
	colorlinks=true,   
	pdfusetitle=true,  
}

\usepackage{diagbox}  
\usepackage{makecell} 
\usepackage{hhline} 
\usepackage{stackengine}

\usepackage[labelfont={small, bf, sf}, textfont={small,sf}]{caption}
\numberwithin{equation}{section} 
\usepackage{authblk}
\usepackage[in]{fullpage}
\usepackage{cite}

\newcommand{\beq}{\begin{equation}}
\newcommand{\eeq}{\end{equation}}
\newcommand{\bes}{\begin{subequations}}
	\newcommand{\ees}{\end{subequations}}
\newcommand{\bea}{\begin{eqnarray}}
\newcommand{\eea}{\end{eqnarray}}

\newcommand{\be}{\begin{equation}}
\newcommand{\ee}{\end{equation}}
\newcommand{\msc}{\mathscr}
\newcommand{\goesto}{\to}

\newcommand{\lie}{\pounds}

\newcommand{\nb}{\mathcal{N}}
\newcommand{\bb}{\mathcal{B}}
\newcommand{\brmod}[2]{\llbracket #1, #2 \rrbracket}

\newcommand{\h}[1]{{\hat{#1}}}
 
\newcommand{\fs}{\mathscr{F}}
\newcommand{\sls}{\mathscr{S}}
\newcommand{\slp}{\mathscr{P}}
\newcommand{\beom}{\mathcal{E}}

\newcommand{\ns}{\mathcal{N}}

\newcommand{\ps}{\mathscr{P}}

\newcommand{\cflx}{\varepsilon}

\newcommand{\hateq}{\mathrel{\mathop {\widehat=} }} 

\newcommand{\scrip}{\mathscr{I}^{+}}
\newcommand{\scri}{\mathscr{I}}

\def\be{\begin{equation}}
\def\ee{\end{equation}}
\def\pullback{\Pi}
\def\pushback{\Upsilon}
\def\ps{{\scalebox{.7}{$\scriptscriptstyle +$}}}
\def\ms{{\scalebox{.7}{$\scriptscriptstyle -$}}}
\def\Sp{S_{\ps}}
\def\Sm{S_{\ms}}

\newcommand{\volume}{\eta}
\newcommand{\volumesmall}{\mu}
\newcommand{\nonaffinity}{\kappa}
\newcommand{\shape}{{\cal {K}}}
\newcommand{\expansion}{\Theta}
\newcommand{\inducedmetric}{q}

\newcount\hh
\newcount\mm
\mm=\time
\hh=\time
\divide\hh by 60
\divide\mm by 60
\multiply\mm by 60
\mm=-\mm
\advance\mm by \time
\def\hhmm{\number\hh:\ifnum\mm<10{}0\fi\number\mm}

\def\be{\begin{equation}}
\def\ee{\end{equation}}
\def\pullback{\Pi}
\def\pushback{\Upsilon}

\newcommand{\ve}{\varepsilon}

\newcommand{\mfp}{{\mathfrak{p}}}
\newcommand{\mfq}{{\mathfrak{q}}}
\newcommand{\mfs}{{\mathfrak{s}}}

\newcommand{\mfh}{{\mathfrak{h}}}
\newcommand{\mfi}{{\mathfrak{i}}}
\newcommand{\phase}{prephase }

\numberwithin{equation}{section}

\usepackage{cleveref}

\crefname{equation}{Eq.}{Eqs.}
\creflabelformat{equation}{#2#1#3}

\crefname{section}{Sec.}{Sec.}
\crefname{appendix}{Appendix}{Appendices}
\crefname{figure}{Fig.}{Figs.}

\crefname{definition}{Def.}{Defs.}
\crefname{prop}{Prop.}{Props.}
\crefname{lemma}{Lemma}{Lemmas}
\crefname{corollary}{Cor.}{Cors.}
\crefname{thm}{Theorem}{Theorems}
\crefname{remark}{Remark}{Remarks}

\begin{document}
	\setlength{\parindent}{10pt}
	
	\title{Horizon phase spaces in general relativity}
	
	\author[1]{Venkatesa Chandrasekaran\thanks{venchandrasekaran@ias.edu}}
	\author[2]{\'Eanna \'E. Flanagan\thanks{eef3@cornell.edu}}
	
	\affil[1]{\small \it School of Natural Sciences, Institute for Advanced Study, 1 Einstein Drive, Princeton, NJ 08540 USA}
	\affil[2]{\small \it Department of Physics, Cornell University, Ithaca, NY, 14853, USA}

	\maketitle

	\begin{abstract}
		We derive a prescription for the phase space of general relativity on
		two intersecting null surfaces using the null initial
                value formulation.  The phase space allows generic
                smooth initial data, and the corresponding boundary
                symmetry group is the semidirect product of the group
		of arbitrary diffeomorphisms of each null boundary which coincide at the corner, 
		with a group of reparameterizations of the null generators.
                The phase space can be consistently extended by acting
                with half-sided boosts that generate Weyl shocks along
                the initial data surfaces. The extended phase space includes the relative boost angle between the null surfaces as part of the initial data.

                We then apply the
                Wald-Zoupas framework to compute gravitational charges
                and fluxes associated with the boundary symmetries.
                The non-uniqueness in the
		charges can be reduced to two free parameters  by
                imposing covariance and invariance under rescalings of
                the null normals.  We show that the 
                Wald-Zoupas stationarity criterion cannot be used to
                eliminate the non-uniqueness.  The different choices
                of parameters correspond to different choices of
                polarization on the phase space.
                We also derive the symmetry groups and charges for two
                subspaces of the phase space, the first  
                obtained by fixing the direction of the normal
                vectors, and the second by fixing the direction and
                normalization of the  normal vectors.  The second
                symmetry group consists of Carrollian
                diffeomorphisms on the two boundaries.

		Finally we specialize to future event horizons by
                imposing the condition that the area element 
		be non-decreasing and become constant at late times.
                For perturbations about stationary backgrounds
                we determine the independent dynamical degrees of
                freedom by solving the constraint equations along the
                horizons. We mod out by the degeneracy directions of the
                presymplectic form, and apply a similar procedure
                for weak non-degeneracies, to obtain the horizon edge modes and the Poisson structure.
                We show that the area operator of the black hole generates a shift in the relative boost
                angle under the Poisson bracket.

	\end{abstract}

	\tableofcontents

\section{Introduction}

        \subsection{Overview}

        	The phase space of gravitational theories at null boundaries plays an important role in a complete understanding of black holes, gravitational entropy, and quantum gravity. While some of these topics have been studied extensively in holography, from the perspective of quantum information theory, they have not been well understood in general spacetimes from the perspective of semi-classical Lorentzian gravity. Constructing the phase space of gravitational subregions surrounded by horizons,\footnote{Here and throughout the paper, we use the terms horizon, null surface, and null boundary interchangeably.} and characterizing the horizon degrees of freedom, would comprise a fundamental step towards tackling these questions in Lorentzian signature. It would lay the groundwork for making contact with Hilbert spaces and dynamics of subsystems like black holes.

Despite steady progress over the years \cite{Iyer:1994ys, WZ, Hopfmuller:2016scf, Donnay:2016ejv, CFP, Hopfmuller:2018fni, Wieland:2017zkf, Chandrasekaran:2020wwn, Harlow:2019yfa, Chandrasekaran:2021vyu, freidel2021weyl, Chandrasekaran:2021hxc, Odak:2022ndm,Adami:2020amw,Adami:2020ugu,Adami:2021nnf,Adami:2021kvx,Donnay:2019jiz}, there are a myriad of aspects of the null boundary phase space which have yet to be systematically analyzed. The primary tool used in this field is the covariant phase space formalism \cite{Witten:1986qs, Crnkovic1987, Crnkovic:1987tz, Ashtekar1991,
		LeeWald1990, Wald:1993nt, Iyer:1994ys}.
In order to do so, one has to specify
the field configuration space by imposing boundary conditions.
These boundary conditions can involve some fixed
background structures on the boundary, for which different choices are
possible, and they determine the symmetry group.
The other main aspect is the characterization of the charges of the
boundary degrees of freedom associated with the symmetries.  This requires a choice of decomposition of
the boundary pullback of the presymplectic potential into boundary, corner and flux
terms \cite{Harlow:2019yfa,Chandrasekaran:2020wwn}. The presence of
background fields typically also leads to anomalies
\cite{Hopfmuller:2018fni, Chandrasekaran:2020wwn}.  
	
	But what boundary conditions should one choose? How does this
        formalism fit in with the action formulation of the theory?
        How should one perform the decomposition? Should one allow
        anomalies? Some aspects of these questions can be answered in
        a first-principles manner, but others are riddled with
        ambiguities. These questions have been tackled in
        several ways \cite{Reisenberger:2007pq,Reisenberger:2007pq,WZ,Hawking:2016sgy,
          Chandrasekaran:2021hxc, freidel2021weyl}, but to some extent
        using ad-hoc principles, and, at least in the context of
        finite null boundaries, have not been answered at a
        sufficiently general level.
        Moreover, in all of these cases,
        when the null boundary phase space is studied it is typically
        in the context of a single null boundary. But in order to fully understand the
        independent physical degrees of freedom we have to cast the problem in
        terms of intersecting null surfaces in the null initial value
        formulation.

	        Our goals in this paper are to make concrete progress
                on the Lorentzian problem by providing a general
                prescription for constructing the phase space of
                 general relativity on intersecting null surfaces, and
                 deriving boundary symmetries and charges. We also
                 compute the Poisson structure in terms of the independent degrees of freedom obtained by solving the constraints, restricted for simplicity to perturbations of eternal black holes.
                More specifically:

	\begin{itemize}
		\item We adopt the null initial value formulation of
                  general relativity on two intersecting null surfaces, and extend the space of initial data by including a degree of freedom analogous to the lapse function, and by including an edge mode generated by acting with half-sided boosts (Secs.\ \ref{sec:initialvalue} and \ref{halfsidedboosts}).

        \item For the \phase space\footnote{A \phase space is equipped with a degenerate presymplectic form; the physical phase space is obtained by modding out by degeneracy directions.} of arbitrary initial data on the two null surfaces, the symmetry group consists of arbitrary diffeomorphisms of the two null boundaries that coincide on the corner, in a semi-direct product with independent reparameterizations of the null generators of the respective boundaries (Sec.\ \ref{sec:phasespaces}).

	\item Gravitational charges associated with the symmetries are uniquely determined by specifying a decomposition of the pullback of the symplectic potential to the boundary into boundary, corner and flux terms.  We impose on this decomposition the requirements of boundary covariance, invariance under rescaling of the null normal, and invariance under choice of auxiliary normal.  These requirements determine the decomposition up to two parameters, which correspond to different choices of polarization on the \phase space.
We discuss why the Wald-Zoupas stationarity method for fixing the decomposition cannot be applied.
                  Finally we compute the Wald-Zoupas charges and
                  fluxes, which are the most general set of
                  charges/fluxes possible for a null boundary (Secs.\ \ref{sec:decomp_charge} and \ref{sec:nonunique}).

      		\item 
                  Along
                  the way we also derive the symmetry groups and charges for two
                subspaces of the \phase space, the first  
                obtained by fixing the direction of the normal
                vectors, and the second by fixing the direction and
                normalization of the  normal vectors.  The second
                symmetry group consists of Carrollian
                diffeomorphisms on the two boundaries (Secs.\ \ref{sec:phasespaces} and \ref{sec:decomp_charge}).

	      \item Lastly, we restrict to perturbations of eternal
                black holes, and compute the phase space of the
                intersecting left and right future horizons by solving
                the constraints and modding out by degeneracy
                directions.  We obtain the Poisson brackets in terms
                of the resulting independent canonically conjugate
                variables, consisting of corner and bulk terms
                (generalizing the similar single-horizon computation
                of \cite{Hawking:2016sgy}).
                In the corner piece we show that there exists a null
                boundary analogue of the Hayward term
                \cite{1993PhRvD..47.3275H}. This term contains, as
                canonically conjugate pairs, the area element of the
                black hole and the half-sided boost degree of freedom,
                which encodes the relative boost angle between the two
                sides of the black hole. We explain how this edge mode
                can be thought of as a dynamical time variable, and is
                the horizon analogue of the crossed product degree of
                freedom discussed in \cite{Chandrasekaran:2022eqq}
                (Sec. \ref{sec:interp}).

        \end{itemize}
	
	We now discuss these points in more detail.

\subsection{General \phase space prescription}
                
Our starting point is the null initial value formulation for vacuum
general relativity for two null
surfaces $\Sp$ and $\Sm$ which intersect in a spacelike two surface
$S_0$.  For example, $S_0$ could be the bifurcation two sphere of an
eternal black hole spacetime and $\Sp$ and $\Sm$ could be the future
horizons.  Initial data on $S_0$, $\Sp$ and $\Sm$ will determine the
metric within the future domain of dependence of $S_0$, up to
diffeomorphisms, and the space of initial data will define the
phase space (up to edge mode degeneracies discussed in Sec.\ \ref{sec:interp}).
Thus, in the black hole example, the phase space we construct is
that of the black hole interior.
Various formulations of the initial value problem in
this context have been given in the literature
\cite{1962JMP.....3..908S,70785209-fc5c-312c-9bd2-7c68ce777f34,1993CQGra..10..773H,Reisenberger:2007pq,Brady:1995na,Mars:2022gsa,Mars:2023hty}.
In Sec.\ \ref{sec:initialvalue} we extend slightly these formulations and
define an equivalence class of tensor fields on $\Sp$, $\Sm$ and
$S_0$ [Eq.\ (\ref{ids1}) below] that is sufficient to determine the future evolution, where the
equivalence relation is related to rescalings of the null normals.

We also extend the space of initial data in two key ways.  First, in
Sec.\ \ref{sec:initialdataextended}, we extend the space of initial
data to include a choice of normalization of the null
normal covector on each null surface. This specification is analogous to specifying the lapse
function in the initial value formulation on spacelike hypersurfaces.  As in
that context, different choices of this information yield metrics on spacetime
which are related by diffeomorphisms.  However, in the double null
context, these diffeomorphisms do not correspond to degeneracy
directions of the presymplectic form of the theory, as we show in
Sec.\ \ref{sec:chargeder}, and so are not true
gauge degrees of freedom.  Therefore the space of initial data must be extended
to include these degrees of freedom in order to serve as a parameterization of the phase space.

Second, in Sec.\ \ref{halfsidedboosts}, we extend the space of initial
data to include ``half-sided boosts''
\cite{Carlip:1993sa,Bousso:2020yxi,2020PhRvD.101d6001B}.
These are boosts about the surface $S_0$ that act only on one of the
two null surfaces but not the other.
They have been found to be important in quantum gravity due to their
connections with generalized entropy and modular flow
\cite{Bousso:2020yxi,2020PhRvD.101d6001B,Chandrasekaran:2022eqq}. A
detailed understanding of these transformations, and their relationship
to the area operator at the level of the gravitational phase space, is
therefore important for precisely connecting the area operator to modular flow in semi-classical gravity.

Starting from smooth solutions,
these transformations generate solutions with discontinuities or
shocks in the Weyl tensor\footnote{These shocks were first studied in
Refs.\ \cite{Bousso:2020yxi,2020PhRvD.101d6001B} in the specialized
contexts where the surface $S_0$ is extremal or marginally trapped.  The half-sided boosts we
define here are generalizations of the ``kink transforms'' and ``left
stretch transformations'' defined there, which were shown to be holographic duals of a
        smooth version of half-sided modular flow on the boundary. In
        particular, our half-sided boosts are well defined for any codimension-two surface.}.
We show that the space of initial data for these more general solutions 
can be characterized by relaxing a
particular continuity condition at the corner $S_0$.  The enlarged space of initial data
then contains a new edge mode which again does not correspond to a
degeneracy direction of the presymplectic form, as we show in Sec. \ref{sec:sf}.  This edge mode can be
understood to be the relative boost angle between a particular
rescaling-invariant null normal defined on $\Sm$ [Eq.\ (\ref{ellinvariant}) below]
and a similar quantity defined on $\Sp$.

We next define a maximal \phase space to be the set of all smooth vacuum metrics
for which $\Sp$ and $\Sm$ are null and $S_0$ is spacelike.  Each such metric determines a unique set of initial data on the two null surfaces.
We show in Sec.\ \ref{sec:phasespaces} that 
the corresponding symmetry group consists of arbitrary diffeomorphisms of
the two null boundaries that coincide on the corner, in a semi-direct
product with the group of arbitrary reparameterizations of the null
generators (or equivalently arbitrary rescalings of the normal covectors).
We call this the horizon Weyl-diffeomorphism group. The corresponding
algebra on a single null hypersurface
was previously studied in Ref.\ \cite{Adami:2021nnf} using a coordinate-dependent method.

We also consider two different subspaces of this maximal \phase space.
We define the restricted horizon \phase space in Sec.\ \ref{sec:rhps} 
by fixing the directions of the normal vectors on the
null surfaces.  The symmetry group is then modified by replacing the
general boundary diffeomorphism component with the Carrollian
diffeomorphisms which preserve this null direction, studied in
Refs.\ \cite{Donnay:2016ejv,Ciambelli:2018xat,Ciambelli2019b,Donnay:2019jiz,Freidel:2022vjq,Freidel:2022bai}.
A yet smaller \phase space can be obtained by fixing the normalization
of the contravariant normals, in which case the symmetry group
is modified by restricting the null generator reparameterizations to be
determined in terms of the Carrollian diffeomorphisms.
We note that our previous analysis of \phase spaces and symmetries on null surfaces
\cite{CFP} considered a further reduction of the \phase space in which the inaffinity of
the null generators was effectively fixed.  This has the effect of
reducing the reparameterizations of the null generators to two
families of supertranslations, affine and Killing \cite{Donnay:2016ejv,CFP}.

\subsection{Gravitational charges and fluxes}

A second key goal of this paper is to compute gravitational charges
associated with symmetry generators and associated with arbitrary
2-surfaces $\partial \Sigma$ in the null boundaries $\Sp$ and $\Sm$.  These charges can
be loosely interpreted as being associated with \phase subspaces
associated with initial data to the future of $\partial \Sigma$.
The general method for computing such charges was developed by Wald
and Zoupas \cite{Wald:1999wa} and generalized further in
Refs.\ \cite{Compere:2020lrt,Harlow:2019yfa,Freidel:2020xyx,Chandrasekaran:2020wwn,Margalef-Bentabol:2020teu,freidel2021weyl,Freidel:2021cjp,Ciambelli:2022vot,Chandrasekaran:2021vyu},
and is reviewed in Sec.\ \ref{sec:gp} below.  The method consists of
taking the expression for the variation of a charge obtained from
Hamiltonian methods, and discarding a certain non-integrable portion of the
expression in order to yield an integrable charge.
Alternative frameworks have been developed recently that directly yield
integrable charge expressions, either by enlarging the \phase space to include embedding degrees of freedom, or by defining ``dressed \phase subspaces'' in which the surface $\delta \Sigma$ is a covariant functional
of the dynamical fields in the theory
\cite{Ciambelli:2021nmv,Freidel:2021dxw,Speranza:2022lxr,Klinger:2023tgi,Carrozza:2022xut}.
It would be interesting to revisit the analysis of this paper using
these alternative frameworks, but we will leave this for future work.

In the Wald-Zoupas framework, gravitational charges are uniquely
determined by specifying a decomposition of the pullback of the
presymplectic potential to the boundary into boundary, corner and flux
terms.  This decomposition is closely related to, but not uniquely
determined by, a specification of a complete action principle for the
theory, including bulk, boundary and corner terms.  It is determined by a choice of presymplectic potential
or polarization on the \phase space.
We review in Sec.\ \ref{sec:gp} various criteria that have be used in the literature to
attempt to determine a unique such decomposition, and contrast with our approach in the present paper.

One such criterion is the stationarity requirement of Wald and Zoupas \cite{WZ}, that 
        the flux term should vanish identically on
	stationary solutions, for all allowed field variations.
This criterion has been used successfully in many contexts to single out unique decompositions, including 
BMS charges in vacuum general relativity \cite{WZ} and in
Einstein-Maxwell \cite{Bonga:2019bim} and in the analysis of finite
null boundaries of  
\cite{CFP}.  However, we show that it cannot be applied in
our context, due to the lack of any decompositions for which it is satisfied.
This occurs because the symmetries themselves are time dependent in a
certain sense, and thus their fluxes would not be expected to vanish in stationary
	situations, which is a purely kinematical effect.
	We argue therefore that in the most general contexts, the stationarity requirement
	does not provide useful constraints on the free parameters in
        the decomposition of the presymplectic potential.

A second criterion is the Dirichlet condition advocated in
Ref.\ \cite{Chandrasekaran:2020wwn}, which imposes that the flux
vanishes when Dirichlet-like boundary conditions are imposed.  This
criterion was used to obtain a unique decomposition in 
Ref.\ \cite{Chandrasekaran:2020wwn}, but we do not impose it here
since it is incompatible with covariance requirements which we do
impose.

In this paper, we do not aim a priori to eliminate all the non-uniqueness
in the decomposition of the presymplectic potential. Instead, we take
	the perspective that there are certain fundamental conditions the
	decomposition should satisfy, and determine the resulting
        freedom. 
        The key condition we impose is boundary
	covariance, that all the quantities that enter the construction should
	transform covariantly or be anomaly-free.  In other
	words, they should not depend on any background structures other than
	those which enter into the definition of the \phase space.
	The philosophy here is that anomalies should not
	be forced into the theory if they can otherwise be eliminated by a
	different choice of decomposition, and only those anomalies which
	cannot be eliminated this way ought to be considered to be
        fundamental. For example,
        Brown and Henneaux implicitly showed the existence of
	such a fundamental anomaly in the boundary symmetry algebra         in
	$\text{AdS}_3$ \cite{brown1986}, which 
	ultimately stems from the need to introduce a foliation near the
	boundary to renormalize the symplectic current. No such
        fundamental anomalies will arise in this paper.
	
	A consequence of the requirement of boundary covariance in our
        context, and of an addition requirement described in Sec.\ \ref{sec:eps},
        	is that all quantities in the formalism should be
	invariant under rescalings of the null normals, and under changes in
	the choice of auxiliary null vector or rigging vector on the
        boundary (see Appendix \ref{app:nullreview}).
	We compute in Sec.\ \ref{sec:decomp_charge} the most general decomposition of the symplectic
        potential consistent with these requirements, and find a three parameter freedom\footnote{For general values of our parameters, our flux term
        does not solely consist of a specific type of polarization,
        such as Dirichlet, but is rather a complicated combination
        of several different types of polarizations, including
        Dirichlet and York \cite{Rignon-Bret:2023fjq}.}.        We
        advocate that this freedom should not be viewed as 
        an ambiguity, but rather that it reflects different choices
        of polarizations on the phase space, akin to the usual freedom in choosing the
        configuration variables and conjugate momenta in Newtonian
        mechanics
        \cite{Odak:2021axr,Odak:2022ndm,Rignon-Bret:2023fjq,Freidel:2020xyx,Freidel:2020svx,Freidel:2020ayo}.             The different polarizations correspond to different
        choices of variational principle (see Secs.\ \ref{covr}, \ref{sec:cnu} and \ref{sec:nonunique}).        

        Given  our general decomposition of the presymplectic
        potential into corner, boundary and flux terms, we compute in
        Sec.\ \ref{sec:decomp_charge}        the
        corresponding gravitational charges and fluxes, which are
        the most general set of charges possible for a null boundary.
        We also give the corresponding results for the two
        smaller phase spaces discussed above.

\subsection{Physical phase space and Poisson structure}

	In order to determine the actual physical degrees of
        freedom of the theory, we construct in Sec.\ \ref{sec:interp}
        the physical phase space of the intersecting
        null boundaries.
	To make the analysis more tractable,
        we specialize
        to the restricted \phase space in which the direction of the null
        generators is fixed.
        We also restrict to
        perturbations of eternal black holes, allowing
        arbitrary
        non-stationary perturbations.  The specialization to 
        event horizons implies 
        that the area element is non-decreasing, and that the shear and
        expansion decay to zero at late times.

        The computation
        involves four stages, starting from the
        prephase space of on-shell solutions:
        \begin{itemize}
          \item Modding out
        by degeneracies associated with bulk diffeomorphisms, which
        yields the space of initial data discussed above.
        \item Suitably fixing the
        rescaling freedom to obtain independent coordinates for the
        initial data, that parameterize solutions to the constraints
        which have been integrated out.
        \item Modding out by
        additional degeneracies of the presymplectic form associated
        with edge or corner modes.  In addition to degeneracies, we
        also encounter directions which are weakly but not strongly non-degenerate, which require a slightly
        novel treatment based on an approach of Prabhu, Satischandrand
        and Wald \cite{Prabhu:2022zcr} (see Appendix \ref{app:zero}).
        \item Enumerating the final independent degrees of
          freedom, consisting of bulk and corner or edge modes,
          and computing the corresponding Poisson brackets.
          The result also allows us to characterize the different polarizations
        of the degrees of freedom. 
        \end{itemize}

	This provides a complete characterization of the horizon
        degrees of freedom.
        The bulk degrees of freedom contain the usual spin 2 mode on
        each null surface, together with two scalar modes on each null
        surface which obey a slightly unusual algebra given
        by Eqs.\ (\ref{pb44}) and (\ref{pb45a}) -- (\ref{pb45c}).
        Roughly speaking, one can think of these two scalar modes as
        parameterizing the two scalings of the normal covector and
        normal vector, on each surface.

        The corner degrees of freedom contain a canonically
        conjugate pair of spin 2 variables, 
        given by Eq.\ (\ref{spin2corner}).
        They
        also contain, as canonically conjugate pairs, the area
        element of the black hole and the half-sided boost degree of
        freedom, which encodes the relative boost angle between the
        two sides of the black hole [Eq.\ (\ref{pb45d}) below]. 
        This pair corresponds to a null boundary
        analogue of the Hayward corner term in the gravitational
        action \cite{1993PhRvD..47.3275H}.
          This result is not particularly surprising, given the well-known
          result that the area term is canonically conjugate to the
          rotation angle around the Euclidean black hole conical
          defect \cite{Carlip:1993sa}, and the fact that the
          rotation angle analytically continues to the boost angle
          about the bifurcation surface in Lorentzian signature.
          However, the derivation given here is purely Lorentzian,
          which makes it amenable to generalizations beyond the bifurcation surface of an
          eternal black hole, as well as to perturbative canonical
          quantization.  
The boost edge mode can also be thought of as a dynamical time
variable which allows us to define dressed observables as in Ref.\ \cite{Harlow:2021dfp}, and is the
horizon analogue of the crossed product degree of freedom discussed in
\cite{Chandrasekaran:2022eqq}.   This means that locations to the
future of the initial data surface can be gauge invariantly specified
by giving the boost degree of freedom together with some other data.

	\subsection{Notation}
	
	Lower case Roman indices $a,b, \ldots$ from the start of the alphabet
	denote spacetime tensors, whereas lower case Roman indices $i,j,
	\ldots$ from the middle of the alphabet denote tensors intrinsic to
	the null surface ${\cal N}$. Capital Roman indices $A, B, C, \ldots$
	will denote tensors on the two surface $S_0$ where two null surfaces intersect.
	
	As this paper was being completed we learned of a similar analysis by
	Odak, Rignon-Bret and Speziale which has considerable overlap with our results and
	which is being submitted simultaneously to the arxiv \cite{ORS}.
	There is also some overlap (particularly in Secs.\ \ref{halfsidedboosts} and
        \ref{sec:interp}) with the recent work
	by Ciambelli, Freidel and Leigh \cite{Ciambelli:2023mir}.
	Alternative approaches to the gravitational phase space at null
	boundaries can be found in
	Refs.\ \cite{Reisenberger:2007pq,Reisenberger:2007pq,
          Adami:2020amw,Adami:2020ugu,Adami:2021nnf,Adami:2021kvx}.

	\section{General prescriptions for computing gravitational charges at finite boundaries}
	\label{sec:gp}
	
	\subsection{Review: charges determined uniquely by specification of boundary
		and corner fluxes, or \phase space polarization}
	\label{covr}
	
	In generally covariant theories with specified boundary conditions, the covariant phase space
	formalism \cite{Witten:1986qs, Crnkovic1987, Crnkovic:1987tz, Ashtekar1991,
		LeeWald1990, Wald:1993nt, Iyer:1994ys} can be used to compute
	symmetries, as well as global and localized charges.
	The general method was developed by Wald and Zoupas \cite{Wald:1999wa},
	which has since been extended and now exists in several slightly
	different versions
	\cite{Compere:2020lrt,Harlow:2019yfa,Freidel:2020xyx,Chandrasekaran:2020wwn,Margalef-Bentabol:2020teu,freidel2021weyl,Freidel:2021cjp,Ciambelli:2022vot,Chandrasekaran:2021vyu}.
	Here we will briefly review the prescription and notation
	of Refs.\ \cite{Chandrasekaran:2020wwn,Chandrasekaran:2021vyu}, which emphasizes
	which quantities need to be specified in order to determine unique
	gravitational charges.  We will use the notation (CFSS,XX) to mean
	Eq.\ (XX) of Chandrasekaran, Flanagan, Shehzad and Speranza (CFSS)
	\cite{Chandrasekaran:2021vyu}.
	
	\subsubsection{Action principle}
	\label{sec:actionprinciple}
	
	The starting point is an oriented manifold $M$ with boundary $\partial M$ which
	can be expressed as a union of different components
	\be
	\partial M = \cup_j \nb_j,
	\ee
	where the orientations of $\nb_j$ are the induced orientations.
	We assume that each boundary component $\nb_j$ has a boundary
	$\partial \nb_j$ which consists of a codimension two surface $\bb_j$
	or {\it corner} where it intersects $\nb_{j+1}$, and another corner
	$\bb_{j-1}$ where it intersects $\nb_{j-1}$.  We choose the
	orientations of $\bb_j$ so that for any two-form $\rho$ we have
	\be
	\label{orien}
	\int_{\partial \nb_j} \rho = \int_{\bb_j} \rho - \int_{\bb_{j-1}}
	\rho.
	\ee
	
	We are given a field configuration space $\fs$ of the dynamical fields
	$\phi$ which includes
	specification of boundary conditions at each $\nb_j$.
	The on-shell subspace of $\fs$ is
	the \phase space $\sls$ of the theory, which consists of solutions
	of the equations of motion.  The actual phase space
	$\slp$ is obtained from $\sls$ by modding out the degeneracy directions of the
	presymplectic form \cite{Wald:1999wa,Harlow:2019yfa}.
        In this paper the modding out will occur in two stages.  The
        first stage involves modding out by bulk diffeomorphisms, and
        yields a space of initial data on the two null surfaces.  The
        second stage involves modding out by additional degeneracy
        directions related to edge modes in the initial data, and
        generates the phase space $\slp$.
	We will mostly work with the \phase space $\sls$
        or with the space of initial data.

	We write the action functional of the theory as
	\be
	\label{action1}
	S = \int_M L' + \sum_j \int_{\nb_j} \ell_j' + \sum_j \int_{
		\bb_j} {\hat c}_j,
	\ee
	in terms of a bulk Lagrangian $L'$, boundary actions $\ell_j'$ and corner
	actions ${\hat c}_j$.  We choose a presymplectic potential $\theta'$ which satisfies
	\be
	\delta L' = E(\phi) \delta \phi + d \theta',
	\ee
	where $E(\phi)=0$ are the equations of motion.  The pullback of
	${\underline \theta'}$ of $\theta'$ to the boundary component $\nb_j$ is decomposed as
	\be
	\label{decompos11}
	{\underline \theta'} = - \delta \ell_j' + d \beta_j' + \beom_j,
	\ee
	in terms of the boundary term $\ell_j'$, an exact {\it corner term} $d \beta_j'$ and
	a {\it boundary flux} $\beom_j$.  
	We also define the {\it corner flux} ${\hat \cflx}_j$ via
	\be
	\label{blde0}
	{\underline \beta}'_{j} - {\underline \beta}'_{j+1} = - \delta {\hat c}_j
	+ d {\hat \gamma}_j + {\hat \cflx}_j,
	\ee
	where on the left hand side the underlines denote pullbacks to the corner $\bb_j$.
	This equation defines the corner flux up to the addition of the exact term $d {\hat \gamma}_i$ which will
	not play any role below.
	Then the variation of the action (\ref{action1}) can be written in
	terms of the boundary and corner fluxes as \cite{Chandrasekaran:2021vyu}
	\be
	\label{deltaS}
	\delta S = \sum_j \int_{\ns_j} \beom_j +  \sum_j \int_{\bb_j} {\hat \cflx}_j
	\ee
	when the bulk equations of motion are satisfied.
	
	Finding a well posed variational principle now involves in general two
	elements (see the discussion by Harlow and Wu \cite{Harlow:2019yfa}):
	\begin{itemize}
		\item A specialization of the definition of the field configuration
		space $\fs$ (i.e. the choice of boundary conditions) and of the
		decomposition (\ref{decompos11}) to ensure the vanishing of the
		boundary and corner flux terms in the variation (\ref{deltaS}), for
		certain of the boundary components $j$ (typically the timelike boundaries).
		
		\item A specification of data $\mfs$ on certain of the
		boundary components (typically the spacelike boundaries) that are
		sufficient to determine a unique solution in $\sls$, ie a physical
		state. We denote by $\fs_\mfs$ the corresponding subspace of the
		field configuration space $\fs$.  Then the variational principle
		is well posed in $\fs_\mfs$ if
		the remaining corner and
		boundary terms in the variation (\ref{deltaS}) vanish for
		field configurations $\phi$ and variations $\delta \phi$
		restricted to $\fs_\mfs$.
	\end{itemize}
	It is necessary to have these two different types of elements, since
	the variation in the action should be nonzero when one varies from one
	physical state to another.
	For example, for vacuum general relativity with timelike and spacelike
	boundaries, one typically fixes the induced metric on the timelike
	boundaries, and this data becomes part of the definition of the field
	configuration space $\fs$.  The data $\mfs$ can be taken to be the
	induced metric on the initial and final spacelike boundaries, which
	generally should yield a unique solution \cite{1962PhRv..126.1864B,Bartnik:1993gh}.
	Other kinds of boundary conditions are studied in Ref.\ \cite{Odak:2021axr}.

	Null boundary components complicate this well known framework.
	For vacuum general relativity one can certainly define field
	configuration spaces $\fs$ and decompositions (\ref{decompos11}) 
	which make the boundary flux $\beom$ vanish.  These include
	Dirichlet boundary conditions in which one fixes the induced
	metric on the boundary \cite{Chandrasekaran:2020wwn,ORS}.
	Lehner, Myers, Poisson and Sorkin \cite{Lehner2016}
	have derived the form of the boundary and corner actions $\ell_j$ and
	${\hat c}_j$ required to yield  $\beom_j = {\hat \cflx}_j
	=0$, for timelike, null and spacelike boundaries and various corner
	combinations of these boundaries, assuming that the induced metric is
	fixed on all the boundaries.  Another possibility is the 
	York boundary condition in which the conformal induced metric
	and expansion are fixed \cite{1988bqc..book..246Y,Odak:2021axr,Rignon-Bret:2023fjq,Witten:2018lgb,ORS}.
	However both of these boundary conditions require restricting physical
	degrees of freedom on the null boundary.
	
	An alternative approach to variational principles with null boundaries
	is to treat the null boundary as part of the specification of the
	state $\mfs$, as for spacelike boundaries.
	This is natural since null boundary components can form (pieces of)
	good initial data surfaces, at least for massless fields.
	Thus, rather than
	specifying boundary conditions in the definition of the field
	configuration space $\fs$, one specifies certain data $\mfs$ on the null
	boundaries that are sufficient to determine unique solutions when one
	combines all the past and future boundaries, and formulates an action
	principle within $\fs_\mfs$.
	A complication here is
	that one wants to specify half the degrees of freedom on the null
	surface, like one does on spacelike surfaces where one can specify a
	field but not its normal derivative, but initial data on a
	null surface consists of fields without their normal derivatives.
	In particular, specifying the induced metric on all the null
	boundaries overdetermines the solution and so is not viable within
	this approach.
	Nevertheless it is possible to find appropriate data $\mfs$ for vacuum
	general relativity with two intersecting null boundaries, as we discuss in Sec.\ \ref{sec:nonunique}.

	\subsubsection{Definitions of symmetries and charges}
	\label{sec:dsc}
	
	Infinitesimal symmetries are represented by vector fields $\xi^a$ on
	$M$, with corresponding spacetime Lie derivatives $\lie_\xi$.  The
	corresponding vector fields on $\sls$ will be denoted by ${\hat \xi}$,
	and the corresponding \phase space Lie derivative will be denoted by
	$L_{\hat \xi}$.  These Lie derivatives can be written using Cartan's
	magic formula in terms of exterior derivatives and contractions,
	applied on spacetime $M$ and on \phase space $\sls$:
	\be
	\lie_\xi = i_\xi d + d i_\xi, \ \ \ \ \
	L_{\hat \xi} = I_{\hat \xi} \delta + \delta I_{\hat \xi}.
	\ee
	The spacetime vector field $\xi^a$ can depend on the
	fields $\phi$, and if so the quantity $\delta \xi^a$ is the one form on
	$\sls$ which takes values in vector fields on spacetime.  Its hatted
	version ${\hat {\delta \xi}}$ instead takes values in vector fields on $\sls$.
	The anomaly operator on differential forms on \phase space and
	spacetime is defined as
	\be
	\label{anom}
	\Delta_{\hat \xi} = L_{\hat \xi} - \lie_\xi - I_{{\hat {\delta \xi}}}.
	\ee
	The Lagrangian $L'$ and presymplectic potential $\theta'$ are not
	assumed to be covariant, but their anomalies (\ref{anom}) are assumed
	to be given in terms of quantities $b'$, $\lambda'$ as
	\begin{subequations}
		\label{eqn:blambdadef}
		\begin{eqnarray}
		\Delta_{\h\xi} L' &=& d\Delta_{\h\xi} b' \label{eqn:b'} \\
		\Delta_{\h\xi} \theta' &=& \Delta_{\h\xi}\delta b' + d\Delta_{\h\xi}
		\lambda'.
		\label{eqn:noncovtheta}
		\end{eqnarray}
	\end{subequations}

	We next discuss what quantities need to be defined in order to
	determine gravitational charges and fluxes, which do not quite
	coincide with the quantities needed for a variational principle.
	Suppose we are given a spatial slice $\Sigma$ whose boundary $\partial \Sigma$
	lies in one particular boundary component $\nb_{\bar j}$.
	Then the items needed are \cite{Chandrasekaran:2021vyu}
	\begin{itemize}
		\item A choice
		(\ref{decompos11}) of decomposition of the
		presymplectic potential on the boundary component $\nb_{\bar j}$
		only, that defines the boundary flux $\beom_{\bar j}$. 
		\item To allow for non-covariant quantities, we require a refinement
		of the definition (\ref{blde0}) of corner contributions, as follows.
		We choose decompositions of the quantities $\beta_j'$ on $\nb_j$ of
		the form\footnote{Taking a pullback of the
			relation (\ref{eqn:bldecomp}) to a corner, and using that $\lambda'$ is required
			to be continuous from one boundary component $\nb_j$ to the next
			yields that the quantities defined here are related to those defined
			in Eq.\ (\ref{blde0}) by
			\be
			\label{rels}
			{\hat c}_j = {\underline c}'_j - {\underline c}'_{j-1}, \ \ \ \
			{\hat \gamma}_j = {\underline \gamma}'_j - {\underline
				\gamma}'_{j-1}, \ \ \ \
			{\hat \cflx}_j = {\underline \cflx}_j - {\underline \cflx}_{j-1}.
			\ee
			Thus, specifying the decomposition (\ref{eqn:bldecomp}) is equivalent
			to  choosing splittings of the corner actions ${\hat c}_j$
			into contributions from the two adjacent boundaries $\nb_j$ and
			$\nb_{j-1}$, and choosing extensions of those contributions into
			the interior of the boundaries.}
		\beq \label{eqn:bldecomp}
		\beta_j' - {\underline \lambda}' = - \delta c_j'
		+d\gamma_j' + \cflx_j,
		\eeq
		in terms of corner terms $c_j'$, 
		exact terms $d \gamma_j'$ and corner fluxes $\cflx_j$.  Here the
		underline denotes a pullback to $\nb_j$. To define charges on $\nb_{\bar j}$,
		only the decomposition (\ref{eqn:bldecomp}) for $j = {\bar j}$ is needed.
		In our application of the formalism in this paper, the decomposition
		(\ref{eqn:bldecomp}) will not be needed since we exclude
		anomalies\footnote{\label{anom1}When there are no anomalies, the last term on the
			right hand side of Eq.\ (\ref{eqn:Hxi}) vanishes, and the charge depends only
			on $\ell'$ and $\beta'$.  See also the seventh column of Table 2 of
			CFSS, where modifications to the corner flux affect the charges only
			through an anomaly.}.

	\end{itemize}

	The last steps in obtaining the charge associated with a symmetry
	$\xi^a$ are as follows.  We define the
	Noether current $J_\xi'$ by 
	\beq
	J_\xi' = I_{\h\xi}\theta' - i_\xi L' - \Delta_{\h\xi} b',
	\label{eqn:noethercurrent}
	\eeq
	and choose a Noether charge $Q_\xi'$ that satisfies
	$J_\xi' = dQ_\xi'$.  
	Then, given a spatial slice $\Sigma$ whose boundary $\partial \Sigma$
	lies in $\nb_{\bar j}$, the charge is defined to be
	\be
	\label{hfinal}
	{\tilde H}_\xi = \int_{\partial \Sigma} {\tilde h}_\xi,
	\ee
	where [Eqs.\ (CFSS,2.30) and (CFSS,2.46)]
	\beq
	\label{eqn:Hxi}
	{\tilde h}_\xi = Q_\xi' + i_\xi \ell' - I_{\h\xi}\beta' - \Delta_{\hat \xi} c'.
	\eeq   
	Here the orientation of $\Sigma$ is the induced orientation given by regarding it as the boundary of its past, and the orientation of $\partial \Sigma$ is that induced from $\Sigma$.
	Also here and below for simplicity we drop the
	subscripts ${\bar j}$ that specify the particular boundary component.

	There are a variety of ambiguities or choices that must be made in this prescription
	that correspond to transformations that act on the variables in the
	formalism.  In Table 2 of CFSS we list eight such transformations.
	However, we show there that none of these
	modify the charges (\ref{hfinal}) or the symplectic form of
	the theory, except for those that modify the
	boundary flux $\beom$ or corner flux $\cflx$.
	Thus the charges are uniquely determined\footnote{This is
		also true of the algebra satisfied by the charges, by
		Eqs.\ (CFSS,2.48), (CFSS,3.6) and (CFSS,3.10).} by choices of
	$\beom$ and $\cflx$ on $\nb_{\bar j}$.  Moreover the corner flux
	$\cflx$ is only needed when there are anomalies, see footnote
	\ref{anom1} above.

	Choices of boundary and corner fluxes are also 
	closely related to choices of \phase space
	polarizations\footnote{We are using 
		the term polarization to be synonymous with a choice of symplectic
		potential.  This is a slight abuse of terminology since the definition
		of a choice of polarization  -- a foliation by
		Lagrangian submanifolds -- is equivalent to a choice of
		symplectic potential locally but not globally.},
	that is, one-forms $\Xi$ on $\sls$ whose variation 
	$\delta \Xi$ is the presymplectic form $\Omega$ of the theory
	\cite{Hopfmuller2017a,Odak:2021axr,Odak:2022ndm,Rignon-Bret:2023fjq}. 
	More specifically, suppose that we are given a spatial slice $\Sigma$ whose boundary
	$\partial \Sigma$ lies in $\nb_{\bar j}$.  Then a natural choice of
	polarization
	is determined by specifying those boundary fluxes and corner fluxes
	from the action variation (\ref{deltaS})
	that are to
	the past of $\Sigma$ and flipping the sign (see Appendix \ref{app:pol}):
	\be
	\label{polarizationanswer}
	\Xi = - \sum^\prime_j \int_{\ns_j} \beom_j
	- \sum^\prime_j \int_{\bb_j} {\hat \cflx}_j
	- \int_{\Delta \ns_{\bar j}} \beom_{\bar j}.
	\ee
	Here the primed sums include only the boundaries and corners to the
	past of $\Sigma$, and $\Delta \ns_{\bar j}$ is that portion of $\ns_{\bar j}$
	to the past of $\partial \Sigma$.
	When there are no anomalies, charges on $\nb_{\bar j}$ are determined
	by the boundary flux $\beom_{\bar j}$, which also determines the last
	term in the polarization (\ref{polarizationanswer}).

	Of the information that is needed to
	determine the boundary flux $\beom$ and thence the charges, a portion
	determines the presymplectic form $\Omega$, and within that
	presymplectic form a further portion determines the polarization $\Xi$.
	Given a presymplectic potential
	$\theta'$, the boundary flux is determined up to the boundary canonical
	transformations\footnote{See columns five and six of Table 2 of CFSS.}
	\be
	\beom \to \beom + \delta B - d
	\Lambda.
	\ee
	Here the $\Lambda$ freedom is not fixed by the action
	principle\footnote{The action principle (\ref{deltaS}) 
		determines the boundary and corner fluxes only up to the
		transformations $\beom_j \to \beom_j + d \Lambda_j, {\hat \cflx}_j \to
		{\underline \Lambda}_j - {\underline \Lambda}_{j-1}$.
		See however the discussion in the last paragraph of Sec. 2 of CFSS
		for a circumstance in which this freedom can be fixed by the action
		principle.}, and fixing 
	it determines the presymplectic form $\Omega$.
	The $B$ freedom is fixed by the choice of boundary action $\ell$, and
	it determines the choice of 
	polarization $\Xi$ for a given $\Omega$.

	Finally we remark that our prescription (\ref{eqn:Hxi}) for gravitational
	charges accommodates field-dependent symmetries $\xi^a = \xi^a[\phi]$,
	but has 
	the property that the charge
	${\tilde H}_\xi$
	for a given field configuration $\phi$ depends on the symmetry only through
	its value at $\phi$:
	\be
	{\tilde H}_\xi[ \phi, \xi^a[\ldots] ] = \left. {\tilde H}_\xi[ \phi, \xi^a] \right|_{\xi
		= \xi[\phi]}.
	\ee
	In particular it does not depend on the variation $\delta \xi^a[\phi]$.
	On the other hand the amount (CFSS,2.48) by which it fails to satisfy
	the expected relation for the generator of a symmetry does depend on
	$\delta \xi^a[\phi]$.  There are other prescriptions for defining
	charges for field dependent symmetries which have different
	properties
	\cite{Odak:2022ndm,Adami:2020amw,Adami:2020ugu,Adami:2021nnf,Adami:2021kvx}.
	In particular the interplay between anomalies and 
	field dependence has been explored in Ref. \cite{Odak:2022ndm}.  
	In this paper neither anomalies nor field dependence will arise.

	\subsection{Survey of possible general criteria for restricting
		choice of  boundary and corner fluxes}
	\label{sec:survey}
	
	We now turn to reviewing various criteria that can be used
	to restrict or determine the choices of boundary and corner fluxes
	$\beom$ and $\cflx$, and thus determine the algebra of charges.
	One of these criteria (Sec.\ \ref{sec:eps} below) is new.
	
	\subsubsection{Wald-Zoupas stationarity requirement}\label{sec:stat_general}
	
	In the original Wald-Zoupas construction, the main criterion used to
	choose the flux term $\mathcal{E}(\phi, \delta \phi)$
	was the requirement that it should vanish
	on stationary solutions $\phi$ for all variations $\delta
	\phi$, for some suitable notion of ``stationary''. This was partly
	motivated by the intuition that $\mathcal{E}(\phi, \lie_{\xi}\phi)$
	corresponds to the flux associated to the symmetry $\xi^a$,
	and on a stationary solution there ought not to be any
	flux. In \cite{WZ}, which specialized to null infinity, a solution was
	considered to be stationary if it admitted a global timelike Killing
	vector field. In \cite{CFP}, in the context of general null surfaces, 
	the definition of stationary was relaxed to the requirement that there
	exist a vector field $\tau^a$ which is tangent to the boundary and
	which satisfies the Killing equation on the boundary and to first order in
	deviations off the boundary.  In both of these contexts the
        requirement successfully determined a unique flux definition.

	We will show explicitly in Sec.\ \ref{sec:stationarity} below that it is impossible to
	satisfy the Wald-Zoupas stationarity requirement in the context of the
	maximal or restricted \phase spaces defined in this paper.
        Thus it is not possible to use it to determine a unique
        decomposition.
        The failure of the stationarity requirement occurs because
        our context admits symmetries which
	themselves are time dependent in a certain sense, and therefore one would not expect
	vanishing fluxes for these symmetries even on stationary backgrounds.
	Therefore we do not impose any requirement related to
	stationarity in our analysis.

	\subsubsection{Boundary and corner fluxes of Dirichlet form}
	
	A second criterion that can be used is to impose that the boundary and
	corner fluxes $\beom$ and $\cflx$ have Dirichlet form, that is, that
	they are expressible in the form $\beom = P \delta Q$ and $\cflx = p
	\delta q$, where the
	variables $Q$ and $q$ are intrinsic quantities on the boundaries and
	the corners \cite{Chandrasekaran:2020wwn,Chandrasekaran:2021vyu}.  For
	example in general relativity at a null boundary, $Q$ consists of the
	induced 
	metric and the contravariant components of the normal.
	Defining a field configuration subspace ${\bar \fs}$ for the action
	principle by imposing Dirichlet boundary conditions $Q = $ constant
	and $q =$ constant then yields
	vanishing fluxes $\beom= \cflx = 0$ and a well posed variational principle.
	
	Chandrasekaran and Speranza \cite{Chandrasekaran:2020wwn} gave several
	arguments for the naturalness of imposing that the fluxes have
	Dirichlet form.  However they also showed that for general relativity
	at null boundaries, the Dirichlet requirement is incompatible with the
	vanishing of anomalies (see Sec. \ref{sec:anomalies} below).  In this
	paper we will not impose the Dirichlet requirement, but we will impose
	the vanishing of anomalies.

	\subsubsection{Absence of anomalies}
	\label{sec:anomalies}
	
	Another general criteria that has often been imposed is covariance or lack of anomalies \cite{Wald:1999wa,Compere:2020lrt,Harlow:2019yfa,Freidel:2020xyx,Chandrasekaran:2020wwn,Margalef-Bentabol:2020teu,freidel2021weyl,Freidel:2021cjp,Ciambelli:2022vot,Chandrasekaran:2021vyu}. This means that the decomposition should
	not depend on any background structures or fields, so that the action of diffeomorphisms
	on spacetime (which transforms all fields and structures) should
	coincide with their action on \phase space (which transforms all the
	dynamical fields) \cite{Harlow:2019yfa}.  An exception is background
	structures that enter into the definition of the field configuration
	space, since they will be preserved by definition by the symmetries
	which preserve the configuration space. 
	For general differential forms $\omega$ on \phase space $\sls$, the
	condition for covariance is just that the anomaly operator (\ref{anom})
	acting on the differential form should vanish:
	\be
	\Delta_{\hat \xi} \omega =0.
	\ee
	For quantities $h_\xi$ that
	depend on a symmetry $\xi^a$, the appropriate condition is instead (see
	footnote 7 of CFSS) that
	\be
	\Delta_{\hat \xi} h_\zeta - h_{\Delta_{\hat \xi}\zeta} =0.
	\ee

	There are some circumstances, particularly involving asymptotic
	boundaries, where it is not possible to avoid anomalies, for example
	the famous analysis by Brown and Henneaux on $\text{AdS}_3$
	\cite{1986CMaPh.104..207B}. Our viewpoint is that anomalies should be
	avoided whenever it is possible to do so.  Accordingly in this paper
	we impose the lack of anomalies as a requirement on our analysis.
	The relevant background quantities on which dependence
	is forbidden will be a choice of a particular null normal for the null
	surface (since the \phase space depends only on the equivalence class
	of null normals under rescalings), or a choice of a particular
	auxiliary null vector $n^a$ (see Sec.\ \ref{sec:eps} below for more details).

	As is well known, the signature of the presence of anomalies is that the algebra of
	charges differs from the Lie algebra of diffeomorphisms on spacetime
	\cite{1986CMaPh.104..207B,Haco:2018ske,Hawking:2016sgy},
	demonstrated most generally by Chandrasekaran and Speranza \cite{Chandrasekaran:2020wwn}.
	There are two quantities that govern the non-covariance in the formalism of
	Sec.\ \ref{covr}, which following
	Ref.\ \cite{Odak:2022ndm} we denote by
	\begin{subequations}
		\begin{eqnarray}
		a_{\hat \xi} &=& \Delta_{\hat \xi} ( \ell' + b' + d c'),\\
		A_{\hat \xi} &=& - \Delta_{\hat \xi} \cflx.
		\end{eqnarray}
	\end{subequations}
	Then the anomalies in the flux, charges and charge algebra are given
	by \cite{Chandrasekaran:2020wwn,Chandrasekaran:2021vyu,Odak:2022ndm}
	\bes
	\bea
	\label{fluxanom}
	\Delta_{\hat \xi} \beom &=& \delta a_{\hat \xi} - a_{\hat {\delta
			\xi}} + d A_{\hat \xi}, \\
	\label{chanom}
	\Delta_{\hat \zeta} {\tilde h}_\xi - {\tilde h}_{\Delta_{\hat \zeta}
		\xi} &=& i_\xi a_{\hat \zeta} + I_{\hat \xi} A_{\hat \zeta}, \\
	\{\tilde H_\xi, \tilde H_\zeta\} &=&  \tilde H_{\brmod{\xi}{\zeta}} + \tilde
	K_{\xi,\zeta} \label{eqn:tilHtilH},
	\eea
	\ees
	where
	\be
	\tilde K_{\xi,\zeta} = \int_{\partial\Sigma}\left(
	i_\xi a_\h\zeta - i_\zeta a_\h\xi\right) \label{eqn:tilK}
	\ee
	and ${\brmod{\xi}{\zeta}}$ is the adjusted Lie bracket defined for
	field-dependent symmetry vector fields by (CFSS,2.3).  
	Here Eq.\ (\ref{fluxanom}) follows from Eqs.\ (CFSS,2.12), (CFSS,2.43) and (CFSS,2.7b),
	while Eq.\ (\ref{chanom}) is Eq.\ (CFSS,B.12).

	\subsubsection{Impose that the algebra of charges does not a have a central extension}
	\label{sec:anomalies1}
	
	Another criterion that is sometimes imposed to restrict choices in the formalism
	is to require that the algebra of charges reproduce the Lie algebra of
	diffeomorphisms on spacetime, without any central extension.   As we
	have seen above, this follows from the absence of anomalies.

	\subsubsection{Charges associated with a \phase subspace should not
		depend on the choice of full \phase space}  
	\label{sec:eps}
	
	The charges (\ref{hfinal}) defined in Sec.\ \ref{covr} above are
	associated with a cross section $\partial \Sigma$ of one of the
	components of the boundary.  One can associate with this cross section
	a subspace $\sls_{\partial \Sigma}$ of the \phase $\sls$.  For example, if
	$\partial \Sigma$ is a cut of future null infinity $\scrip$, then the subspace
	$\sls_{\partial \Sigma}$ is
	the set of metrics determined by the data on $\scrip$ to the future of
	$\partial \Sigma$.
	Although the charges (\ref{hfinal}) are defined and computed within the
	full \phase space $\sls$, their physical interpretation is
	loosely\footnote{This interpretation is only heuristic and not precise
		since the action of general symmetries on $\sls_{\partial \Sigma}$ is not
		defined; only those symmetries which preserve $\partial \Sigma$ can be
		symmetries on the subspace.} generators of symmetries on
	$\sls_{\partial \Sigma}$.  Also, since the charges can be expressed in terms of
	fields evaluated on $\partial \Sigma$ from Eq.\ (\ref{hfinal}), they
	can be interpreted as functionals on $\sls_{\partial \Sigma}$.
	
	It can occur that the subspace $\sls_{\partial \Sigma}$ can be embedded in two
	different \phase spaces $\sls$ and $\sls'$.  This arises in
	our \phase space definitions in Sec.\ \ref{sec:phasespaces} below,
	which are defined in terms of data specified on a closed spacelike
	2-surface $S_0$ and on two null surfaces $\Sp$ and $\Sm$ to its future 
	which intersect on $S_0$.  Now suppose that we specify a cross section
	$\partial \Sigma$ of $\Sp$ and associated sub \phase space
	$\sls_{\partial \Sigma}$. One can choose a different 2-surface $S_0'$ on $\Sp$
	between $S_0$ and $\partial \Sigma$, a null surface $\Sm'$ that
	intersects $\Sp$ at $S_0'$, and a corresponding \phase space
	$\sls'$.
	Thus the space $\sls_{\partial \Sigma}$
	can be viewed as a subspace of either $\sls$ or $\sls'$.
	
	In such contexts we suggest adopting as a principle that the charges
	computed starting from $\sls$ or from $\sls'$ should coincide.
	We will apply this principle to \phase spaces of two intersecting
	null surfaces in Sec.\ \ref{sec:pr} below, where it will yield
	nontrivial constraints on the choice of boundary and corner fluxes.

	\subsubsection{Vanishing flux for all boundaries in Minkowski spacetime}
	
	There is another criterion that has been used to inform and restrict
	definitions of the field configuration space $\fs$, rather than the
	choice of decomposition.
	In Ref. \cite{Ashtekar:2021kqj}, Ashtekar, Khera, Kolanowski and
	Lewandowski (AKKL) study the charges and fluxes of symmetries at null
	boundaries.
        Their analysis is similar to the study of Ref.\ \cite{CFP},
        except that they impose the strong restriction that  
	the null boundary must be a non-expanding horizon, that is, one for which
	the expansion and shear vanish identically. Their reason for this
	restriction is that if arbitrary null boundaries are allowed, then
	there are boundary surfaces in Minkowski spacetime that have nonvanishing fluxes.
	For example, a null cone in Minkowski has non-vanishing expansion, and
	thus nonvanishing fluxes by Eq.\ (\ref{fluxnew1}) below. 
	AKKL argue that nonvanishing fluxes in Minkowski
	do not make sense physically, and that restricting to non-expanding
	horizons is the appropriate way to avoid this issue.

	We argue here for an alternative philosophy, that nonvanishing fluxes
	in Minkowski are not always unphysical.
        There do exist contexts in which it is possible to
        consistently impose that fluxes in Minkowski should vanish,
        just as it is sometimes possible to impose the Wald-Zoupas requirement
        that fluxes should vanish for perturbations of stationary
        boundaries.  However, as discussed in
        Sec.\ \ref{sec:stat_general} above, some phase space
        definitions allow ``time dependent symmetries'' for which it
        is not natural or possible to impose that fluxes should vanish
        on stationary backgrounds.  In such contexts we
        argue that similarly it is not natural to impose vanishing of
        fluxes in Minkowski spacetime.

	\subsection{Charge non-uniqueness reflects various possible choices of polarizations}
	\label{sec:cnu}
	
	In most previous works, such as \cite{WZ, Chandrasekaran:2021vyu,
		freidel2021weyl}, the non-uniqueness in the choice of decomposition
	(\ref{decompos11}) of the symplectic potential is regarded as a
	problem. 
	Our view is that the non-uniqueness instead reflects different choices
	of polarization (see Sec.\ \ref{sec:gp} above), as suggested in
	Refs.\ \cite{Odak:2021axr,Odak:2022ndm,Rignon-Bret:2023fjq,Freidel:2020xyx,Freidel:2020svx,Freidel:2020ayo}.    
        In other words, the freedom in decomposition is not something which
	one needs to remedy, but rather reflects different types of boundary
	degrees of freedom, in much the same way
	as we have different polarizations in ordinary Newtonian mechanics.  
	
	The different choices correspond to different possible variational
	principles to describe the dynamics. They map onto different subspaces
	of the \phase space that are obtained when the natural boundary
	conditions associated with the polarizations are imposed
	(for example Dirichlet, Neumann, or York boundary conditions).
	Hence the different choices do not really correspond to ambiguities.
	These points are illustrated by our treatment of two intersecting null
 	surfaces in the remainder of this paper: the charges we
        compute in Sec.\ \ref{sec:decomp_charge} depend on some free
        parameters, and the polarizations and variational principles
        that correspond to those parameter choices are described in
        Sec.\ \ref{sec:nonunique}.

	\section{Initial value formulation of vacuum general relativity with two intersecting null surfaces}
	\label{sec:initialvalue}
	
	In this section we review the null initial value formulation for
	vacuum general relativity for two intersecting null surfaces, based on the
	treatments of Sachs \cite{1962JMP.....3..908S},
        Rendall \cite{70785209-fc5c-312c-9bd2-7c68ce777f34}, Hayward
	\cite{1993CQGra..10..773H} and Reisenberger \cite{Reisenberger:2007pq}.
        We modify these treatments slightly to
	express the result in terms of equivalence classes.
	This initial value formulation will be used to motivate our definitions
	of field configuration space in Sec.\ \ref{sec:phasespaces} below.
	An alternative treatment of a covariant double null initial value
	formulation by Mars and S\'anchez-P\'erez
	which makes use of an auxiliary null normal
	can be found in Refs.\ \cite{Mars:2022gsa,Mars:2023hty}.
	
	The context is that we consider two null surfaces $\Sp$ and $\Sm$ in
	spacetime which intersect in a closed spacelike 2-surface $S_0$, and
	which lie to the future of $S_0$.  We
	want to define the appropriate initial data on these three manifolds
	that will uniquely determine a solution of the vacuum Einstein
	equation to the future up to gauge.  (The analogous quantities for a
	spacelike hypersurface are the specification of the induced 3-metric
	and extrinsic curvature on the hypersurface.)  We will start by
	defining the appropriate set of initial data, then explain its
	interpretation and motivation, and finally cite a theorem on
	uniqueness.
	
	Throughout this paper we use the definitions
	of and notations for geometric quantities defined on a null surface that are reviewed in
	Appendix \ref{app:nullreview}.

	\subsection{Definition of initial data structure}
	\label{sec:ids}

	We assume that we are given a closed 2-manifold $S_0$, and two
	3-manifolds with boundary $\Sp$ and $\Sm$ which intersect at $S_0$, whose
	boundaries are $\partial \Sp = S_0$ and $\partial \Sm = S_0$.
	We denote tensors on $S_0$ with indices $A,B, \ldots$,
	tensors on $\Sp$ with indices $i, j, \ldots$, and tensors on $\Sm$
	with indices $i', j', \ldots$.
	
	We define an {\it initial data representative} on $(S_0, \Sp, \Sm)$ to
	be a specification of the following fields:
	\begin{itemize}
		\item A 2-metric $q_{AB}$ on $S_0$.
		\item A one-form ${\bar \omega}_A$ on $S_0$.
		\item Scalar fields $\Theta_\ps$, $\Theta_\ms$ and $m$ on $S_0$.
		\item A nowhere-vanishing vector field $\ell^i_\ps$ on $\Sp$ which is
		nowhere tangent to $S_0$.
		\item A scalar field $\kappa_\ps$ on $\Sp$.
		\item A conformal equivalence class $[q^\ps_{ij}]$ of metrics on $\Sp$ defined as
		follows.  We consider symmetric tensors $q^\ps_{ij}$ on $\Sp$ of signature $(0,+,+)$ whose
		pullbacks to $S_0$ coincide with $q_{AB}$ and which satisfy
		\be
		q^\ps_{ij} \ell_\ps^j=0.
		\label{mr}
		\ee
		We denote by $[ q^\ps_{ij}]$ the equivalence class under conformal rescalings
		\be
		q^\ps_{ij} \to e^{\vartheta_\ps} q^\ps_{ij}
		\label{crs}
		\ee
		for smooth functions $\vartheta_\ps$ on $\Sp$ with $\vartheta_\ps
		=0$ on $S_0$.
		\item Analogous quantities $\ell^{i'}_\ms$, $[ q^\ms_{i'j'}]$ and
		$\kappa_\ms$ on $\Sm$.
	\end{itemize}
	Two representatives $(q_{AB}, {\bar \omega}_A, m,
	\Theta_\ps, \Theta_\ms, \ell^i_\ps, \kappa_\ps, [q^\ps_{ij}],
	\ell^{i'}_\ms, \kappa_\ms, [q^\ms_{i'j'}]) $ and \\
	\noindent
	$({\tilde q}_{AB}, {\tilde {\bar \omega}}_A, {\tilde m},
	{\tilde \Theta}_\ps, {\tilde \Theta}_\ms, {\tilde \ell}^i_\ps, {\tilde
		\kappa}_\ps, [{\tilde q}^\ps_{ij}],
	{\tilde \ell}^{i'}_\ms, {\tilde \kappa}_\ms, [{\tilde q}^\ms_{i'j'}])$
	will be said to be equivalent if there exist smooth functions
	$\sigma_\ps$ on $\Sp$ 
	and $\sigma_\ms$ on $\Sm$ for which
	\begin{subequations}
		\label{2rescalings}
		\begin{eqnarray}
		{\tilde q}_{AB} & = & q_{AB}, \\
		\label{rescaleomega}
		{\tilde {\bar \omega}}_A &=& {\bar \omega}_A + \frac{1}{2} D_A (\sigma_\ps -
		\sigma_\ms), \\
		\label{rescalem}
		{\tilde m} &=& m + \sigma_\ps + \sigma_\ms, \\
		{\tilde \Theta}_\ps &=& e^{\sigma_\ps} \Theta_\ps, \\
		{\tilde \Theta}_\ms &=& e^{\sigma_\ms} \Theta_\ms, \\
		\label{rescaleellplus}
		{\tilde \ell}^i_\ps &=& e^{\sigma_\ps} \ell^i_\ps, \\
		\label{rescalekappaplus}
		{\tilde \kappa}_\ps &=& e^{\sigma_\ps} ( \kappa_\ps + \lie_{{\vec \ell}_\ps} \sigma_\ps), \\
		{[}{\tilde q}^\ps_{ij}] &=& [ q^\ps_{ij}], \\
		\label{rescaleellminus}
		{\tilde \ell}^{i'}_\ms &=& e^{\sigma_\ms} \ell^{i'}_\ms, \\
		\label{rescalekappaminus}      
		{\tilde \kappa}_\ms &=& e^{\sigma_\ms} ( \kappa_\ms + \lie_{{\vec
				\ell}_\ms} \sigma_\ms), \\
		{[}{\tilde q}^\ms_{i'j'}] &=& [ q^\ms_{i'j'}],
		\end{eqnarray}
	\end{subequations}
	where $D_A$ is any derivative operator on $S_0$.
	Finally we define an initial data structure
	\be
	\label{ids1}
	\mfi = \left[q_{AB}, {\bar \omega}_A, m,
	\Theta_\ps, \Theta_\ms, \ell^i_\ps, \kappa_\ps, [q^\ps_{ij}],
	\ell^{i'}_\ms, \kappa_\ms, [q^\ms_{i'j'}] \right]
	\ee
	to be an equivalence class under this equivalence relation, obtained
	from a representative $(q_{AB}, {\bar \omega}_A, m,
	\Theta_\ps, \Theta_\ms, \ell^i_\ps, \kappa_\ps, [q^\ps_{ij}],
	\ell^{i'}_\ms, \kappa_\ms, [q^\ms_{i'j'}])$.

	\subsection{Interpretation and motivation}
	
	Suppose that we are given a
	spacetime $(M,g_{ab})$ and that $\Sp$ and $\Sm$ are two null surfaces
	in $M$ which intersect in a spacelike 2-surface $S_0$.  Then the
	metric $g_{ab}$ induces an initial data structure $\mfi =
	\mfi[g_{ab}]$ as follows.
	We pick choices of future-directed normal
	covectors on $\Sp$ and $\Sm$, and raise the index with the metric to obtain
	normal null vectors ${\vec \ell}_\ps$ on $\Sp$ and ${\vec \ell}_\ms$ on $\Sm$.  These are
	naturally interpreted as intrinsic vectors $\ell^i_\ps$ and $\ell^{i'}_\ms$,
	and are unique
	up to the rescalings (\ref{rescaleellplus}) and (\ref{rescaleellminus}).
	We define $\kappa_\ps$ and $\kappa_\ms$ to be the associated
	non-affinities defined by Eq.\ (\ref{kappadef}).
	We compute from $g_{ab}$ the induced metrics $q^\ps_{ij}$ and
	$q^\ms_{i'j'}$ on $\Sp$ and $\Sm$ and take the equivalence classes
	$[q^\ps_{ij}]$ and $[q^\ms_{i'j'}]$ under conformal rescaling.
	We define on the 2-surface $S_0$ the fields $\Theta_\ps$ and
	$\Theta_\ms$ to be the expansions given by (\ref{Kdecompos}).
	On $S_0$ we also define $q_{AB}$ to be the induced metric, and $m$ to be
	given by
	\be
	{\vec \ell}_\ps \cdot {\vec \ell}_{\ms} = - e^{m}.
	\label{normdef}
	\ee
	Finally we define the H\'a\'ji\^{c}ek one-form ${\bar \omega}_A$ to be the
	pullback to $S_0$ of the one-form \cite{Gourgoulhon:2005ng,Gourgoulhon:2005ch}
	\be
	\frac{1}{2} e^{-m} (\ell_\ps^b \nabla_a \ell_{\ms\,b} -
	\ell_\ms^b \nabla_a \ell_{\ps\,b}).
	\label{rotoneform}
	\ee
	Because our vector fields ${\vec \ell}_\ps$ and ${\vec \ell}_\ms$
	are defined only on $\Sp$ and $\Sm$ respectively (and $S_0$), 
	the one-form (\ref{rotoneform}) is not uniquely defined, but its
	pullback ${\bar \omega}_A$ which involves
	only derivatives along $S_0$ is well defined.
	The H\'a\'ji\^{c}ek one-form ${\bar \omega}_A$ is related to the
	pullback $\omega_A$ to $S_0$ of the rotation one-form $\omega_i$
	defined in Eq.\ (\ref{twist}) by
	\be
	{\bar \omega}_A = \omega_A - \frac{1}{2} D_A m,
	\label{Hajicek}
	\ee
	from Eqs.\ (\ref{normdef}), (CFP,3.17) and (CFP,3.32).

	The various fields we have defined transform under the rescalings
	of the normals as given by Eqs.\ (\ref{2rescalings})
	\cite{Gourgoulhon:2005ng,CFP}.
	Therefore the equivalence class
	\be
	\left[q_{AB}, {\bar \omega}_A, m,
	\Theta_\ps, \Theta_\ms, \ell^i_\ps, \kappa_\ps, [q^\ps_{ij}],
	\ell^{i'}_\ms, \kappa_\ms, [q^\ms_{i'j'}]\right]
	\ee
	is independent of our choice of these normals, and is the
	initial data structure $\mfi = \mfi[g_{ab}]$ induced by
	the metric.
	
	\subsection{Statement of theorem}
	\label{sec:theorem}
	
	We can now state the main result:
	
	\bigskip
	\noindent
	{\bf Theorem:} Suppose we are given an initial data structure
	$\mfi$ on $(S_0, S_\ps, S_\ms)$.  Then there exists a vacuum solution  $(M,g_{ab})$
	and embeddings of $S_0$, $S_\ps$ and $S_\ms$ into $M$ as
	spacelike and null surfaces, for which the metric is consistent with the
	initial data structure, $\mfi[g_{ab}] = \mfi$.  The
	metric is unique up to gauge\footnote{and up to restriction to
		the intersection of domains of definition.}.

	In Appendix \ref{app:ivf} we show that this theorem follows from
	the harmonic gauge, double null initial value formulation of
	Rendall \cite{70785209-fc5c-312c-9bd2-7c68ce777f34}, translated into
	covariant language.  It should also be possible to derive the theorem
	from the more general covariant double null initial value formalism
	of Mars and S\'anchez-P\'erez \cite{Mars:2022gsa,Mars:2023hty}.
	
	\subsection{Alternative representation of initial data structure}
	\label{sec:idsalternative}
	
	We next define alternative coordinates on the space of initial data
	structures that will be useful in the computations in the remainder of the paper.
	
	The metric $q_{AB}$ on $S_0$
	determines a volume form $\mu_{AB}$ on $S_0$ (up to sign).  This determines a
	unique two form ${\bar \mu}^\ps_{ij}$ on $\Sp$ restricted to $S_0$ whose pullback
	to $S_0$ coincides with $\mu_{AB}$, and which satisfies ${\bar \mu}^\ps_{ij}
	\ell_\ps^j=0$.  We now define ${\bar \mu}^\ps_{ij}$ along all of $\Sp$ by
	demanding that it be Lie transported,
	\be
	\label{mu0def}
	\lie_{{\vec \ell}_\ps} {\bar \mu}^\ps _{ij} = 0.
	\ee
	Finally, given an induced metric $q_{ij}^\ps$ on $\Sp$ we denote 
	by $\mu^\ps_{ij}$ the associated volume form which satisfies $\mu^\ps_{ij} \ell^j_\ps =0$, and we define $\nu_\ps$ on $\Sp$ by
	\be
	\mu^\ps_{ij} = e^{\nu_\ps} {\bar \mu}^\ps_{ij}.
	\ee
	Thus $\nu_\ps$ measures the ratio between the physical volume form and the Lie transported one. It follows from the definition that
	\be
	\left. \nu_\ps  \right|_{S_0} = 0.
	\label{nuS0}
	\ee
	We now define the rescaled metric
	\be
	{\bar q}^\ps_{ij} = e^{-\nu_\ps} q^\ps_{ij}.
	\label{q0def}
	\ee
	This rescaled metric parameterizes the conformal equivalence class
	$[q^\ps_{ij}]$, since it is invariant under the conformal
	transformations (\ref{crs}).  Thus we can use ${\bar q}^\ps_{ij}$ and
	${\bar q}^\ms_{i'j'}$ in the initial data structure (\ref{ids1}) instead of the
	conformal equivalence classes.

	Given the rescaled metric ${\bar q}^\ps_{ij}$ and the remaining
	elements of a representative of the initial data structure, we can
	obtain the remaining quantities on the null surface $\Sp$ as follows.
	We solve the Raychaudhuri equation (\ref{grelations1e}) specialized to
	$R_{ab}=0$ to obtain $\Theta_\ps$ on $\Sp$ given its initial data on $S_0$.
	Note that the shear squared source term in the Raychaudhuri equation
	can be expressed in terms of ${\bar q}^\ps_{ij}$ and its derivative,
	since from Eqs.\ (\ref{q0def}) and (\ref{equiv1}) the factors of $e^{\nu_\ps}$ cancel out:
	\be
	\sigma^\ps_{ij} \sigma^\ps_{kl} q_\ps^{ik} q_\ps^{jl} = \frac{1}{4} 
	\lie_{{\vec \ell}_\ps} {\bar q}^\ps_{ij} \,
	\lie_{{\vec \ell}_\ps} {\bar q}^\ps_{kl} \,
	{\bar q}^{\ps\, ik} {\bar q}^{\ps\, jl}.
	\ee
	Now taking a Lie derivative with respect to ${\vec \ell}_\ps$ of
	Eq.\ (\ref{q0def}), comparing to Eqs.\ (\ref{Kdecompos}) and
	(\ref{grelations1b}) and making use of Eq.\ (\ref{mu0def}) yields
	\bes
	\label{equiv}
	\bea
	\label{equiv0}
	\Theta_\ps&=& \lie_{{\vec \ell}_\ps} \nu_\ps, \\
	\label{equiv1}
	\sigma^\ps_{ij} &=& \frac{1}{2} e^{\nu_\ps} \lie_{{\vec \ell}_\ps}
	{\bar q}^\ps_{ij}.
	\eea
	\ees
	Solving Eq.\ (\ref{equiv0}) using the initial data (\ref{nuS0})
	now yields $\nu_\ps$, and inserting this into Eqs.
	(\ref{equiv1}) and (\ref{q0def}) yields
	the shear $\sigma^\ps_{ij}$ and
	the induced metric $q^\ps_{ij}$.

	\subsection{Definition of extended initial data structure}
	\label{sec:initialdataextended}
	
	Initial data structures of the form (\ref{ids1}) determine a
	vacuum spacetime $(M,g_{ab})$ to the future, which is unique up to
	diffeomorphisms $\psi : M \to M$ which preserve the boundaries $\Sp$
	and $\Sm$, under which the metric transforms as
	$g_{ab} \to \psi_* g_{ab}$. 
	For any such
	diffeomorphism we define the boundary diffeomorphisms induced on the
	boundaries as
	\be
	\label{bds}
	\varphi_\ps = \left. \psi \right|_{S_\ps}, \ \ \ \
	\varphi_\ms = \left. \psi \right|_{S_\ms}, \ \ \ \
	\varphi_0 = \left. \psi \right|_{S_0}.
	\ee
	We can also define additional near-boundary pieces of $\psi$ as follows.
	Since $\psi$ preserves $S_\ps$ and $S_\ms$, the pullback of the
	normal one-forms must effect rescalings of these normals:
	\be
	\label{gammadef}
	\psi_* \ell_{\ps\,a} = e^{\gamma_\ps} \ell_{\ps\,a}, \ \ \ \psi_*
	\ell_{\ms\,a} = e^{\gamma_\ms} \ell_{\ms\,a},
	\ee
	which defines the functions $\gamma_\ps = \gamma_\ps[\psi,
	\ell_{\ps\,a}]$ and $\gamma_\ms = \gamma_\ms[\psi, \ell_{\ms\,a}]$ on $S_\ps$
	and $S_\ms$.  Under rescalings
	\be
	\label{rescalings2}
	\ell_{\ps\,a} \to e^{\sigma_\ps} \ell_{\ps\,a}, \ \ \ \ \ell_{\ms\,a}
	\to e^{\sigma_\ms} \ell_{\ms\,a}
	\ee
	of
	the normal one-forms,
	these functions transform as
	\be
	\label{gammalaw}
	\gamma_\ps \to \gamma_\ps + \varphi_{\ps\,*} \sigma_\ps - \sigma_\ps,
	\ \ \ \ 
	\gamma_\ms \to \gamma_\ms + \varphi_{\ms\,*} \sigma_\ms - \sigma_\ms.
	\ee
	
	Specifying the initial data (\ref{ids1}) will restrict the boundary
	diffeomorphisms (\ref{bds}) to be the identity\footnote{Except when boundary diffeomorphisms are symmetries of the initial data.}, but will not restrict the
	pieces $\gamma_\ps$ and $\gamma_\ps$ of the bulk diffeomorphism
	$\psi$.  Those pieces are analogous to the lapse function in the
	initial value formulation of general relativity on spacelike hypersurfaces.
	However, we will see in Sec.\ \ref{sec:chargeder} below that the degrees of freedom
	corresponding to $\gamma_\ps$ and $\gamma_\ms$ are not degeneracy
	directions of the presymplectic form, and therefore must be included
	in the gravitational phase space.  For this reason it is useful to
	consider an extended kind of initial data structure which includes the
	normal covectors.  The definition is
	\be
	\label{ids2}
	\mfh = \left[q_{AB}, {\bar \omega}_A, m,
	\Theta_\ps, \Theta_\ms, \ell_{\ps\,a}, \ell^i_\ps, \kappa_\ps, [q^\ps_{ij}],
	\ell_{\ms\,a}, \ell^{i'}_\ms, \kappa_\ms, [q^\ms_{i'j'}] \right],
	\ee
	where the normal covectors are constrained to satisfy
	\be
	\ell_{\ps\,a} \ell_\ps^a = \ell_{\ms\,a} \ell_\ms^a = 0
	\ee
	and
	\be
	\left. \ell_{\ps\,a} \ell_\ms^a \right|_{S_0} = \left. \ell_{\ms\,a} \ell_\ps^a \right|_{S_0} = -e^m,
	\label{con22}
	\ee
	{\it cf.}\ Eq.\ (\ref{normdef}).  Here the equivalence relation is given by Eq.\ (\ref{2rescalings})
	augmented by the relations
	${\tilde \ell}_{\ps\,a} = e^{\sigma_\ps} \ell_{\ps\,a}$, ${\tilde \ell}_{\ms\,a} =
	e^{\sigma_\ms} \ell_{\ms\,a}$.  Note that $\mfh$ is no longer an
	intrinsic geometric structure since the normal covectors are not
	defined as tensor fields on the boundaries $\Sp$ and $\Sm$.  It is
	defined in the context of a manifold $M$ with boundary components
	$\Sp$ and $\Sm$.
	Specifying an extended initial data structure of the form (\ref{ids2})
	again determines a metric $g_{ab}$ on $M$ which is unique up to bulk diffeomorphisms
	$\psi$, but now the pieces (\ref{bds}) and (\ref{gammadef}) of the bulk
	diffeomorphism are all constrained to be trivial ($\gamma_\ps =
	\gamma_\ms =0$).
	Also, as in Sec.\ \ref{sec:ids}, specifying a metric on $M$ for which $\Sp$
	and $\Sm$ are null and $S_0$ is spacelike determines a unique extended initial data structure
	which we denote by
	\be
	\label{frakhgab}
	\mfh = \mfh[g_{ab}].
	\ee

	\section{Phase spaces and symmetry groups of general relativity with two
		intersecting null boundaries} 
	\label{sec:phasespaces}
	
	In this section we discuss various definitions of 
        phase spaces for
	general relativity at null boundaries.  We extend and modify the
	treatment of Chandrasekaran, Flanagan and Prabhu \cite{CFP}
	(henceforth CFP) in a number of ways:
	
	\begin{itemize}
		\item CFP considered a single null surface, whereas here we instead
		consider two null surfaces $\Sp$ and $\Sm$ that lie to the future of
		the spacelike two-surface $S_0$ where they intersect.  This
		modification is needed so that we can precisely define a
		gravitational phase space.  A single null surface does not provide
		an initial value formulation, and the phase space of two
		intersecting null surfaces does not factor into two symplectically
		orthogonal subspaces, so it is not possible to cleanly define the
		portion of the phase space associated with a single null
		surface\cite{Freidel:2020xyx,Freidel:2020svx}.

		\item CFP constructed a phase space by fixing some degrees of freedom
		which are not degeneracy directions of the presymplectic form. Here
		we include those degrees of freedom, which include (i) allowing nonzero
		variations of the inaffinities $\kappa$, discussed in Appendix H of
		CFP; and (ii) allowing nonzero variations of the contravariant
		normal $\ell^i$, discussed in Refs.\ \cite{Hopfmuller2017a,Chandrasekaran:2020wwn}.
		
		\item The symmetry group of CFP was a subgroup of the diffeomorphism
		group of the null surface.  Here we will find a symmetry group which instead
		is larger than this boundary diffeomorphism group, since the pieces (\ref{gammadef}) of the
		bulk diffeomorphisms will be included in the symmetry transformations.
		
	\end{itemize}

We shall define three different \phase spaces: a maximal horizon \phase
space $\sls$ in Sec.\ \ref{dfcs}, a restricted horizon \phase space
$\sls_\mfp$ in Sec.\ \ref{sec:rhps} in which the direction of the
contravariant normal is fixed, and a smaller \phase space $\sls_\mfq$
in Sec.\ \ref{sec:rhps} which the
magnitude as well as the direction of the contravariant normal are
fixed, previously defined in Appendix H of CFP.
The corresponding phase spaces will be denoted $\slp$, $\slp_\mfp$
and $\slp_\mfq$.  The symmetry groups associated with these phase
spaces are derived in Secs.\ \ref{sec:hwd} and \ref{sec:group1}.  Finally in
Sec.\ \ref{halfsidedboosts} we discuss extensions of the three \phase spaces
associated with half-sided boosts.

	\subsection{Maximal horizon phase space definition}
	\label{dfcs}

	We fix a manifold $M$ with boundary components $S_\ps$ and $S_\ms$ which
	intersect in a two dimensional surface $S_0$.  We define the
	field configuration space $\fs$ to be
	the set of smooth metrics $g_{ab}$ on $M$ for which $S_\ps$ and $S_\ms$ are
	null and $S_0$ is spacelike.  Metrics in $\fs$ thus obey the
	constraints
	\be
	g^{ab} \ell^\ps_a \ell^\ps_b = 0, \ \ \ \ \
	g^{ab} \ell^\ms_a \ell^\ms_b = 0,
	\label{bc00}
	\ee
	on $\Sp$ and $\Sm$.
        
	Next, we define the \phase space $\sls$ to be the on-shell subspace of $\fs$,
	metrics that satisfy the vacuum field equations.  Metrics in
        $\sls$ determine unique extended initial data structures of
        the form (\ref{ids2}).  Thus initial data furnish a partial
        coordinate system on $\sls$.

        Finally, the
	phase space of the theory $\slp$ is $\sls$ modded out by degeneracy directions of the symplectic form
	\cite{Wald:1999wa,Harlow:2019yfa}.  Now extended initial
        data structures of the form (\ref{ids2}) determine the metric in the bulk
        up to diffeomorphisms.  Since these diffeomorphisms correspond
        to degeneracy directions, these
        structures determine unique elements of $\slp$.  However, the
        correspondence is not one-to-one because of additional
        degeneracies in the presymplectic form related to edge or
        corner modes which we discuss in detail in Sec.\ \ref{sec:interp} below.
        Thus the phase space $\slp$ is in one-to-one correspondence
        with the set of initial data structures (\ref{ids2}) modulo
        these edge mode degeneracies.
        The phase space $\slp$ is the largest phase space that one can
	define for two intersecting null surfaces, so we call it the maximal
	horizon phase space.  It was previously considered in Ref.\ \cite{Chandrasekaran:2020wwn}.

	\subsection{Horizon Weyl-diffeomorphism symmetry group}
	\label{sec:hwd}
	
	In this section we derive the symmetry group of the maximal horizon
	phase space, which we will call the horizon Weyl-diffeomorphism group.
	The corresponding algebra [Eq.\ (\ref{groupstructure1}) below] was found previously by
	Adami, Grumiller, Sheikh-Jabbari, Taghiloo, Yavartanoo and Zwikel using a
	different approach \cite{Adami:2021nnf}.

	Bulk diffeomorphisms $\psi : M \to M$ which preserve the field configuration
	space $\fs$ are just those that preserve the boundaries $S_0$,
	$\Sp$ and $\Sm$.  For these diffeomorphisms we define the boundary
	diffeomorphisms $\varphi_\ps$, $\varphi_\ms$ and $\varphi_0$ and scaling
	functions $\gamma_\ps$ and $\gamma_\ms$ via Eqs.\ (\ref{bds})
	and (\ref{gammadef}).  These transformations automatically preserve
	the conditions (\ref{bc00}) on the metric since
	\be
	(\psi_* g^{ab} )\ell^\ps_a \ell^\ps_b = e^{-2 \gamma_\ps} \psi_* (g^{ab}
	\ell^\ps_a \ell^\ps_b) =0.
	\ee
	
	We parameterize elements of the group as
	\be
	(\varphi_\ps,\gamma_\ps, \varphi_\ms, \gamma_\ms),
	\label{cc}
	\ee
	where we omit $\varphi_0$ since it is determined from $\varphi_\ps$ or
	$\varphi_\ms$ by taking the limit to $S_0$.
	Portions of the bulk diffeomorphism not contained in (\ref{cc}) are
	trivial (corresponding to zero charges) as will be shown below.
	Consider first the
	subgroup $\gamma_\ps = \gamma_\ms = 0$.  This subgroup is the subgroup
	of ${\rm Diff}(\Sp) \times {\rm Diff}(\Sm)$ given by the constraint
	that the limits to $S_0$ of $\varphi_\ps$ and $\varphi_\ms$ coincide.
	Equivalently, it can be written as
	\be
	{\rm Diff}(S_0) \times {\rm Diff}'(\Sp) \times {\rm Diff}'(\Sm),
	\label{group:diffeo}
	\ee
	where ${\rm Diff}'(\Sp)$ is the subgroup of the diffeomorphism group of
	$\Sp$ which act as the identity map on $S_0$.

	The full group is the semidirect product of the diffeomorphism
	component (\ref{group:diffeo}) with a group of Weyl rescalings on the two
	boundaries $\Sp$ and $\Sm$ parameterized by $\gamma_\ps$ and
	$\gamma_\ms$.  The group composition law can be obtained by applying
	the definition (\ref{gammadef}) to a composition $\psi_2 \circ \psi_1$
	and is
	\be
	\label{groupstructure}
	(\varphi_{\ps\,2},\gamma_{\ps\,2}, \varphi_{\ms\,2}, \gamma_{\ms\,2})
	\circ (\varphi_{\ps\,1},\gamma_{\ps\,1}, \varphi_{\ms\,1}, \gamma_{\ms\,1})
	=
	(\varphi_{\ps\,3},\gamma_{\ps\,3}, \varphi_{\ms\,3}, \gamma_{\ms\,3}),
	\ee
	where
	\begin{subequations}
		\begin{eqnarray}
		\varphi_{\ps\,3} &=& \varphi_{\ps\,2 } \circ \varphi_{\ps\,1}, \\
		\gamma_{\ps\,3} &=& \gamma_{\ps\,1} + \gamma_{\ps\,2} \circ
		\varphi_{\ps\,1}, \\
		\varphi_{\ms\,3} &=& \varphi_{\ms\,2 } \circ \varphi_{\ms\,1}, \\
		\gamma_{\ms\,3} &=& \gamma_{\ms\,1} + \gamma_{\ms\,2} \circ
		\varphi_{\ms\,1}.
		\end{eqnarray}
	\end{subequations}

	The linearized version of these results is as follows.  We represent
	the bulk diffeomorphism $\psi$ by a vector field $\xi^a$ on $M$, so that
	$\psi_* = 1 + \lie_{\vec \xi} + O(\xi^2)$.
	Since $\psi$ preserves $\Sp$ and $\Sm$ the vector $\xi^a$ must be
	tangent to these boundaries, so
	\be
	\xi^a \ell_{\ps\,a} = \xi^a \ell_{\ms\,a} = 0.
	\label{tangent}
	\ee
	Hence restricted to $\Sp$ and $\Sm$ the symmetry vector field
	corresponds to intrinsic vector fields
	$\chi^i_\ps$ and $\chi^{i'}_\ms$ that generate the boundary symmetries
	$\varphi_\ps$ and $\varphi_\ms$.  These vector fields are
	are constrained to be
	tangent to $S_0$ on $S_0$ and to coincide there.  The linearized
	versions of the definitions (\ref{gammadef}) are
	\be
	\lie_{\vec \xi} \ell^\ps_a = \gamma_\ps \ell^\ps_a, \ \ \ \ \lie_{\vec
		\xi} \ell^\ms_a = \gamma_\ms \ell^\ms_a.
	\label{idents12}
	\ee
	The linearized version of the group composition law
	(\ref{groupstructure}) is
	the Lie bracket
	\be
	\label{groupstructure1}
	\left[ (\chi^i_{\ps\,1},\gamma_{\ps\,1}, \chi^{i'}_{\ms\,1}, \gamma_{\ms\,1}),
	(\chi^i_{\ps\,2},\gamma_{\ps\,2}, \chi^{i'}_{\ms\,2}, \gamma_{\ms\,2})
	\right]
	=
	(\chi^i_{\ps\,3},\gamma_{\ps\,3}, \chi^{i'}_{\ms\,3}, \gamma_{\ms\,3}),
	\ee
	where
	\begin{subequations}
		\label{groupstructure2}
		\begin{eqnarray}
		\chi^i_{\ps\,3} &=& \lie_{{\vec \chi}_{\ps\,1}} \chi^i_{\ps\,2}, \\
		\gamma_{\ps\,3} &=& \lie_{{\vec \chi}_{\ps\,1}} \gamma_{\ps\,2}
		-    \lie_{{\vec \chi}_{\ps\,2}} \gamma_{\ps\,1}
		, \\
		\chi^i_{\ms\,3} &=& \lie_{{\vec \chi}_{\ms\,1}} \chi^i_{\ms\,2}, \\
		\gamma_{\ms\,3} &=& \lie_{{\vec \chi}_{\ms\,1}} \gamma_{\ms\,2}
		-    \lie_{{\vec \chi}_{\ms\,2}} \gamma_{\ms\,1}.
		\end{eqnarray}
	\end{subequations}
	Here we are assuming that the diffeomorphisms are not field-dependent.
	Note that the algebra (\ref{groupstructure1}) contains the universal corner symmetry algebra
	of Refs.\ \cite{Donnelly2016a,Ciambelli:2021vnn} at the corner $S_0$.
	The algebra (\ref{groupstructure1}) truncated to a single null surface agrees
        with that analyzed in Ref.\ \cite{Ciambelli:2023mir}.
	
	Finally we show that the parameterization (\ref{cc}) of the group
	omits only trivial diffeomorphisms whose charges vanish.  This can be
	seen from the expression (\ref{finalH}) below for the charge, which depends
	only on $\gamma$ and $\varphi$ (ie $\chi^i$), and which vanishes for
	all background solutions and for all $\partial \Sigma$ only when $\gamma
	= \chi^i=0$\footnote{Note that we are using a definition of triviality here
		which is that the left hand side of Eq.\ (CFSS,2.49) vanishes, which
		differs from the condition used in CFP and in Wald-Zoupas \cite{Wald:1999wa}
		that the first term on the
		right hand side vanish.  However the two conditions are equivalent in
		the application considered here, since the second term on the right
		hand side of Eq.\ (CFSS,2.49) also vanishes for all backgrounds if and
		only if $\chi^i = \gamma =0$.}.
	
	The Weyl rescalings
	$(\gamma_+, \gamma_-)$ can be understood intuitively as follows. We can think of
	each null boundary as the limit of a family or foliation of spacelike
	hypersurfaces. Different such foliations will induce different null
	normals on the boundary which are all related by rescalings. Since
	these rescalings are determined by the choice of spacetime foliation
	near the null surface, they parameterize bulk degrees of freedom
	which transform under bulk diffeomorphisms $\psi$ rather
	than the intrinsic null surface data which transform under 
	the boundary diffeomorphisms $\varphi_\ps$ and $\varphi_\ms$.

	\subsection{Restricted horizon phase space with fixed direction of contravariant normal}
	\label{sec:rhps}

	We now turn to the smaller phase space that is obtained by requiring
	that the direction of the up-index normals $\ell^a_\ps$ and $\ell^a_\ms$ be fixed.
	
	We start by defining a particular geometric structure on $(S_0, \Sp,
	\Sm)$ which we call a {\it boundary structure}.
	We consider pairs 
	\be
	\label{pairs}
	( {\hat \ell}_\ps^i, {\hat \ell}_\ms^{i'})
	\ee
	of intrinsic vector fields on $\Sp$ and $\Sm$.
	Here we have inserted hats on the vectors to indicate that they are
	not necessarily related to normal covectors by raising the index with
	the metric as in Eq.\ (\ref{rel1}) below.
	We define two such pairs
	$( {\hat \ell}_\ps^i, {\hat \ell}_\ms^{i'})$
	and  $( {\hat \ell}_\ps^{'i}, {\hat \ell}_\ms^{'i'})$
	to be equivalent if they are related by the rescalings
	\bes
	\label{equiv22}
	\bea
	{\hat \ell}_\ps^{'i} &=& e^{\sigma_\ps} {\hat \ell}_\ps^{i}, \\
	{\hat \ell}_\ms^{'i'} &=& e^{\sigma_\ms} {\hat \ell}_\ms^{i'},
	\eea
	\ees
	for some smooth functions $\sigma_\ps$ and $\sigma_\ms$ on
	$\Sp$ and $\Sm$.
	We define
	\be
	\label{pdef}
	\mfp = [ {\hat \ell}_\ps^i, {\hat \ell}_\ms^{i'}]
	\ee
	to be the equivalence class under this equivalence relation.  This is
	the desired boundary structure.
	
	As in CFP, a metric $g_{ab}$ on $M$ in $\fs$
	determines a unique boundary structure
	\be
	\mfp = \mfp[g_{ab}].
	\label{frakpgab}
	\ee
	One picks null normals $\ell_{\ps a}$
	and $\ell_{\ms a}$ on $\Sp$ and $\Sm$, raises the indices with
	the metric to obtain
	the vectors $\ell_\ps^a = g^{ab} \ell_{\ps\,b}$,
	and $\ell_\ms^a = g^{ab} \ell_{\ms\,b}$, regards these as
	intrinsic vectors $\ell_\ps^i$ and $\ell_\ms^{i'}$, and then takes the  
	equivalence class $[ \ell_\ps^i, \ell_\ms^{i'}]$.  The result is
	independent of which normal one forms are chosen, by the
	equivalence relation (\ref{equiv22}).
	
	We can also obtain a boundary structure 
	$
	\mfp = \mfp[ \mfh ] 
	$
	from an extended initial data structure $\mfh$ of the form
	(\ref{ids2}).  We choose a
	representative 
	$(q_{AB}, {\bar \omega}_A, m,
	\Theta_\ps, \Theta_\ms, \ell_{\ps\,a}, \ell^i_\ps, \kappa_\ps, [q^\ps_{ij}],
	\ell_{\ms\,a}, \ell^{i'}_\ms, \kappa_\ms, [q^\ms_{i'j'}] )$ of the
	equivalence class, discard all the fields except 
	$(\ell^i_\ps, \ell^{i'}_\ms )$,
	and take the
	equivalence class under the equivalence relation (\ref{equiv22}).
	Then, given a metric $g_{ab}$, the boundary and extended
	initial data structures (\ref{frakpgab}) and (\ref{frakhgab}) induced by metric are related by
	\be
	\label{poh}
	\mfp \left[ \mfh[g_{ab}] \right] = \mfp[g_{ab}].
	\ee

	Finally, given the manifold $M$ and the boundaries $(S_0, \Sp, \Sm)$,
	and given a choice of boundary structure $\mfp$,
	we define the field configuration space ${\msc F}_\mfp$ to be the set of
	metrics $g_{ab}$ on $M$ for which $\Sp$ and $\Sm$ are null and 
	for which $\mfp[g_{ab}] = \mfp$.
	The space $\fs_\mfp$ is a subspace of the field configuration space $\fs$ 
	defined in Sec.\ \ref{dfcs} above.
	Metrics $g_{ab}$ in $\fs_\mfp$ obey the constraints
	\be
	\label{bc01}
	{\hat \ell}^{[a}_\ps g^{b]c} \ell_{\ps\,c} =0, \ \ \ \ \
	{\hat \ell}^{[a}_\ms  g^{b]c} \ell_{\ms\,c}=0,
	\ee
	on $\Sp$ and $\Sm$,
	for any choice of representative of
	$({\hat \ell}_\ps^i, {\hat \ell}_\ms^{i'})$ of $\mfp$, regarded as spacetime vector
	fields, and for any choices of normal covectors $\ell_{\ps\,a}$
	and $\ell_{\ms\,a}$ on $\Sp$ and $\Sm$.
	These constraints are stronger than the corresponding constraints
	(\ref{bc00}) for the field configuration space $\fs$.
	We define the corresponding on-shell subspace or \phase space
	$\sls_\mfp$ and phase space ${\mathscr P}_\mfp$ as in Sec.\ \ref{dfcs}.
	The phase space ${\mathscr P}_\mfp$ is in one-to-one correspondence
	with the set of extended initial data structures $\mfh$
	subject to the restriction
	\be
	\label{restrict1}
	\mfp[ \mfh ] = \mfp,
	\ee
	and modulo the edge mode issues discussed in Sec.\ \ref{sec:interp}.
        
	For a given choice of normalization of the normal one-forms
	$\ell_{\ps\,a}$ and $\ell_{\ms\,a}$, and given a metric $g_{ab}$,
	we can choose the representative $(\ell_\ps^a, \ell_\ms^a)$ of $\mfp$ to make
	\be
	\ell_\ps^a = g^{ab} \ell_{\ps\,b}, \ \ \ \ \ell_\ms^a = g^{ab}
	\ell_{\ms\,b}.
	\label{rel1}
	\ee
	Then the normal one forms $\ell_{\ps\,a}$ and $\ell_{\ms\,a}$
	are independent of $g_{ab}$, and are fixed up to 
	the rescaling (\ref{rescalings2}), while the normal vectors
	$\ell_\ps^a$ and $\ell_\ms^a$ do depend on the metric.
	Here we have dropped the hats to indicate the specific normal vectors
	which depend on the metric in this way.
	We shall adopt this convention throughout the paper, for both
	phase spaces $\slp$ and $\slp_\mfp$.  The rescaling parameterized
	by $\sigma_\ps$ and $\sigma_\ms$ then acts on both sets of normals, as
	in Eqs.\ (\ref{2rescalings}) and (\ref{rescalings2}).

	Finally we note that there is an alternative, smaller phase space
	definition, given in Appendix H of CFP, which we now briefly review.
	We replace the pairs (\ref{pairs}) with
	collections of tensor fields
	\be
	( {\hat \ell}_\ps^i, \ell_{\ps\,a}, {\hat \ell}_\ms^i, \ell_{\ms\,a}),
	\label{qrep}
	\ee
	where $\ell_{\ps\,a}$ and $\ell_{\ms\,a}$ are normal one-forms.  These fields are restricted
	to satisfy
	\be
	\label{con3}
	\left. {\hat \ell}_\ps^a \ell_{\ms\,a} \right|_{S_0} = \left. {\hat \ell}_\ms^a \ell_{\ps\,a} \right|_{S_0}.
	\ee
	The equivalence relation (\ref{equiv22}) is replaced by 
	\bes
	\label{equiv22small}
	\bea
	{\hat \ell}_\ps^{\prime\,a} &=& e^{\sigma_\ps} {\hat \ell}_\ps^{a}, \ \ \ \ \ell^{\prime}_{\ps\,a} =
	e^{\sigma_\ps} \ell_{\ps\,a}, \\
	{\hat \ell}_\ms^{\prime\,a} &=& e^{\sigma_\ms} {\hat \ell}_\ms^{a}, \ \ \ \ \ell^{\prime}_{\ms\,a} =
	e^{\sigma_\ms} \ell_{\ms\,a},
	\eea
	\ees
	and the boundary structure (\ref{pdef}) is replaced by the definition
	\be
	\label{qdef}
	\mfq = [ {\hat \ell}_\ps^i, \ell_{\ps\,a}, {\hat \ell}_\ms^{i'}, \ell_{\ms\,a}].
	\ee
	We define the corresponding field configuration space $\fs_\mfq$,
	\phase space $\sls_\mfq$ and phase space ${\mathscr P}_\mfq$ as
	above.  Metrics in $\fs_\mfq$ obey the constraints
	\be
	{\hat \ell}_\ps^a = g^{ab} \ell_{\ps\,b}, \ \ \ \ {\hat \ell}_\ms^a = g^{ab}
	\ell_{\ms\,b}
	\label{rel2}
	\ee
	for any choice of representative of $\mfq$ of the form (\ref{qrep}).

	The three \phase spaces we have defined are related by
	\be
	\sls_\mfq \subset \sls_\mfp \subset \sls,
	\ee
	where $\mfp = \mfp[\mfq]$ is obtained from $\mfq$ by a definition
	analogous to that discussed before Eq. (\ref{poh}).
	Within the smallest \phase space $\sls_\mfq$, the normal covectors
	and vectors are fixed, up to the rescalings (\ref{equiv22small}) which
	do not affect Eqs.\ (\ref{rel2}), and the metric $g_{ab}$ is
	constrained to satisfy Eqs.\ (\ref{rel2}).  In the larger \phase space
	$\sls_\mfp$, the metric is constrained to fix the directions of the
	right hand sides of Eqs.\ (\ref{rel2}), but not the scalings, from
	Eqs.\ (\ref{bc01}). Finally,
	within the maximal \phase space $\sls$ both the directions and
	scalings can be freely chosen, so there is no corresponding constraint on the metric.

	\subsection{Restricted horizon phase space symmetry group: Carrollian diffeomorphisms}
	\label{sec:group1}
	
	The symmetry group of the restricted horizon phase space $\slp_\mfp$ consists of symmetries
	\be
	(\varphi_\ps,\gamma_\ps, \varphi_\ms, \gamma_\ms)
	\label{cc11}
	\ee
	of the form (\ref{cc}), with the following restrictions.  The
	diffeomorphisms
	$(\varphi_\ps, \varphi_\ms)$ on $\Sp$ and $\Sm$ must preserve $S_0$
	and their restrictions to $S_0$ must coincide,
	and there must exist functions $\beta_\ps$ and $\beta_\ms$ on $\Sp$
	and $\Sm$ for which
	\be
	\label{rhdef}
	\varphi_{\ps\,*} {\hat \ell}^i_\ps = e^{\beta_\ps} {\hat \ell}^i_\ps, \ \ \ \ 
	\varphi_{\ms\,*} {\hat \ell}^{i'}_\ms = e^{\beta_\ms} {\hat \ell}^{i'}_\ms.
	\ee
	To see this we note that the group is a subgroup of the group
	discussed in Sec.\ \ref{sec:hwd},
	and that the bulk diffeomorphism $\psi$
	must preserve the boundary structure,
	\be
	\psi_* \mfp = \mfp.
	\ee
	From the definition (\ref{pdef}) it follows that there exists
	functions $\beta_\ps$ and $\beta_\ms$ for which $\psi_*$ acts as a
	rescaling of the form (\ref{equiv22}):
	\begin{subequations}
		\begin{eqnarray}
		\psi_* {\hat \ell}^i_\ps &=& e^{\beta_\ps} {\hat \ell}^i_\ps, \\
		\psi_* {\hat \ell}^{i'}_\ms &=& e^{\beta_\ms} {\hat \ell}^{i'}_\ms.
		\end{eqnarray}
	\end{subequations}
	The relations (\ref{rhdef}) now follow from the definitions (\ref{bds}).

	The corresponding Lie algebra is given by Eqs.\ (\ref{groupstructure1}) and (\ref{groupstructure2}),
	but with the additional constraint that 
	the pairs of vector
	fields $(\chi^i_\ps, \chi^{i'}_\ms)$
	satisfy
	the linearized versions of Eqs. (\ref{rhdef}):
	\be
	\lie_{{\vec \chi_\ps}} {\hat \ell}^i_\ps  = \beta_\ps {\hat \ell}^i_\ps, \ \ \ \ \ 
	\lie_{{\vec \chi_\ms}} {\hat \ell}^{i'}_\ms  = \beta_\ms {\hat \ell}^{i'}_\ms.
	\label{iidd1}
	\ee
	In coordinates $(u,\theta^1,\theta^2)$ on $\Sp$ for which ${\vec
		\ell}_\ps = \partial_u$, the symmetry vector field takes the form
	\be
	\label{gensym}
	{\vec \chi}_\ps = f(u,\theta^A) \partial_u + Y^A(\theta^B) \partial_A,
	\ee
	where the function $f$ vanishes on $S_0$.
        The corresponding algebra was first studied in Ref. \cite{Donnay:2016ejv}.
        The symmetry (\ref{gensym}) generalizes the
	symmetries of CFP and of Sec.\ 2.5  of
        Ref.\ \cite{Donnay:2016ejv}, where the function $f$ took the specific form $f(u,\theta^A) =
	\alpha(\theta^A) - \beta(\theta^A) u$, corresponding to affine and
	Killing supertranslations.
	
	We note that if we use the specific up index normals (\ref{rel1}) that
	depend on the metric, then the functions $\beta_\ps$ and $\beta_\ms$ defined by
	Eq.\ (\ref{iidd1})
	and $\gamma_\ps$ and $\gamma_\ms$ defined by Eq.\ (\ref{idents12})
	are not independent
	when restricted to $S_0$.
	It follow from Eqs.\ (\ref{rel1}) that on $S_0$ we have
	\be
	\ell^a_\ps \ell_{\ms\,a}= \ell^a_\ms \ell_{\ps\,a},
	\label{cv}
	\ee
	and taking a Lie derivative
	with respect to a symmetry ${\vec \xi}$ and using the definitions
	(\ref{iidd1}) and (\ref{idents12})
	yields
	\be
	\left.  \beta_\ps \right|_{S_0}  + \left.\gamma_\ms \right|_{S_0} =
	\left.  \beta_\ms \right|_{S_0}  + \left.\gamma_\ps \right|_{S_0} =
	\lie_\chi m,
	\label{const11}
	\ee
	where we define $m$ by the value of Eq.\ (\ref{cv}) being $-e^m$,
	cf. Eq.\ (\ref{normdef}).
	The result (\ref{const11}) is invariant under rescalings, from
	the rescaling laws (\ref{rescalem}) for $m$, (CFP,4.12) for
	$\beta_\ps$ and $\beta_\ms$, and the linearized versions of
	(\ref{gammalaw}) for $\gamma_\ps$ and $\gamma_\ms$.

	If we restrict attention to one null boundary $\Sp$, the group of
	boundary diffeomorphisms $\varphi_\ps$ that satisfy (\ref{rhdef})
	have been called {\it Carrollian diffeomorphisms} \cite{Ciambelli2019b,Donnay:2019jiz,Freidel:2022vjq,Freidel:2022bai}, since they
	preserve the fiber bundle structure of a Carroll structure\footnote{A
		Carroll structure is a three dimensional manifold $M$ together with a
		tensor field $q_{ij}$ of rank $(0,+,+)$ and a nowhere vanishing vector
		field $n^i$ with $q_{ij} n^j=0$.}.

	For the smaller phase space $\slp_\mfq$, the symmetries consist of
	the pairs of boundary diffeomorphisms $(\varphi_\ps, \varphi_\ms)$ as
	described above, while the functions $\gamma_\ps$ and $\gamma_\ms$ are
	no longer freely specifiable and instead are given in terms of the
	boundary diffeomorphisms by
	\be
	\label{iidd}
	\gamma_\ps = \beta_\ps, \ \ \ \ \gamma_\ms = \beta_\ms,
	\ee
	as derived in CFP.
	The constraint (\ref{const11}) then reduces to 
	\be
	\left. \beta_\ps\right|_{S_0} + \left. \beta_\ms \right|_{S_0} =
	\lie_\chi m.
	\label{const11a}
	\ee

\subsection{Half-sided boosts and extensions of the phase spaces to include shocks}
\label{halfsidedboosts}

In this subsection we describe extensions of the three \phase spaces
$\sls$, $\sls_\mfp$ and $\sls_\mfq$ defined above to include shocks
generated by ``half-sided boosts'' acting on the initial data
\cite{Carlip:1993sa,Bousso:2020yxi,2020PhRvD.101d6001B}.
We will denote the extended \phase spaces and corresponding phase spaces with primes, as 
$\sls'$, $\sls_\mfp'$, $\sls_\mfq'$ and $\slp'$, $\slp_\mfp'$ and $\slp_\mfq'$.
These phase space extensions will play an important role in the
symplectic form of the theory, which we will compute in Sec.\ \ref{sec:interp}.

\subsubsection{Normal boosts}

We start by defining the action of boosts and half-sided boosts on
initial data. There is some ambiguity in how boosts are defined on
non-stationary boundaries, related to the choice of parameter used
along null generators \cite{Ciambelli:2023mir}, but we will
see that this ambiguity does not affect the definition of the
extension of the \phase space.

We initially focus attention on the null surface $\Sp$ and drop
the $+$ subscripts for simplicity.  As described in the previous
subsections, linearized symmetries in the maximal horizon phase space $\slp$ are
described by arbitrary smooth vector field $\chi^i$ on $\Sp$, together
with functions $\gamma$ which determine how the normal covector is
rescaled by the symmetry, according to Eq.\ (\ref{gammadef}).
Within the restricted horizon phase space $\slp_\mfp$, the vector
field $\chi^i$ is restricted by
the condition that there should exist a function $\beta$ on $\Sp$ for
which [cf.\ Eq.\ (\ref{iidd1})]
\be
\lie_\chi \ell^i = \beta \ell^i,
\label{iidd111}
\ee
while in the smaller phase space $\slp_\mfq$ there is the further
restriction (\ref{iidd}).

Now, for all three phase spaces, given a specification of an initial data structure we define 
the class of boost symmetries by imposing three conditions.  These are (i) the vector field is of the form
\be
\chi^i = f \ell^i
\label{boostdef1}
\ee
for some function $f$; (ii) the vector field satisfies the relation
  \be
  \lie_\ell ( \lie_\ell + \kappa + \alpha_0 \Theta ) \chi^i =0;
  \label{boostdef2}
  \ee
  and (iii) the function $\gamma$ is given by
\be
\label{iidd23}
\gamma = \beta.
\ee
  Here $\alpha_0$ is a parameter whose value can be varied to modify
  the definition of the class of symmetries.  The choice $\alpha_0=0$
  corresponds to the use of affine parameter along generators to
  define boosts; nonzero choices of $\alpha_0$ and their motivation
are discussed in Sec.\ \ref{sec:fluxder} and Appendix \ref{app:boostflux} below.
The symmetries (\ref{boostdef2}) belong to the symmetry algebras of all three phase
spaces\footnote{\label{caveat}More precisely, they belong to the symmetry algebras
for a single null surface.  For a pair of intersecting null surface
they are not symmetries, since they do not preserve the surface $S_0$.
By contrast, the half-sided boosts which we define below are symmetries
for pairs of intersecting surfaces, since they preserve $S_0$.}, since from Eq.\ (\ref{boostdef1}) the condition (\ref{iidd111}) is
automatically satisfied with
\be
\label{boost3}
\beta = - \lie_\ell f.
\ee
In addition, the definition (\ref{boostdef2}) is invariant under the rescaling
(\ref{2rescalings}) of the normal, if $\chi^i$ is taken to be rescaling
invariant.
Thus the class of
symmetries is well defined given an initial data equivalence class of
the form (\ref{ids2})\footnote{The definition of boost symmetries depends on
the choice of state or of point in phase space, for all three phase
spaces discussed in this paper.  However it is state-independent for
the smaller phase space treated in CFP \cite{CFP} when $\alpha_0=0$,
since the inaffinity $\kappa$ is included in the definition of the
phase space in that case.}.

The physical interpretation of these symmetries is clarified when we
specialize to a choice of rescaling of the normal to impose
\be
\kappa + \alpha_0 \Theta  =0,
\label{CFLcondt}
\ee
following Ciambelli, Freidel and Leigh (CFL) \cite{Ciambelli:2023mir}
for the case $\alpha_0=1/2$ \footnote{CFL picked this choice of parameter value
in order to make the boost charges be monotonic.  This is discussed
further in Appendix \ref{app:boostflux}.}. 
We then choose a parameter $u$ along the null generators for which
${\vec \ell} = \partial_u$, called ``dressing time'' by CFL.
The general solution of Eqs. (\ref{boostdef1}), (\ref{boostdef2}) and
(\ref{boost3}) is then  of the form
\be
{\vec \chi} = b(\theta^A) \left[ u - u_0(\theta^A) \right]
\frac{\partial}{\partial u}.
\label{prop1}
\ee
The symmetry also acts on the normal covector via pullback as
\be
\ell_a \to (1 - b) \ell_a,
\label{prop2}
\ee
from Eqs.\ (\ref{gammadef}), (\ref{iidd23}) and (\ref{boost3}).
The properties (\ref{prop1}) and (\ref{prop2}) show that 
the symmetry is a (linearized) angle-dependent boost with rapidity parameter $b$ about
the surface $B$ given by $u = u_0(\theta^A)$.
We will restrict attention to the case where the surface $B$ lies in
$\Sp$, that it, it is to the future of $S_0$.

\subsubsection{Half-sided boosts}
\label{halfsided}

We define half-sided boosts by implementing the diffeomorphism to the
future of the null surface that intersects $\Sp$ at $B$, but not to
the past. This definition corresponds to modifying 
the symmetry vector field (\ref{prop1}) to become
\be
   {\vec   {\chi}} = f H(f) \partial_u,
   \label{hsb33}
\ee
where $H$ is the Heaviside step function, and $f = b (u - u_0)$ with
$b > 0$.  As before we take $\gamma = \beta$, where now
\be
\beta = - \lie_\ell[ f H(f)] = - H(f) \lie_\ell f = - H(f) b(\theta).
\ee
Some explicit examples of half-sided boosts which demonstrate the
relationship between the actions of the diffeomorphism on the boundary
and in the bulk are given in Appendix \ref{app:boosts}.
From Eqs.\ (\ref{symt}) we can now compute how the initial data
transforms, in the scaling convention where the normal covector is kept fixed.
We obtain that 
\bes
\label{transformed1}
\bea
\delta \ell^i &=& 0, \\
\delta \kappa &=& - \alpha_0 H(f) \lie_\ell( \Theta f) + b(\theta^A)
\delta[ u - u_0(\theta^A)], \\
\delta \Theta &=& H(f) \lie_\ell (f \Theta),\\
\delta {\bar q}_{ij} &=& f H(f) \lie_\ell {\bar q}_{ij}.
\eea
\ees
We see that for stationary backgrounds with $\Theta = \lie_\ell {\bar q}_{ij}=0$,
the only change to the initial data is the insertion of a delta
function in the inaffinity.  Because the symmetry vanishes to the past
of the surface $B$, it vanishes in particular on $S_0$ and so is a
well defined symmetry on the three phase spaces $\slp$, $\slp_\mfp$
and $\slp_\mfq$ for intersecting null surfaces (see footnote
\ref{caveat} above).

When evolved into the future, the transformed initial data
(\ref{transformed1}) gives rise to a solution with a shock (see
Appendix \ref{app:boosts}).  The shock can be removed using a
diffeomorphism, so it can be pure gauge if the null surface $\Sp$ is
treated as lying in the interior of the spacetime.  It is not pure
gauge however in the
framework used here, where $\Sp$ is part of the boundary of spacetime.
We will see explicitly in Sec.\ \ref{sec:interp} that the half-sided boosts do
not correspond to degeneracies of the presymplectic form.

We note that our definition of half-sided boosts does not require that
the boost surface $B$ be an extremal surface, unlike the similar kink
transform defined in
Refs.\ \cite{Bousso:2020yxi,2020PhRvD.101d6001B} \footnote{The
extremal assumption might be necessary for half-sided boosts to 
be generated by the area operator.  We derive this correspondence
in Sec.\ \ref{sec:sf} below for the special case of stationary backgrounds, but
the general case is an open question.}.

\subsubsection{Phase space extensions}
\label{sec:pse}

From the discussion in Sec.\ \ref{sec:initialdataextended}, general initial data can be represented
in terms of equivalence classes of the form (\ref{ids2}).  In these
equivalence classes, we can choose to replace the inaffinity $\kappa_\ps$
with a function $\alpha_\ps$ on $\Sp$ which contains the same information, and
similarly for $\kappa_\ms$.  The function $\alpha_\ps$ is defined by
\bes
\label{alphadef0}
\bea
\lie_{\ell_\ps} \alpha_\ps &=& \kappa_\ps, \\
\label{alphazero}
\left. \alpha_\ps \right|_{S_0} &=& 0.
\eea
\ees
Under the rescaling transformation (\ref{2rescalings}) it transforms
as
\be
\alpha_\ps \to \alpha_\ps + \sigma_\ps - \sigma_{\ps0},
\ee
where $\sigma_{\ps0}$ is the function on $\Sp$ defined by
$\lie_{\ell_\ps} \sigma_{\ps0}=0$ and $\sigma_{\ps0} = \sigma_\ps$ on $S_0$.
In particular the value of $\alpha_\ps$ on $S_0$ is invariant,
consistent with the relation (\ref{alphazero}).

Consider now a half-sided boost of the form (\ref{prop1}), modified by
the replacement
\be
\label{enlargelimit}
u_0(\theta^A) \to \varepsilon u_0(\theta^A)
\ee
where $\varepsilon>0$ is an infinitesimal parameter and we have chosen
$u=0$ on $S_0$.  With this replacement the boost surface $B$
approaches $S_0$.  Then, one of the effects of the transformation
(\ref{transformed1}) of the initial data is to generate a discontinuous jump in
$\alpha_\ps$ near the surface $S_0$.  If we now take the pointwise
limit $\varepsilon \to 0$ of the initial data functions, the result
lies in an enlarged function space defined by relaxing the boundary
condition (\ref{alphazero}) and allowing
$\left. \alpha_\ps \right|_{S_0}$ to be arbitrary,
and similarly for $\left. \alpha_\ms \right|_{S_0}$.
This larger function space defines
our enlarged \phase spaces $\sls'$, $\sls_\mfp'$ and $\sls_\mfq'$.
In Sec.\ \ref{sec:interp} below we will calculate the presymplectic form
on the enlarged \phase spaces by carefully taking the limit
(\ref{enlargelimit}), and then mod out by degeneracy directions
to compute the phase spaces $\slp'$, $\slp_\mfp'$ and $\slp_\mfq'$.
The enlarged phase spaces allow two different shocks in the initial
data, that should be thought of as propagating along null surfaces that are
just to the future of $\Sp$ and $\Sm$.
In Appendix \ref{app:enlarge} we describe analogous definitions of
phase space enlargements involving shocks in the simpler context of a
free scalar field in 1+1 dimensions, to give some intuition for the
more complicated gravitational case considered here.

The parameters $\left. \alpha_\ps \right|_{S_0}$ and $\left. \alpha_\ms \right|_{S_0}$ of
the extended \phase space can be interpreted as boost rapidity
parameters or boost angles in the following sense.
Within the original \phase space $\sls$, it is possible to  
completely fix the rescaling freedom by imposing the
conditions\footnote{Note that this specialization
is not compatible with the convention (\ref{conv1}), and therefore the
transformation properties (\ref{symt}) and (\ref{transformed1})
do not apply when (\ref{altconv}) is used.} 
\be
m = \kappa_\ps = \kappa_\ms = \Xi = 0,
\label{altconv}
\ee
where we have decomposed the
H\'a\'ji\^{c}ek one-form (\ref{rotoneform}) as
$
   {\bar \omega}_A = D_A \Xi + \epsilon_{AB} D^B \Gamma.
$
The normal vectors $L_\ps^i$ and $L_\ms^{i'}$ with this convention are
related to those in an arbitrary convention by
\be
\label{ellinvariant}
L_\ps^i = \exp\left[ - \frac{1}{2} m - \Xi - \alpha_\ps \right] \ell_\ps^i,
\ \ \  \ \ 
L_\ms^i = \exp\left[ - \frac{1}{2} m + \Xi - \alpha_\ms \right] \ell_\ms^{i'}.
\ee
These expressions are invariant under the rescalings (\ref{2rescalings}).
It follows from the constraints (\ref{con22}) and (\ref{alphazero})
that on $S_0$ we have ${\vec L}_\ps \cdot {\vec L}_\ms = -1$.
On the extended \phase space
$\sls'$
we have instead that
\be
\left. {\vec L}_\ps \cdot {\vec L}_\ms \right|_{S_0} = - \exp \bigg[
  \left. \alpha_\ps \right|_{S_0} +   \left. \alpha_\ms \right|_{S_0} \bigg].
\ee
Half-sided boosts shift the parameters
$\left. \alpha_\ps \right|_{S_0}$ and $\left. \alpha_\ms
\right|_{S_0}$ and thus rescale the vectors ${\vec L}_\ps$ and ${\vec
  L}_\ms$.  In Sec.\ \ref{sec:sf} below we will show that the
combination $  \left. \alpha_\ps \right|_{S_0} +   \left. \alpha_\ms
\right|_{S_0}$ of these parameters is physical, while the combination
$  \left. \alpha_\ps \right|_{S_0} -   \left. \alpha_\ms
\right|_{S_0}$ is gauge and can be set to zero\footnote{Note that in
Sec.\ \ref{sec:sf} $\left. \alpha_{\ps} \right|_{S_0}$ is denoted by
$b_\ps$ and $\left. \alpha_{\ms} \right|_{S_0}$ is denoted by $- b_\ms$.}.

	\section{Properties of field variations}
	\label{sec:bcs}
	
	In this section we will derive some conditions on field variations in
	the phase spaces $\slp$, $\slp_\mfp$ and $\slp_\mfq$ that will
	be useful in the remainder of the paper.
	
	\subsection{Boundary conditions on the metric perturbation}

	We write the varied metric as
	\be
	g_{ab} + \delta g_{ab} = g_{ab} + h_{ab}.
	\label{vary}
	\ee
	We assume that both the original and varied metric lie in one of the
	configuration spaces $\fs$, $\fs_\mfp$ or $\fs_\mfq$, so that they both satisfy
	one of the conditions (\ref{bc00}), (\ref{bc01}) or (\ref{rel2}).
	For the maximal field configuration space $\fs$, it follows
	immediately that
	\be
	h^{ab} \ell_{\ps\,a} \ell_{\ps\,b} = 0, \ \ \ \ \ h^{ab} \ell_{\ms\,a}
	\ell_{\ms\,b} = 0.
	\label{condts11}
	\ee
	These conditions are automatically satisfied
	for a metric perturbation of the form
	\be
	h_{ab} = \lie_\xi g_{ab}
	\label{metricsym}
	\ee
	generated by a representative $\xi^a$ of a symmetry in the
	horizon Weyl-diffeomorphism symmetry algebra discussed in
	Sec.\ \ref{sec:hwd}.  This follows from
	\be
	\ell^a \ell^b \nabla_a \xi_b = \ell^a \nabla_a (\ell^b \xi_b) -
	(\ell^a \nabla_a \ell^b) \xi_b = (\lie_{\vec \ell} - \kappa) (\ell^b
	\xi_b)=0,
	\ee
	where we have dropped the $\pm$ subscripts for simplicity and used Eqs.\ (\ref{kappadef}) and (\ref{tangent}).
	
	For the restricted field configuration space $\fs_\mfp$, we
	obtain instead of Eqs.\ (\ref{condts11}) the stronger conditions
	\be
	\label{condts12}
	{\hat \ell}^{[a}_\ps h^{b]c} \ell_{\ps\,c} =0, \ \ \ \ \
	{\hat \ell}^{[a}_\ms  h^{b]c} \ell_{\ms\,c}=0.
	\ee
	These follow from taking a variation of Eq.\ (\ref{bc01}) and
	noting that the variation does not affect the normal vectors or covectors there.
	For symmetries $\xi^a$ in the algebra discussed in Sec.\ \ref{sec:group1}, the
	conditions (\ref{condts12}) are automatically satisfied for variations of the
	form (\ref{metricsym})\footnote{This can be seen as follows.  From the constraint (\ref{bc01})
		and dropping the $\pm$ subscripts
		it follows that $g^{ab} \ell_b = - \Psi {\hat \ell}^a$ for some $\Psi$.  Now acting with $\lie_\xi$
		and using Eqs.\ (\ref{metricsym}), (\ref{idents12}) and (\ref{iidd1}) gives that
		$h^{ab} \ell_b = \left[ \lie_\chi \Psi + \Psi (\beta - \gamma) \right]
		{\hat \ell}^a$ from which Eq.\ (\ref{condts12}) follows.}.  
	
	For the smallest field configuration space $\fs_\mfq$, the conditions
	are obtained from varying Eqs.\ (\ref{rel2}) and are yet stronger:
	\be
	\label{condts12a}
	h_{ab} \ell_\ps^b = 0, \ \ \ \ \
	h_{ab} \ell_\ms^b = 0.
	\ee
	These conditions are automatically
	satisfied by variations (\ref{metricsym}) for symmetries ${\vec \xi}$
	in the corresponding algebra, as shown in Appendix D of CFP.

	\subsection{General identities involving extended initial data structures}
	
	As discussed in Secs.\ \ref{dfcs} and \ref{sec:group1} above, elements of the phase
	spaces $\slp$, $\slp_\mfp$ and $\slp_\mfq$ can be
	parameterized in terms of extended initial data structures of the form
	(\ref{ids2}).  We next derive some useful identities related to
	variations of these structures.

	Given an extended initial data structure $\mfh$ and a varied
	structure $\mfh + \delta \mfh$, there are two separate
	rescaling freedoms: the background scaling parameterized by
	$(\sigma_\ps, \sigma_\ms)$, and the scaling of the varied structure
	parameterized by $(\sigma_\ps + \delta \sigma_\ps, \sigma_\ms + \delta \sigma_\ms)$.
	Following Ref.\ \cite{Chandrasekaran:2020wwn} we fix the freedom $(\delta \sigma_\ps, \delta \sigma_\ms)$ by
	demanding that
	\be
	\label{conv1}
	\delta \ell_{\ps\,a} = 0, \ \ \ \ \ \delta \ell_{\ms\,a} = 0.
	\ee
	This does not entail any loss of generality
	and is consistent with the convention discussed around Eq.\ (\ref{rel1}).
	From now on in this section we will drop the $+$ and $-$
	subscripts since the same identities apply to both $\Sp$ and $\Sm$.
	For the maximal phase space $\slp$,
	varying the definition $\ell^a = g^{ab} \ell_b$ gives
	\be
	\label{deltaell}
	\delta \ell^a = - h^{ab} \ell_b.
	\ee
	It follows from the constraint (\ref{condts11}) that $\delta \ell^a \ell_a=0$,
	so the variation of the normal is an intrinsic vector $\delta \ell^i$.
	Within the restricted phase space $\slp_\mfp$ it follows from
	Eqs.\ (\ref{condts12}) that $\delta \ell^i \propto \ell^i$, and we define the field $\Psi$
	by
	\be
	\delta \ell^i = \Psi \ell^i.
	\label{Psidef}
	\ee
	This field is invariant under rescalings.
	Within the smaller phase space $\slp_\mfq$ we have instead
	\be
	\delta \ell^i = \Psi = 0,
	\ee
	from Eq.\ (\ref{rel2}).
	We choose the variation of the auxiliary normal to be
	\be
	\delta n_a = - \frac{1}{2} (h^{bc} n_b n_c) \ell_a - (h^{bc} n_b
	\ell_c) n_a
	\ee
	in order to preserve the relations (\ref{auxnormal}).

	We define $h_{ij}$ to be the pullback to the surface of $h_{ab}$,
	which gives
	\be
	\label{deltaqdef}
	h_{ij} = \delta q_{ij},
	\ee
	from Eq.\ (\ref{vary}) since since variations commute with pullbacks.
	Writing the relation (\ref{deltaell}) in the form $g_{ab} \delta
	\ell^b = - h_{ab} \ell^b$, taking the pullback to the surface and
	using Eq.\ (CFP,3.7) gives the identity
	\be
	\label{con11}
	q_{ij} \delta \ell^j=- h_{ij} \ell^j.
	\ee
	Next, defining the trace of the metric perturbation
	\be
	h = g^{ab} h_{ab},
	\label{htrace}
	\ee
	we have from the definition (\ref{volumesmall})
	that the variation of the volume forms $\eta_{ijk}$ and $\mu_{ij}$
	are given by
	\be
	\label{addi1}
	\delta \eta_{ijk} = \frac{1}{2} h \eta_{ijk}
	\ee
	and
	\be
	\label{deltamu}
	\delta \mu_{ij} = \frac{1}{2} h \mu_{ij} + \eta_{ijk} \delta \ell^k.
	\ee
	We can relate the trace $h$ to the intrinsic trace $q^{ij} h_{ij}$
	using formula (\ref{qabdef}) for $q_{ab}$ and Eqs.\ (CFP,3.7) and
	(\ref{deltaell}).  This yields
	\be
	h = q^{ab} h_{ab} - 2 n^a \ell^b h_{ab} = q^{ij} h_{ij} + 2 \delta
	\ell^i n_i.
	\label{hformula}
	\ee
	Next, taking a variation of the formula
	(CFP,3.28) for the expansion $\Theta$ gives
	\be
	\label{deltaTheta}
	\delta \Theta = \delta ({\hat D}_i \ell^i) = (\delta {\hat D}_i)
	\ell^i + {\hat D}_i \delta \ell^i = \frac{1}{2} \lie_\ell h + {\hat
		D}_i \delta \ell^i.
	\ee
	Here ${\hat D}_k$ is the divergence operator defined in Sec.\ 3.3 of CFP.

	\subsection{Transformation properties under rescaling and under changes of
		auxiliary normal}
	
	The transformation properties of the background quantities under
	the rescaling (\ref{rescale00}) of the null normal are given in Eq.\ (\ref{contransformc}),
	together with
	\bes
	\label{contransformd}
	\bea
	n_i &\to& e^{-\sigma} n_i, \\
	q^{ij} &\to & q^{ij}, \\
	q_i^{\ j} &\to & q_i^{\ j}, \\
	\varpi_i &\to& \varpi + q_i^{\ j} D_j \sigma,
	\eea
	\ees
	from Eqs.\ (\ref{auxnormal}), (\ref{contransformc7}), (\ref{twist}), (\ref{qabdef}), (\ref{qmixed})
	and (\ref{ptwist}).
	By varying these relations and using the convention (\ref{conv1}) we find
	that the varied quantities transform as
	\begin{subequations}
		\label{Vcontransformc}
		\begin{eqnarray}
		\label{Vcontransformc1}
		\delta n^i &\to& e^{-\sigma} \delta n^i, \\
		\label{Vcontransformc2}  
		\delta \ell^i &\to& e^{\sigma} \delta \ell^i, \\
		\label{Vcontransformc3}
		h_{ij} &\to&  h_{ij}, \\
		\label{Vcontransformc5}
		\delta \nonaffinity &\to& e^{\sigma} (\delta \nonaffinity +
		\lie_{\delta \ell} \sigma),\\
		\label{Vcontransformc6}
		\delta \expansion &\to& e^{\sigma} \delta \expansion,\\
		\delta \sigma_{ij} &\to& e^{\sigma} \delta \sigma_{ij},\\
		\label{Vcontransformc7}
		\delta {\shape}_i^{\ j} &\to& e^{\sigma} \left[\delta {\shape}_i^{\ j} +
		D_i\sigma \delta \ell^j  \right],\\
		\delta \varpi_i &\to& \delta \varpi_i.
		\end{eqnarray}
	\end{subequations}

	The other type of transformation we will be concerned with is the
	transformation \cite{Chandrasekaran:2020wwn}
	\be
	n^a \to n^a + \Delta n^a
	\ee
	of the background auxiliary null normal, of the form (\ref{Deltandef}).
	Since $\Delta n^a \ell_a=0$ we can regard $\Delta n^i$ as an intrinsic
	vector, even though $n^i$ itself is not defined.
	The down index auxiliary normal transforms as $n_a \to n_a + q_{ab} \Delta n^b +
	\ell_a (n_b \Delta n^b)$, from Eq.\ (\ref{qabdef}), and taking the
	pullback and using Eq.\ (CFP,3.7) gives
	\be
	n_i \to n_i + q_{ij} \Delta n^j.
	\label{tz1}
	\ee
	The projection tensors $q^{ij}$ and $q_i^{\ j}$ transform as
	\bes
	\label{tz2}
	\bea
	q^{ij} &\to& q^{ij} + \ell^i \Delta n^j + \ell^j \Delta n^i, \\
	q_i^{\ j} &\to& q_i^{\ j} + q_{ik} \Delta n^k \ell^j,
	\eea
	\ees
	from Eqs.\ (\ref{qabdef}) and (\ref{qmixed}).
	Finally the transformation property of the projected rotation one-form
	$\varpi_i$ can be derived from Eqs.\ (\ref{twist}), (\ref{ptwist}),
	(\ref{qmixed}), (\ref{grelations1a2}), and (\ref{Kident}) and is
	\be
	\label{tz3}
	\varpi_i \to \varpi_i + (\kappa q_{ij} - K_{ij}) \Delta n^j.
	\ee

	\section{Gravitational charges and fluxes}\label{sec:decomp_charge}
	
	In this section we compute gravitational charges and fluxes for the
	three symmetry algebras discussed in Sec.\ \ref{sec:phasespaces},
	using the general formalism reviewed in Sec.\ \ref{sec:gp}.
	Throughout this section we will specialize to cuts $\partial \Sigma$
	of $\Sp$ and drop the $+$ subscripts on the various fields.
	
	\subsection{Criteria for restricting choices of boundary and corner fluxes}
	\label{sec:pr}
	
	As discussed in Sec.\ \ref{sec:survey} above, we will impose various
	criteria on the 
	choices
	(\ref{decompos11}) and
	(\ref{eqn:bldecomp})
	of boundary and corner fluxes
	in order to make the gravitational charges as unique as possible.  We
	will see that they are unique up to a two parameter freedom.
	
	The first criteria is that the decompositions should be independent of
	the choice of normalization of the null vectors $\ell^i$ and
	$\ell^{i'}$, as in CFP.  Any such dependence would give rise to
	anomalies which we exclude.
	Note that because we are working in the maximal \phase space $\sls$ in which no
	boundary conditions have been imposed, this criterion is more
	stringent than if we were working in a smaller \phase space, for
	example where the induced metric is fixed.  Our criterion thus
	excludes the boundary term $\ell \sim \kappa \eta_{ijk}$ which is
	standard in that context \cite{Lehner2016}; see
	Sec.\ \ref{sec:actionprinciple} above for 
	further discussion.

	The second criteria is that the
	decompositions should be independent of the choice of auxiliary
	normals $n_i$ and $n_{i'}$ on $\Sp$ and $\Sm$, as
	discussed in the introduction and in 
	Ref.\ \cite{Chandrasekaran:2020wwn}.
	These normals are
	necessary to define certain quantities that appear in the
	presymplectic potential (\ref{theta000}), as discussed in Appendix \ref{app:nullreview}, so excluding such dependence gives nontrivial 
	constraints.   
	
	One might think that excluding dependence on the auxiliary normals is
	necessary to avoid anomalies, just as for the rescaling freedom.
	However this is not actually the case, since our \phase space definitions
	$\sls$, $\sls_\mfp$ and $\sls_\mfq$ depend on a choice of two surface
	$S_0$.  It follows that on the surface $\Sp$, for each choice of normal $\ell^i_\ps$
	there is a unique preferred auxiliary normal $n_{\ps\, i}$, defined
	as follows.  We define a function $u$ on $\Sp$ by setting $u=0$ on
	$S_0$, and by solving along each generator of $\ell^i_\ps$ on $\Sp$
	the ordinary differential equation
	\be
	(\lie_{\ell_\ps} + \kappa_\ps) u = 1.
        \label{udef11}
	\ee
	The function $u$ then transforms under rescalings by $u \to
	e^{-\sigma_\ps} u$.  We set $n_{\ps\,i} = -D_i u/(1 -\kappa_\ps u)$,
        where the factor of $(1-\kappa \ps u)$ enforces the normalization condition (\ref{auxnormal}) from Eq.\ (\ref{udef11}),
        and similarly define
	$n_{\ms\,i'}$ on   
	$\Sm$.  The auxiliary normals $n_{\ps\,i}$ and $n_{\ms\,i'}$ are thus
	determined by the definitions of the \phase spaces $\sls$,
	$\sls_\mfp$ and $\sls_\mfq$, and so any dependence on these definitions
	will not induce any anomalies\footnote{There will be anomalies if we
		use the definitions of auxiliary normals associated with $\sls$, but
		consider symmetry vector fields $\xi^a$ associated with a different
		\phase space $\sls'$ which do not preserve $S_0$ and instead preserve a
		different two-surface $S_0'$.} for symmetries in
	the corresponding symmetry algebras.

 	Rather than using the principle of vanishing anomalies, we now
 	invoke instead the principle discussed in Sec.\ \ref{sec:eps} above, which
	mandates that for given subspace $\sls_{\partial \Sigma}$ of \phase space,
	charges should not depend on the choice of \phase space $\sls$
	which contains it.  In the present context this forbids any dependence
	on the choice of $S_0$, and so it justifies our requiring that
        the boundary and corner fluxes do not depend the choices of
        auxiliary normals.

	\subsection{Noether charge}
        
	We now turn to computing the charges.
	We start by considering the maximal \phase space $\sls$ and the Noether
	charge (\ref{eqn:noethercurrent}).  The Noether charge two-form is given by
	\be
	Q'_{\xi\,\, ij} = - \frac{1}{16 \pi} \volume_{ijk} q^k,
	\label{qij}
	\ee
	where
	\be
	q^c = g^{cd} \lie_\xi \ell_d + \lie_\xi \ell^c - 2 \xi^b
	\nabla_b \ell^c.
	\label{qcdef}
	\ee
	This expression was derived in CFP in the context of a smaller \phase
	space, but that derivation is valid also in the present context.
	All of the terms on the right hand side of Eq.\ (\ref{qcdef}) are intrinsic
	vectors, which justifies the notation in Eq.\ (\ref{qij}).
	The first term can be written using Eqs.\ (\ref{idents12}) as $\gamma \ell^c$.
	In the second term, both ${\vec \xi}$ and ${\vec \ell}$ are intrinsic
	vectors on $\Sp$, and so it can be written as $\lie_\chi \ell^k$.
	Finally the third term can be written as $-2 \chi^l {\cal K}_l^{\ k}$
	as argued in CFP.  Combining these results yields
	\be
	Q'_{\xi \, ij} = \frac{1}{16 \pi} \eta_{ijk} \left[ 2 \chi^l {\cal
		K}_l^{\ k} - \gamma \ell^k - (\lie_\chi \ell)^k \right].
	\label{noetherans}
	\ee
	If we specialize to the restricted \phase space $\sls_\mfp$, then from Eq.\ (\ref{iidd1}) the Noether
	charge reduces to 
	\be
	Q'_{\xi \, ij} = \frac{1}{16 \pi} \eta_{ijk} \left[ 2 \chi^l {\cal
		K}_l^{\ k} - \gamma \ell^k - \beta \ell^k \right].
	\ee
	Specializing further to $\sls_\mfq$ and using the identity
	(\ref{iidd}) gives the expression given in CFP:  
	\be
	Q'_{\xi \, ij} = \frac{1}{8 \pi} \eta_{ijk} \left[ \chi^l {\cal
		K}_l^{\ k} - \beta \ell^k \right].
	\ee

	\subsection{General decomposition of presymplectic potential}

	We use the standard expression for the presymplectic potential for
	vacuum general relativity \cite{WZ}
	\be
	\theta'_{abc}  = \frac{1}{16\pi} \varepsilon_{abc}{}^d~  ( g^{ef} \nabla_{d}h_{ef} - \nabla^{e}h_{de} ). \label{Presymplectic form}
	\ee
	For the maximal \phase space $\sls$, the pullback of this
	expression to  
	a null surface was derived in Ref.\ \cite{Chandrasekaran:2020wwn}
	using the convention (\ref{conv1})
	and is given by
	\begin{eqnarray}
	\label{theta000}
	\theta'_{ijk} &=& \frac{\eta_{ijk}}{16 \pi} \left[ 2 \delta \kappa 
	+ \lie_\ell h + \frac{1}{2} \Theta h + h_{ij}
	\sigma_{kl} q^{ik} q^{jl} +     {\hat D}_i \delta \ell^i +
	2 \kappa  n_i \delta \ell^i
	- \Theta n_i \delta \ell^i -2 \varpi_i \delta \ell^i
	\right]. \ \ \ \ 
	\end{eqnarray}
	Here some of the notations are defined in Appendix \ref{app:nullreview} and
	Sec.\ \ref{sec:bcs},
	and ${\hat D}_i$ is the divergence operator associated with the volume
	form $\eta_{ijk}$.  This expression must be invariant 
	under the rescaling (\ref{rescale00}) of the normal,
	since it was derived from the expression (\ref{Presymplectic form})
	which is explicitly invariant.  The invariance can also be checked
	directly using Eqs.\ (\ref{contransformd}),
	(\ref{Vcontransformc}) and (\ref{contransformc}).   
	Similarly it is invariant under the transformation (\ref{Deltandef}) of the
	auxiliary normal, as can be checked from Eqs.\ (\ref{tz1}),
	(\ref{tz2}) and (\ref{tz3}). 
	Therefore it is well defined one-form on the \phase space $\sls$.

	Our goal now is to decompose this presymplectic potential into an
	exact term, a total variation, and a flux term, as in Eq.\ (\ref{decompos11}).
	As discussed in Sec.\ \ref{sec:pr} above, we will restrict attention to
	decompositions which are invariant under rescalings and changes of
	auxiliary normal.
	By trial and error using the various quantities defined in Appendix
	\ref{app:nullreview}, we find that there is a unique three parameter family of
	such decompositions which we now describe.
	
	We define the boundary term $\ell'_{ijk}$ by
	\be
	\label{ellf}
	16 \pi \ell'_{ijk} = c_3 \volume_{ijk}  \Theta,
	\ee
	where $c_3$ is an arbitrary constant.  We exclude
	terms proportional to $\kappa \volume_{ijk}$ used in Ref.\ \cite{Chandrasekaran:2020wwn}
	since they are not invariant under rescaling.  
	We also exclude terms proportional to
	$\volume_{ijk} ( \kappa - \lie_\ell \, {\rm ln} \, \expansion)$, which
	are rescaling invariant, since they are ill-defined in the region of
	\phase space where $\Theta =0$.  Taking a variation of the boundary
	term (\ref{ellf}) and using the
	identities (\ref{addi1}) and (\ref{deltaTheta}) gives
	\be
	16 \pi \delta \ell'_{ijk} = \volume_{ijk} \left[  
	\frac{1}{2} c_3 h \Theta
	+    \frac{1}{2} c_3 \lie_\ell h 
	+ c_3 {\hat D}_i \delta \ell^i \right].
	\label{deltaellf}
	\ee
	
	Next we define the corner term $\beta'_{ij}$ by
	\be
	\label{cornerdef}
	16 \pi \beta'_{ij} = \left( \frac{1}{2} c_3 + c_2 \right) h \mu_{ij} +
	\left( c_3 + c_5 \right) \eta_{ijk} \delta \ell^k,
	\ee
	where $c_2$ and $c_5$ are additional arbitrary constants.
	Taking an exterior derivative using the identities (\ref{addi2}) and
	(CFP,3.25) gives
	\be
	\label{dbeta}
	16 \pi (d \beta')_{ijk}  = \volume_{ijk} \left[ \left( \frac{1}{2}
	c_3+ c_2\right)
	(\lie_\ell h  + h \Theta) + (c_3 + c_5) {\hat D}_l \delta \ell^l
	\right].
	\ee

	Now combining Eqs.\ (\ref{theta000}), (\ref{deltaellf}) and (\ref{dbeta}) with the general form
	(\ref{decompos11}) of the decomposition yields that the flux term
	is
	\begin{eqnarray}
	\label{fluxnew}
	{\cal E}_{ijk} &=& \frac{\eta_{ijk}}{16 \pi} \left[ 2 \delta \kappa +
	(1-c_5) {\hat D}_i \delta \ell^i + (1 - c_2) \lie_\ell h + \left( \frac{1}{2}-c_2\right) \Theta h + h_{ij}
	\sigma_{kl} q^{ik} q^{jl} 
	\right. \nonumber \\
	&& \left. + 2 \kappa  n_i \delta \ell^i - \Theta n_i \delta \ell^i -2 \varpi_i \delta \ell^i  \right].
	\end{eqnarray}
	Note that this flux does not have the Dirichlet
	form\footnote{Note that the perturbations $h_{ij}$ and $\delta \ell^i$
		are not independent 
		because of the constraint (\ref{con11}).
		It follows that the flux (\ref{ddd}) depends on $(\pi^{ij}, \pi_i)$ only through
		the equivalence class $[\pi^{ij}, \pi_i]$, where we define the
		equivalence relation $({\tilde \pi}^{ij}, {\tilde \pi}_i) \sim
		(\pi^{ij}, \pi_i)$ if there exists a $\chi^i$ for which
		\be
		{\tilde \pi}^{ij} = \pi^{ij} + \ell^{(i} \chi^{j)}, \ \ \ \ \
		{\tilde \pi}_i = \pi_i + q_{ij} \chi^j.
		\ee
		For the variables $\pi^{ij}$ and $\pi_i$ defined in Ref.\ 
		\cite{Chandrasekaran:2020wwn}, the equivalence class is invariant
		under changes $\Delta n^i$ of the auxiliary normal, with the
		transformation vector $\chi^i$ given by
		\be
		16 \pi \chi^i = 2 q^{ij} \sigma_{jk} \Delta n^k - (2 \kappa + \theta)
		\Delta n^i + (\sigma_{jk} \Delta n^j \Delta n^k) \ell^i.
		\ee
	}
	\be
	\label{ddd}
	{\cal E}_{ijk} = \pi^{ij} h_{ij} + \pi_i \delta \ell^i.
	\ee
	that was used in Ref.\ \cite{Chandrasekaran:2020wwn}.
	It is not possible to achieve Dirichlet form while maintaining
	invariance under the two transformations.
	If we specialize to the restricted \phase space $\sls_\mfp$, the
	flux expression is given by using Eq.\ (\ref{Psidef})
	in Eq.\ (\ref{fluxnew}):
	\begin{eqnarray}
	\label{fluxnew1}
	{\cal E}_{ijk} &=& \frac{\eta_{ijk}}{16 \pi} \left[ 2 \delta \kappa 
	+ (1-c_5) \lie_\ell \Psi + (2 - c_5) \Theta \Psi
	+ (1 - c_2) \lie_\ell h + \left( \frac{1}{2}-c_2\right) \Theta h \right.
	\nonumber \\
	&& \left. + h_{ij}
	\sigma_{kl} q^{ik} q^{jl}  - 2 \kappa \Psi 
	\right].
	\end{eqnarray}
	The corresponding result for the \phase space $\sls_\mfq$ is given by setting $\Psi = 0$.

	Turn now to the choice of corner flux, which is defined via a decomposition 
	of the form (\ref{eqn:bldecomp}) of $\beta' - \lambda'$.  Note that
	$\lambda'$ vanishes here by Eq.\ (\ref{eqn:noncovtheta}) since our
	presymplectic form (\ref{Presymplectic form}) is covariant.
	The only corner term $c_{ij}'$ allowed by the two invariances is a
	term proportional to $\mu_{ij}$.  Terms proportional to $\kappa
	\mu_{ij}$, $\Theta \mu_{ij}$ or $m \mu_{ij}$ are
	disallowed by rescaling invariance from
	Eqs.\ (\ref{contransformc}). Terms proportional to $D_{[i} n_{j]}$ or
	$D_{[i} \varpi_{j]}$ are disallowed by invariance under choice of
	auxiliary normal, by Eqs.\ (\ref{contransformd}).  We therefore take
	\be
	16 \pi c_{ij}' = c_6 \mu_{ij}
	\ee
	for some constant $c_6$.  We take $\gamma'=0$ since there are no
	one-forms allowed by the invariances.  The resulting corner
	flux from Eqs.\ (\ref{eqn:bldecomp}) and (\ref{deltamu}) is then
	\be
	16 \pi \cflx_{ij} = \left( \frac{1}{2} c_3 + c_2 + \frac{1}{2} c_6
	\right) h \mu_{ij} + (c_3 + c_5 + c_6) \eta_{ijk} \delta \ell^k.
	\label{cornerfluxval}
	\ee
	We note that the choice of corner term $c'_{ij}$ does not affect the
	charge expression (\ref{eqn:Hxi}) since its anomaly is vanishing.
	Therefore the charges will not depend on the parameter $c_6$.
	On the other hand the polarization (\ref{polarizationanswer}) will
	depend on this choice.

	\subsection{Derivation of charges}
	\label{sec:chargeder}
	
	We can now derive an expression for the quasi-local charges ${\tilde H}_\xi$ by substituting the expressions
	(\ref{noetherans}), (\ref{ellf}) and (\ref{cornerdef}) into the
	general formula given by Eqs.\ (\ref{hfinal}) and (\ref{eqn:Hxi}),
	using the fact that $c'_{ij}$
	has no anomaly.  Evaluating the corner term $-I_{\hat \xi} \beta'_{ij}$ in
	Eq.\ (\ref{eqn:Hxi}) requires replacing the metric perturbation $h_{ab}$ with $\lie_\xi g_{ab}$
	to obtain $-
	\beta'(g_{ab}, \lie_\xi g_{ab})$.
	For the trace $h$ we obtain
	\be
	\label{hval}
	h = 2 \nabla_a \xi^a = 2 {\hat D}_i \chi^i + 2 \gamma,
	\ee
	where we have used Eqs.\ (CFP,3.27) and (\ref{idents12}).
	Using this together with the identities (\ref{symt}) gives
	\begin{eqnarray}
	\label{finalH}
	{\tilde H}_\xi &=& \frac{1}{16 \pi} \int_{\partial \Sigma} \eta_{ijk}
	\left[ 2 \chi^l {\cal
		K}_l^{\ k} - (1 + 2 c_2 - c_5) \gamma \ell^k - (1 + c_3 + c_5)
	(\lie_\chi \ell)^k \right. \nonumber \\
	&& \left. + c_3 \Theta \chi^k - (c_3 + 2 c_2) \ell^k {\hat D}_l \chi^l \right].
	\end{eqnarray}
This expression is invariant under rescaling of the normal, from
Eqs.\ (\ref{contransformc}).  It is also actually independent of the
parameter $c_3$, since the terms  
	proportional to $c_3$ are exact\footnote{This follows from the
		general transformation theory of Ref.\ \cite{Chandrasekaran:2021vyu},
		since changes in $c_3$ corresponds to changing the boundary term
		$\ell'$ by an exact form and the corner term $\beta'$ by a total
		variation, keeping the symplectic potential $\theta'$ and boundary
		flux ${\cal E}$ fixed.  This transformation corresponds the following
		combination of transformations listed in Table 2 of
		Ref.\ \cite{Chandrasekaran:2021vyu}: $a = - d \sigma$, $\nu = \delta
		\sigma$ and $e = \sigma$, where $\sigma = c_3 \mu / (16 \pi)$.}:
	\be
	- d(i_\chi \mu)_{ijk} = \eta_{ijk} \left[ \Theta \chi^k - \lie_\chi
	\ell^k - \ell^k {\hat D}_l \chi^l \right].
	\ee
	Thus one can set $c_3=0$ in Eq.\ (\ref{finalH}), although alternative
	forms of the expression can be obtained by taking $c_3=-1-c_5$ or $c_3
	= - 2 c_2$.  Hence the charges depend on the two parameters $c_2$ and $c_5$.
	
	There are a number of special cases of the general charge expression (\ref{finalH}).
	First, specialized to supertranslations $\chi^i  = f \ell^i$ it reduces to
	\be
	{\tilde H}_{\xi} =
	\frac{1}{16\pi} \int_{\partial \Sigma} \eta_{ijk} \ell^k \left[ 2 \kappa f + (1-2c_2+ c_5) \lie_\ell f - 2 c_2 \Theta f - (1 + 2 c_2 - c_5) \gamma \right].
	\label{chargenew3}
	\ee
	Second, if we consider the restricted \phase space $\sls_\mfp$,
	we can replace the term $\lie_\chi \ell^k$ in Eq.\ (\ref{finalH}) with $\beta \ell^k$ from Eq.\ (\ref{iidd1}).
	Third, if we specialize further to the \phase space $\sls_\mfq$ we can
	use Eq.\ (\ref{iidd}) to replace $\gamma$ with $\beta$.
	Taking $c_3 = -2 c_2$ to simplify the final answer then yields
	\be
	{\tilde H}_{\xi} =
	\frac{1}{8\pi} \int_{\partial \Sigma} \eta_{ijk}  \left[ \chi^l {\cal K}_l^{\ k} - \beta \ell^k - c_2 \Theta \chi^k \right].
	\label{chargenew4}
	\ee
	The charge expression (CFP,6.27) is the special case of this result when $c_2=1$.

	We next discuss some of the properties of the charges,
	and some criteria that can be used to restrict the choices of the constants $c_2$, $c_3$ and $c_5$,
	paralleling the discussion in a more general context given in Sec.\ \ref{sec:survey}.
	
	\begin{itemize}
		
		\item We have imposed that the decomposition (\ref{decompos11}) 
		and the resulting charges be invariant under the
		symmetry (\ref{rescale00}) of rescaling of the null normal $\ell^i$, and
		under the symmetry of changing the choice (\ref{Deltandef}) of auxiliary
		normal $n^a$.

		\item Wald and Zoupas proposed using a stationarity criterion to
		reduce the ambiguity in the decomposition of the presymplectic
		potential \cite{Wald:1999wa}.  They require that the flux ${\cal E}$
		should vanish for all perturbations about a stationary background,
		when $\Theta$ and $\sigma_{ij}$ vanish.
		In the special case of a \phase space in which
                $\delta \kappa=0$
                is imposed, it is possible to satisfy this criterion by imposing
		$c_2 = 1$.  This was the approach followed in 
		the analysis of the $\delta \kappa=0$ \phase space of CFP
		who additionally took $c_3=-2$.  It is not possible to
 		satisfy the Wald-Zoupas stationarity condition in any of the three larger \phase
 		spaces used here (as noted previously for $\sls_\mfq$ in Appendix H of CFP).
		See Sec.\ \ref{sec:stationarity} below for further discussion.

		\item As noted above, the charges (\ref{finalH}) do not depend on $c_3$, 
		and so this parameter
		is unimportant.  It parameterizes a kind of gauge freedom in the
		formalism (see Refs.\ \cite{Chandrasekaran:2020wwn,Chandrasekaran:2021vyu} for a general discussion).
		
		\item The Dirichlet condition suggested by
		Chandrasekaran and Speranza \cite{Chandrasekaran:2020wwn} imposes
		that the flux $\beom$ be of the form
		$p dq$ where the variables $q$ are geometric quantities that are
		intrinsic to the surface.  This requires breaking the rescaling
		symmetry.  Their analysis corresponds to the parameter choices
		$c_3=0$, $c_2=1$, $c_5=1$, together with adding to the right hand
		side of Eq.\ (\ref{ellf}) a term $-2 \kappa \eta_{ijk}$ and adding
		the corresponding variation to the flux (\ref{fluxnew}).

		\item As noted above, there is an additional set of terms which
		respect the two invariances which could be added to the decomposition (\ref{decompos11}).
		Specifically one could add $- 2 c_4 (\kappa - \lie_\ell \ln \Theta) \eta_{ijk}$ 
		to the expression (\ref{ellf}) and $2 c_4 \lie_\ell( \delta \Theta
		/\Theta)$ to the expression (\ref{cornerdef}), where $c_4$ is an
		arbitrary constant, with corresponding changes to the flux (\ref{fluxnew}).
		This gives rise to terms involving $\ln \Theta$ in the charges.
		These terms are well defined everywhere in \phase space except for
		configurations with $\Theta =0$, and in particular for stationary backgrounds.
		In that limit the charge prescription is not well
		defined, and cannot be made so by any suitable limiting procedure.
		For this reason we restrict attention in this paper to the case
		$c_4=0$ \footnote{However we do note that in applications to black hole
			horizons, in so called stationary-to-stationary transitions, that 
			the initial ``stationary'' region is usually approximated as
			stationary since the corrections to stationarity are exponentially
			small at early times.  In this context the $\ln \Theta$ terms in the charges
			are well defined and moreover the exponentially small corrections now
			give order unity corrections to the charges and must be retained.}.

		\item The free parameter $c_2$ in the charge formula
		(\ref{chargenew4}) is closely related to
		recent proposals to define the entropy of a dynamical black hole,
		involving adding a correction to the area term proportional to the
		time derivative of the area \cite{WaldZhang,Rignon-Bret:2023fjq}.

	\end{itemize}

	\subsection{Derivation of fluxes}
	\label{sec:fluxder}

	We now turn to the evaluating the flux, which from Eq. (B.11) of 
	\cite{Chandrasekaran:2021vyu} is given by
	\be
	d {\tilde h}_\xi = I_{\hat \xi} {\cal E} - \Delta_{{\hat \xi}} ( \ell'
	+ b' + d c') - i_\xi( L' + d \ell').
	\ee
	The last term here will not contribute since $\xi^a$ is tangent to the
	boundary at $\Sp$.  The second term will also not contribute since
	there are no anomalies in our construction.  Thus we need to evaluate
	the flux expression (\ref{fluxnew}) at $h_{ab} = \lie_\xi g_{ab}$.
	Using the expressions (\ref{hval}) for $h$ and (\ref{symt}) for $\delta
	\ell^i$, $h_{ij}$ and $\delta \kappa$ gives
	\bea
	\label{gflux}
	16 \pi (d {\tilde h}_\xi)_{ijk} &=& \eta_{ijk} \bigg[ 2 \lie_\chi
	\kappa + (c_5 - 2 c_2 -1) (\lie_\ell \gamma + \Theta \gamma)
	+ (1 + c_5 - 2 c_2) {\hat D}_l \lie_\ell \chi^l  \ \ \ \ \  \nonumber \\
	&&  + (2 - 2 c_2) \lie_\chi \Theta
	+ (1- 2 c_2) \Theta {\hat D}_l\chi^l 
	+ \lie_\chi q_{ij}
	\sigma_{kl} q^{ik} q^{jl} \nonumber \\
	&& - (2 \kappa n_l - \Theta n_l - 2 \varpi_l) \lie_\ell \chi^l
	\bigg].
	\eea
	Here we have also used the identity
	\be
	\lie_\ell {\hat D}_i \chi^i= {\hat D}_i \lie_\ell \chi^i+ \lie_\chi \Theta,
	\ee
	obtained from Eqs.\ (CFP,3.25), (CFP,3.28) and
	(\ref{grelations1c}).

	If we instead consider the restricted \phase space $\sls_\mfp$,
	the flux (\ref{gflux}) can be simplified using Eqs.\ (\ref{iidd1})
	and (\ref{symt2a}) which yields
	\bea
	\label{fluxans2}
	16 \pi (d {\tilde h}_\xi)_{ijk} &=& \eta_{ijk} \bigg[ 2 \lie_\chi
	\kappa -2 \beta \kappa - 2 \lie_\ell \beta
	+ (1- 2 c_2) \Theta {\hat D}_l\chi^l  - \beta \Theta \ \ \ \ \  \nonumber \\
	&&  
	+ (2 - 2 c_2) \lie_\chi \Theta + \lie_\chi q_{ij}
	\sigma_{kl} q^{ik} q^{jl}
	+ (1 + 2 c_2 - c_5) (\lie_\ell + \Theta) \Psi
	\bigg].
	\eea
	The corresponding result for $\sls_\mfq$ is given by setting $\Psi=0$,
	from Eqs.\ (\ref{symt2a}) and (\ref{iidd}).

We now turn to a discussion of a special class of symmetries, the
boosts and half-sided boosts discussed in Sec.\ \ref{halfsidedboosts}
above.  Ciambelli, Freidel and Leigh (CFL) \cite{Ciambelli:2023mir}
showed that the charges associated with these symmetries increase
monotonically along the null surface, if one chooses the parameter
$\alpha_0$ entering the definition (\ref{boostdef2}) of the boost symmetry to have the
value $\alpha_0=1/2$.  In Appendix \ref{app:boostflux} we generalize
that analysis to the context of this paper, where the charges and
fluxes depend on the parameters $c_2$ and $c_5$.  We show that the
required value of $\alpha_0$ to ensure monotonicity is
\be
\label{vvv}
\alpha_0 = 1 - c_2,
\ee
in agreement with CFL in the case considered there of $c_2 = 1/2$.

	\subsection{Stationarity properties of charges and fluxes}\label{sec:stationarity}

	In this section, we show explicitly that the Wald-Zoupas stationarity
	requirement cannot be satisfied in the context of our \phase
	space $\sls_\mfq$.  We specialize to field perturbations of the form (\ref{metricsym}) generated by symmetries.
	We specialize the rescaling freedom to
	make $\kappa =0$, and choose coordinates $(u,\theta^A)$ on $\Sp$ such
	that ${\vec \ell} = \partial_u$ and $u=0$ on $S_0$.  For a stationary
	background the Killing field will then have the form\footnote{This follows from the fact that the surface gravity associated with the Killing field is constant over the null surface \cite{Ashtekar:2001jb,CFP}, so we can rescale $\tau^i$ to make the surface gravity unity.  Writing $\tau^i = u \ell^i$ for some function $u$ on the surface then yields $(\lie_\ell + \kappa) u=0$, which when $\kappa=0$ yields Eq.\ (\ref{tauform}).}       
	\be
        \label{tauform}
	\tau^i = [u - u_0(\theta^A) ] (\partial_u)^i
	\ee
	for some function $u_0$.  We consider a general symmetry ${\vec \chi}$
	of the form (\ref{gensym}).  From Eqs.\ (\ref{iidd}) and (\ref{iidd1}) we
	obtain $\beta = \gamma = -f_{,u}$, and then Eq.\ (\ref{symt4}) gives
	$\delta \kappa = f_{,uu}$.  The result (\ref{hval}) yields $h = 2 D_A Y^A$ and now
	substituting into the flux expression (\ref{fluxnew1}) with $\Psi=0$
	using $\Theta = \sigma_{ij} =0$ by stationarity gives
	\be
	\beom_{ijk} = \frac{\eta_{ijk}}{8 \pi} f_{,uu}(u,\theta^A).
	\ee
	This flux cannot be made to vanish for all symmetries, for any values
	of the free parameter $c_2$.  It does vanish for symmetries of the
	form $f = \alpha(\theta^A) - \beta(\theta^A) u$, which is precisely
	the subgroup of symmetries studied in CFP.
	Any more general time dependence of the symmetries yields nonzero fluxes.
	
	It does not appear to be possible to generalize the Wald-Zoupas
	criterion to make it generally applicable to situations with time
	dependent symmetries.  In the case of the \phase space $\sls_\mfq$, if we consider the fluxes $\beom(\phi,\delta \phi)$ of
	general field perturbations and not those (\ref{metricsym}) generated
	by symmetries, then demanding that the 
	term $(1-c_2) \lie_\ell h$ in the flux (\ref{fluxnew1})
	vanish does select the value $c_2=1$.  However, in this case the first
	term in (\ref{fluxnew1}) is generically nonzero as shown above, so it
	is not possible to demand that $\beom(\phi,\delta \phi)$ vanish.  
	An alternative approach is to restrict attention to 
	the fluxes 
	$\beom(\phi,\lie_\xi
	\phi)$ generated by symmetries, as advocated by
	Rignon-Bret \cite{Rignon-Bret:2023fjq}.  In this case one could hope
	to give a general prescription for defining ``time independent''
	symmetries\footnote{For example, for the particular case of the \phase space
		$\sls_\mfq$, a suitable definition of time independent
		symmetry is
		given by the condition [cf.\ Eq.\ (CFP,4.11b)]
		\be
		\lie_\ell \beta + \kappa \beta - \lie_\chi
		\kappa=0,
		\ee
		which is the rescaling invariant version of the condition
		$f_{,uu}=0$.}, and to demand only that $\beom(\phi,\lie_\xi \phi)$
	vanish for such symmetries.  This approach might be useful
	for clarifying how physically reasonable stationarity requirements are
	compatible with the framework being used.  However, for the purpose
	of restricting the choice of decomposition (\ref{decompos11}), the
	approach does not seem to be useful, at least in the context of
	the \phase space
	$\sls_\mfq$.  This is because
	the term $(1-c_2) \lie_\ell h$ in the flux (\ref{fluxnew1})
	vanishes for all symmetries on stationary backgrounds as shown above,
	and so no such modified version of the 
	Wald-Zoupas criterion can restrict the value of $c_2$.

 	\section{Independent degrees of freedom and Poisson brackets}		
	\label{sec:interp}

        In this section we derive independent coordinates on the
        space of initial data structures of the form (\ref{ids2}) by fixing the rescaling degrees of
        freedom.  We then compute the presymplectic form and mod out
        by its degeneracy directions to obtain the physical phase
        space,
        and derive the Poisson brackets.
        The Poisson brackets exhibit the well known relation
        \cite{1993PhRvD..47.3275H,Carlip:1993sa,Lehner2016,Hopfmuller2017a}
        between
        transformations generated by the area functional and the
        degree of freedom corresponding to half-sided boosts,
        discussed in Sec.\ \ref{halfsidedboosts}.
        An alternative approach to the symplectic structure and
        Poisson brackets for a pair of intersecting null surfaces has
        been given by Reisenberger \cite{Reisenberger:2007pq,Reisenberger:2007ku}.
        Poisson brackets associated with a single null surface have
        been derived in Refs.\ \cite{Hawking:2016sgy,Ciambelli:2023mir}.
                
        For simplicity our analysis will be subject to three restrictions:
	
	\begin{itemize}
		
		\item We will consider only the restricted \phase space
		$\sls_\mfp$ in which the variation of the normal is proportional to the normal,
		$\delta \ell^i \propto \ell^i$, together with its extension
                  $\sls'_\mfp$ discussed in Sec.\ \ref{halfsidedboosts}.
		
		\item We will restrict attention to perturbations about stationary
		backgrounds in which $\Theta = \sigma_{ij} = 0$.
		
		\item We will further restrict the \phase space under consideration by
		imposing conditions appropriate for an event horizon, namely that
		the expansion $\Theta$ is always positive and the area element
		becomes constant at late times.  This will be implemented by
		specifying a boundary condition for solutions of the
                Raychaudhuri equation at
		late times instead of at an initial time.
		
	\end{itemize}

	We start in Sec.\ \ref{sec:ic} by first specifying a set of independent
	coordinates on the space of initial data satisfying the constraint (\ref{restrict1}), by fixing the
        rescaling freedom and solving the
	constraint equations along the null surface.  Then, in Sec.\ \ref{sec:sf},
	we determine the form of the presymplectic form in these
	coordinates, both for the \phase space $\sls_\mfp$ and its
        extension $\sls_\mfp'$.  Modding out by degeneracy directions
        then yields the physical phase spaces $\slp_\mfp$ and
        $\slp_\mfp'$, and allows us to compute Poisson brackets.
        Finally in Sec.\ \ref{sec:area} we discuss
        the action of the area operator.

	\subsection{Independent phase space coordinates on
		stationary backgrounds obtained by solving constraints}
	\label{sec:ic}

	As discussed in Sec.\ \ref{sec:group1}, the restricted phase space
	$\slp_\mfp$ is in one-to-one correspondence with the set of
	extended initial data structures $\mfh$ of the form (\ref{ids2}),
	subject to the restriction (\ref{restrict1}) and modulo edge
        mode issues.  We now define a set of
	independent coordinates on this space of initial data, deferring
        the edge modes to Sec.\ \ref{sec:sf}.
	
	The starting point is the set of variables
	\be
	\label{list}
	\left( q_{AB}, {\bar \omega}_A, m,
	\Theta_\ps, \Theta_\ms, \ell_{\ps\,a}, \ell^i_\ps, \kappa_\ps, [q^\ps_{ij}],
	\ell_{\ms\,a}, \ell^{i'}_\ms, \kappa_\ms, [q^\ms_{i'j'}] \right)
	\ee
	in a representative of the equivalence class $\mfh$.
	First, instead of the conformal equivalence
	class $[q^\ps_{ij}]$ of the induced metric on $\Sp$, we use the
	variable ${\bar q}^\ps_{ij}$ defined in Sec.\ \ref{sec:idsalternative},
	which contains the same information, and similarly on $\Sm$.
	Next, we pick a representative
	\be
	\left( \ell_{\ps\,0}^{a}, \ell_{\ms\,0}^a \right)
	\label{list2}
	\ee
	of the boundary structure $\mfp$.
	It follows from the  
	constraint (\ref{restrict1}) that the normal vectors in (\ref{list}) are related
	to those in (\ref{list2}) by rescalings.  That is, there exist functions
	$\rho_\ps$ on $\Sp$ and $\rho_\ms$ on $\Sm$ for which
	\bes
	\label{rhodef}
	\bea
	\ell_{\ps}^{a} &=& e^{\rho_\ps} \ell_{\ps\,0}^{a}, \\
	\ell_{\ms}^{a} &=& e^{\rho_\ms} \ell_{\ms\,0}^{a}.
	\eea
	\ees
	The variables $\rho_\ps$ and $\rho_\ms$ transform under the rescalings
	(\ref{2rescalings}) as $\rho_\ps \to \rho_\ps + \sigma_\ps$, $\rho_\ms
	\to \rho_\ms + \sigma_\ms$.
	The list of variables (\ref{list}) now
	reduces to
	\be
	\label{list1}
	\left( q_{AB}, {\bar \omega}_A, m,
	\Theta_\ps, \Theta_\ms, \ell_{\ps\,a}, \rho_\ps, \kappa_\ps, {\bar q}^\ps_{ij},
	\ell_{\ms\,a}, \rho_\ms, \kappa_\ms, {\bar q}^\ms_{i'j'} \right).
	\ee

	Next, it will be convenient to use as free data the variables
	\be
	\label{tildekappadef}
	{\tilde \kappa}_\ps = e^{-\rho_\ps} \kappa_\ps - \lie_{{\vec \ell}_{\ps\,0}} \rho_\ps, \ \ \ \ \
	{\tilde \kappa}_\ms = e^{-\rho_\ms} \kappa_\ms - \lie_{{\vec \ell}_{\ms\,0}} \rho_\ms,
	\ee
	instead of the inaffinities $\kappa_\ps$ and $\kappa_\ms$.
	These variables are invariant under rescalings and are just the inaffinities
	computed with respect to the normal vectors $\ell_{\ps\,0}^i$ and $\ell_{\ms\,0}^{i'}$ rather than
	$\ell_{\ps}^i$ and $\ell_{\ms}^{i'}$.
	We now fix choices of normal covectors $\ell^0_{\ps\,a}$ and $\ell^0_{\ms\,a}$
	that satisfy
	\be
	\left. \ell^0_{\ps\,a} \ell^a_{\ms\,0}  \right|_{S_0} = \left. \ell^0_{\ms\,a} \ell^a_{\ps\,0}  \right|_{S_0} = -1
	\label{con55}
	\ee
	and
	use the rescaling freedom (\ref{2rescalings}) to enforce
	\be
	\label{fixell}
	\ell_{\ps\,a} = \ell^0_{\ps\,a}, \ \ \ \ \ \ell_{\ms\,a} = \ell^0_{\ms\,a}.
	\ee
	This fixes the rescaling freedom.  It then follows from
	Eqs.\ (\ref{con22}), (\ref{rhodef}) and (\ref{con55})   
	that
	\be
	\left. \rho_\ps \right|_{S_0} = \left. \rho_\ms \right|_{S_0} = m,
	\label{con66}
	\ee
	so we can omit $m$ from our list of free data.
	We also note that the expansions $\Theta_\ps$ and $\Theta_\ms$ on $S_0$
	provide initial data that determines solutions of the Raychaudhuri
	equations on $\Sp$ and $\Sm$.  However when we impose late time
	boundary conditions on the Raychaudhuri equations as discussed above,
	$\Theta_\ps$ and $\Theta_\ms$ are no longer free data and instead are
	determined in terms of the other variables (see below for more
	details).  With these choices and conventions the free data reduce to
	\be
	\label{lists}
	\left( q_{AB}, {\bar \omega}_A, {\tilde \kappa}_\ps, \rho_\ps, {\bar q}^\ps_{ij},
	{\tilde \kappa}_\ms, \rho_\ms, {\bar q}^\ms_{i'j'} \right).
	\ee

	We can reconstruct from the data (\ref{lists}) the full set of data by solving the 
	Raychaudhuri equation (\ref{grelations1e}) with $R_{ab}=0$.  Let us focus on $\Sp$ and drop the $+$
	subscripts for simplicity, and use coordinates
	$(u,\theta^A)$ with $u = 0$ on $S_0$ and ${\vec \ell}_0 = \partial_u$.
	Then the induced metric is given by $q_{ij} = e^\nu {\bar q}_{ij}$
	from Eq.\ (\ref{q0def}), and the expansion is $\Theta = e^\rho \nu'$.
	Here $\nu$ satisfies
	\be
	\nu'' - {\tilde \kappa} \nu' = - \frac{1}{2} (\nu')^2 - \frac{1}{4}
	{\bar q}'_{ij}  {\bar q}'_{kl} {\bar q}^{ik} {\bar q}^{jl},
	\label{ray1}
	\ee
	and we denote Lie derivatives with respect to ${\vec \ell}_0$ by primes.
	Given the freely specifiable data ${\tilde \kappa}$ and ${\bar q}_{ij}$ on $\Sp$
	we can solve this equation
	to determine $\nu$ on $\Sp$, given the initial condition $\nu =0$ at $u=0$
	and the final condition that $\nu \to $ a constant in the far future.
	The general solution up to second order in the shear is
	\begin{eqnarray}
	\label{raysoln}
	\nu(u) &=& \int_{0}^u d{\bar u} e^{{\alpha}({\bar u})} \int_{\bar u}^\infty d {\bar {\bar u}}
	e^{-\alpha({\bar {\bar u}})} \sigma^2({{\bar {\bar u}}}) \nonumber \\
	&& + \frac{1}{2}
	\int_{0}^u d{\bar u} e^{\alpha({\bar u})} \int_{\bar u}^\infty d {\bar {\bar u}}
	e^{\alpha({\bar {\bar u}})}
	\int_{{\bar {\bar u}}}^\infty d {\hat u} e^{-{\alpha}({\hat u})}
	\int_{{\bar {\bar u}}}^\infty d {\hat {\hat u}} e^{-{\alpha}({\hat {\hat u}})}
	\sigma^2({{\hat u}}) \sigma^2({{\hat {\hat u}}}),
	\end{eqnarray}
	where we have defined [cf.\ Eq.\ (\ref{alphadef0}) above]
	\bes
	\bea
	\sigma^2(u) &=& \frac{1}{4}
	{\bar q}'_{ij}  {\bar q}'_{kl} {\bar q}^{ik} {\bar q}^{jl}, \\
	\label{alphadef}
	\alpha(u) &=&  \int_0^u d {\bar u} \, {\tilde \kappa}({\bar u}).
	\eea
	\ees

 	\subsection{Symplectic form and Poisson brackets}	
	\label{sec:sf}
	
	In this section we compute the polarization (\ref{polarizationanswer})
	and corresponding symplectic form and Poisson brackets in
        terms of the variables defined in  
	the last section.

	We start by expressing the flux (\ref{fluxnew1}) in terms of these variables,
	continuing to specialize to $\Sp$ and dropping the $+$ subscripts.
	Our background quantities are 
	$q_{AB}$ and ${\bar \omega}_A$ on $S_0$, and $\rho$, ${\tilde \kappa}$ and ${\bar q}_{ij}$ on $\Sp$.
	From these we obtain the additional background quantities
	\begin{subequations}
		\begin{eqnarray}
		\kappa &=& e^\rho( {\tilde \kappa} + \rho'),\\
		q_{ij}  &=& e^\nu {\bar q}_{ij} ,\\
		\Theta &=& e^\rho \nu', \\
		\sigma_{ij} &=& {1 \over 2} e^{\nu+\rho} {\bar q}'_{ij},
		\end{eqnarray}
	\end{subequations}
	where primes denote Lie derivatives with respect to ${\vec \ell}_0$
	and we have used Eqs.\ (\ref{q0def}), (\ref{equiv}), (\ref{rhodef}) and (\ref{tildekappadef}).
	Varying these equations and also using Eqs.\ (\ref{Psidef}),
	(\ref{deltaqdef}), (\ref{hformula}),  (\ref{rhodef})
	and (\ref{auxnormal})  
	gives 
	\begin{subequations}
		\begin{eqnarray}
		\delta \kappa &=& e^\rho ( \rho' + {\tilde \kappa}) \delta \rho +
		e^\rho ( \delta \rho' + \delta {\tilde \kappa}), \\
		h_{ij} &=& e^\nu \delta {\bar q}_{ij} + e^\nu {\bar q}_{ij} \delta \nu,\\
		\label{hval1}
		h &=& 2 \delta \nu + {\bar q}^{ij} \delta {\bar q}_{ij} - 2 \delta \rho,\\
		\Psi &=& \delta \rho.
		\end{eqnarray}
	\end{subequations}
	Here in Eq.\ (\ref{hval1}) the second term has vanishing Lie derivative
	with respect to $\ell^i$ from Eq.\ (\ref{mu0def}), and coincides on
	$S_0$ with $q^{AB} \delta q_{AB}$.
	Finally we define $\volume_{0\,ijk} = e^{-\nu} \volume_{ijk}$ which is
	independent of $\nu$. Inserting these results and definitions into the expression (\ref{fluxnew1}) for
	the symplectic potential and using Eq.\ (\ref{rhodef})
	now gives
	\begin{eqnarray}
	16 \pi {\cal E}_{ijk} &=& \volume_{0\,ijk} e^{\nu+\rho} \bigg[ (2 - 2 c_2) \delta
	\nu' + \left(\frac{1}{2} -  c_2\right) \nu' (2 \delta \nu
	+ {\bar  q}^{ij} \delta  {\bar q}_{ij} )
	+ \frac{1}{2} {\bar q}'_{ij}
	\delta {\bar q}_{kl} {\bar q}^{ik}
	{\bar q}^{jl}  \nonumber \\
	&& 
	+ (1 + 2c_2 - c_5) (\delta \rho' + \delta \rho \nu') + 2
	\delta {\tilde \kappa} \bigg].
	\end{eqnarray}

	We next specialize to the coordinate system $(u,\theta^A)$ defined in the
	last section, where $u=0$ on $S_0$ and ${\vec \ell}_0 = \partial_u$.
	In this coordinate system the induced metric only has angular
	components, so we can replace ${\bar q}_{ij}$ with ${\bar q}_{AB}$
	everywhere. We also make the unimodular decomposition
	\be
	{\bar q}_{AB}(u,\theta^A) = \sqrt{q(\theta)} {\hat q}_{AB}(u,\theta^A)
	\ee
	where ${\rm det} \, {\hat q}_{AB}=1$, and where the lack of dependence of
	the prefactor $\sqrt{q}$ on $u$ follows from the definition of
	${\bar q}_{ij}$ given in Sec.\ \ref{sec:idsalternative}.
	Evaluated at $u=0$ this expression reduces to the induced metric
	$q_{AB}$ on $S_0$.
	Integrating over the null surface $\Sp$ now gives for the contribution
	$\Xi_\ps = - \int_{\Sp} \beom$
	to the polarization (\ref{polarizationanswer})
	\bea
	\label{pol1}
	\Xi_\ps&=& \frac{1}{16 \pi} \int_0^\infty du \int d^2\theta \sqrt{q} e^{\nu + \rho} \bigg[
	(2 - 2 c_2) \delta
	\nu' +
	\left(\frac{1}{2} -  c_2\right) \nu' (2 \delta \nu
	+ \delta q / q  )
	\nonumber \\&&
	+ \frac{1}{2} {\hat q}'_{AB}
	\delta {\hat q}_{CD} {\hat q}^{AC}
	{\hat q}^{BD}
	+ (1 + 2c_2 - c_5) (\delta \rho' + \delta \rho \nu') + 2
	\delta {\tilde \kappa} \bigg],
	\eea
	where we used that the orientation of $\Sp$ is opposite to that given
	by the volume form $\eta_{0\,ijk}$.

	We now eliminate $\nu$ in favor of $\rho$, $q$ and ${\hat q}_{AB}$ using
	the general Raychaudhuri solution (\ref{raysoln}).
	We work to linear order in the variations $\delta \rho$, $\delta q$ and $\delta
	{\hat q}_{AB}$.  We also expand the background to linear order about a
	stationary solution with $\rho={\tilde \kappa} =0$, ${\hat q}_{AB}(u,\theta^A) =
	{\hat q}_{0\,AB}(\theta^A)$.
	The terms involving $\nu'$ in the second and fourth terms in Eq.\ (\ref{pol1}) can then be dropped,
	since they are second
	order in deviations from the stationary solution.
	Similarly in the first term we can drop the $e^{\nu + \rho}$ 
	factor and then the $u$ integral can be carried out.
	Using Eqs.\ (\ref{nuS0}) and (\ref{raysoln}) and switching to a
	coordinate $U = \ln u$, the result is
	\bea
	\label{pol2}
	\Xi_\ps 
	&=& \frac{1}{16 \pi} \int_{-\infty}^\infty dU \int d^2\theta \sqrt{q}
	\bigg[\,
	\frac{1}{2} {\hat q}_{AB\,,U}
	(\delta {\hat q}_{CD} + c_0 \delta {\hat q}_{CD\,,U})
	{\hat q}_0^{AC}
	{\hat q}_0^{BD} \nonumber \\
	&& 
	+ (1 + 2 c_2 - c_5) e^\rho \delta \rho_{,U} + 2 e^{\rho + U} \delta {\tilde \kappa}\, \bigg],\ \ \ \ \ 
	\eea
	where $c_0 = 2 - 2 c_2$.
	We see that changing $c_2$ away from the value $c_2=1$ modifies the
	standard expression for the presymplectic potential for the graviton modes
	on the horizon.  Since the correction term is a total variation, it
	does not affect the presymplectic form $\Omega = \delta \Xi$.

	We now turn to deriving explicit expressions for the presymplectic
	potential (\ref{polarizationanswer}) and corresponding symplectic form
	$\Omega = \delta \Xi$, including the corner contributions, that show
	the pairs of conjugate variables.
	Taking a variation of the expression (\ref{pol2}) and diagonalizing
	using the variable redefinitions (\ref{alphadef}) and 
	\be
        \label{zetadef0}
	\zeta = \rho + \frac{1}{c_7} \alpha,
	\ee
	where
\be
c_7 = 1 + 2 c_2 - c_5,
\ee
	gives
	\bea
	\label{pol222}
	\delta \Xi_\ps 
	&=& \frac{1}{16 \pi} \int_{-\infty}^\infty dU \int d^2\theta \sqrt{q}
	\bigg[\,
	\frac{1}{2} \delta {\hat q}_{AB\,,U} \wedge
	\delta {\hat q}_{CD}
	{\hat q}_0^{AC}
	{\hat q}_0^{BD}
	+ c_7 \delta \zeta \wedge \delta \zeta_{,U}
	- c_7^{-1} \delta \alpha \wedge \delta \alpha_{,U}
	\bigg]\ \ 
	\nonumber \\ &&
	+ \frac{1}{16 \pi} \int d^2\theta \bigg\{ \sqrt{q} \, 
	\delta \zeta(\infty) \wedge \delta \alpha(\infty)
	+ \delta \sqrt{q} \wedge \left[ c_7 \delta \zeta(\infty) - c_7 \delta \zeta(-\infty) + \delta \alpha(\infty) \right] \bigg\}. \ \ \ \ 
	\eea
	We re-insert the $+$ subscripts in
	the expression 
	(\ref{pol2}), and add the corresponding expression for the integral
	along $\Sm$.   We also add the corner term at $S_0$
	\be
	- \frac{(c_5 - 2 c_2)}{16 \pi} \int d^2 \theta \, \delta \sqrt{q} \wedge  (\delta \zeta_\ps + \delta \zeta_\ms) 
	\ee
	given by Eqs.\
	(\ref{rels}), (\ref{polarizationanswer}) and
	(\ref{cornerfluxval})\footnote{Note that the same corner contributions
		at $S_0$ are obtained from $\Sp$ and $\Sm$, 
		since the volume forms $\mu_{ij}$ have opposite signs but the induced
		orientations of $S_0$ also have opposite signs.}, using that $\rho =
	\zeta$ on $S_0$.

	We assume that the variables
	$\zeta_\ps$ and $\zeta_\ms$ have finite limits as $U \to \pm \infty$.
	We make the change of variables from $\zeta_\ps$ and $\zeta_\ms$ to the
	even and odd combinations\footnote{\label{uvdef}Here we define coordinates
        $u$ on $\Sp$ and $v$ on $\Sm$ by $u = v =0$ on $S_0$, ${\vec
          \ell}_{\ps0} = \partial_u$ and ${\vec \ell}_{\ms0} = \partial_
        v$.  We use the condition $u=v$ to identify points on $\Sp$
        and $\Sm$, and to define the linear combinations (\ref{zetaeo})
        and similar linear combinations below.}
	\be
        \label{zetaeo}
	\zeta_{\rm e} = \frac{\zeta_\ps + \zeta_\ms}{\sqrt{2}}, \ \ \ \
	\zeta_{\rm o} = \frac{\zeta_\ps - \zeta_\ms}{\sqrt{2}}.
	\ee
	We define the variables ${\bar \zeta}_{\rm e}$, ${\bar \zeta}_{\rm o}$, $\Delta
	\zeta_{\rm e}$, $\Delta \zeta_{\rm o}$ in terms of the limiting values
	\be
        \label{zetabardef}
	\zeta_{\rm e}(U = \pm \infty, \theta^A) = {\bar \zeta}_{\rm e}(\theta^A) \pm \Delta
	\zeta_{\rm e}(\theta^A)/2, \ \ \ \ \zeta_{\rm o}(U = \pm \infty,
	\theta^A) = {\bar \zeta}_{\rm o}(\theta^A) \pm \Delta
	\zeta_{\rm o}(\theta^A)/2.
	\ee
	From the constraint (\ref{con66})
        and the definitions (\ref{alphadef}) and (\ref{zetadef0})
        it follows that \be
        {\bar
			\zeta}_{\rm o} =
	\Delta \zeta_{\rm o}/2.
        \ee
	We also fix a smooth function $g(U)$ with
	\be
	g(U=\pm \infty) = \pm 1/2
	\label{gcondt}
	\ee
	and define ${\tilde \zeta}_{\rm e} = \zeta_{\rm e} - g \Delta
	\zeta_{\rm e} - {\bar \zeta}_{\rm e}$, and similarly for ${\tilde
		\zeta}_{\rm o}$.  These
	quantities have zero limits as $U \to \pm \infty$.
	We similarly decompose the metric variables as
	\bes
	\bea
	{\hat q}_{\ps\, AB} + {\hat q}_{\ms\, AB} &=& \sqrt{2} \left[
	{\bar {\hat q}}_{{\rm e}\,AB} + \Delta {\hat q}_{{\rm e}\,AB}\, g(U) + {\tilde
		{\hat q}}_{{\rm e}\,AB}(U) \right], \\
	{\hat q}_{\ps\, AB} - {\hat q}_{\ms\, AB} &=& \sqrt{2} \left[
	{\bar {\hat q}}_{{\rm o}\,AB} + \Delta {\hat q}_{{\rm o}\,AB} \, g(U) + {\tilde
		{\hat q}}_{{\rm o}\,AB}(U) \right],
	\eea
	\ees
	and obtain from the continuity condition at $S_0$ that ${\hat {\bar
			q}}_{{\rm o}\,AB} = \Delta {\hat q}_{{\rm o}\,AB}/2$.
	Finally we similarly decompose the $\alpha$ variables
	using $\alpha(-\infty) = 0$ from the definition (\ref{alphadef})
	as
	\bes
	\bea
	(\alpha_{\ps} + \alpha_{\ms})/\sqrt{2} &=& 
	{\bar \alpha}_{{\rm e}} [1 + 2 g(U)] + {\tilde
		\alpha}_{{\rm e}}(U), \\
	(\alpha_{\ps} - \alpha_{\ms})/\sqrt{2} &=& 
	{\bar \alpha}_{{\rm o}} [1 + 2 g(U)] + {\tilde
		\alpha}_{{\rm o}}(U) .
	\eea
	\ees

	The final result in terms of these variables is
	\be
	\label{Omegafinal}
	\Omega = \frac{1}{32 \pi} \int_{-\infty}^\infty dU \int d^2 \theta \sqrt{q}
	\, \omega_{\rm bulk}(U,\theta^A)  +\frac{1}{32 \pi} \int d^2 \theta \sqrt{q}
	\, \omega_{\rm corner}(\theta^A),
	\ee
	where
	\bea
	\label{finalbulk}
	\omega_{\rm bulk} &=&
	{\hat q}_0^{AC} {\hat q}_0^{BD} \left[
	\delta {\tilde {\hat q}}_{{\rm e}\,AB,U} \wedge   \delta {\tilde
		{\hat q}}_{{\rm e}\,CD} +
	\delta {\tilde {\hat q}}_{{\rm o}\,AB,U} \wedge   \delta {\tilde
		{\hat q}}_{{\rm o}\,CD}
	\right]
	\nonumber \\
	&&
	-2 c_7 \delta {\tilde \zeta}_{{\rm e},U} \wedge \delta {\tilde  \zeta}_{\rm e}
	-2 c_7 \delta {\tilde \zeta}_{{\rm o},U} \wedge \delta {\tilde  \zeta}_{\rm o}
	+2 c_7^{-1} \delta {\tilde \alpha}_{{\rm e},U} \wedge \delta {\tilde  \alpha}_{\rm e}
	+2 c_7^{-1} \delta {\tilde \alpha}_{{\rm o},U} \wedge \delta {\tilde  \alpha}_{\rm o}
	\eea
	and
	\bea
	\label{finalcorner}
	\omega_{\rm corner} &=&
	  \delta \Delta {\hat q}_{{\rm e}\,AB} \wedge
	\left[ \delta {\bar {\hat q}}_{{\rm e}\,CD} + 2 \int dU g'
	\delta {\tilde {\hat q}}_{{\rm e}\,CD} \right]
	{\hat q}_0^{AC} {\hat q}_0^{BD}
	\nonumber \\ &&
	+  \delta \Delta {\hat q}_{{\rm o}\,AB} \wedge
	\left[ 2 \int dU g'
	\delta {\tilde {\hat q}}_{{\rm o}\,CD} \right]
	{\hat q}_0^{AC} {\hat q}_0^{BD} \nonumber \\
	&&
	- 2 c_7  \delta \Delta \zeta_{\rm e} \wedge
	\left[ \delta {\bar \zeta}_{\rm e} + 2 \int dU g'
	\delta {\tilde \zeta}_{\rm e} \right]
	- 4 c_7 \delta {\bar  \zeta}_{\rm o} \wedge \left[
          2 \int dU g'
	\delta {\tilde \zeta}_{\rm o} \right] 
	\nonumber \\
	&&
	+ 4 c_7^{-1} \delta {\bar \alpha}_{\rm o} \wedge \left[
	2 \int dU g'
	\delta {\tilde \alpha}_{\rm o} \right] 
	+ 4 c_7^{-1}  \delta {\bar \alpha}_{\rm e} \wedge
	\left[  2 \int dU g'
	\delta {\tilde \alpha}_{\rm e} \right]
	\nonumber \\
	&&
	+ 2 (2 \delta {\bar \zeta}_{\rm e} + \delta \Delta \zeta_{\rm e}) \wedge
	\delta {\bar \alpha}_{\rm e} 
	+ 8 \delta {\bar \zeta}_{\rm o} \wedge
	\delta {\bar \alpha}_{\rm o}
        \nonumber \\	&&
	+ 2 \sqrt{2} \frac{\delta \sqrt{q}}{\sqrt{q}} \wedge \left[
	c_7 \delta \Delta \zeta_{\rm e} + 2 \delta {\bar \alpha}_{\rm e}   -
	(c_5 - 2 c_2) \left( \delta {\bar \zeta}_e - \frac{1}{2}  \delta
	\Delta \zeta_{\rm e} \right)  
	\right].
	\eea

        So far this calculation has been in the context of the \phase
        space $\sls_\mfp$ defined in Sec.\ \ref{sec:rhps}.  We now discuss
        the symplectic form on the extension $\sls_\mfp'$ of this \phase
        space associated with half-sided boosts, discussed in
        Sec.\ \ref{halfsidedboosts}.  As explained there,
        we can enlarge the \phase space by making the replacements in
        the initial data 
        \bes
        \bea
        \kappa_\ps(u,\theta^A) &\to& \kappa_\ps(u,\theta^A) +
        b_\ps(\theta^A) \delta (u - \varepsilon ),\\
        \kappa_\ms(u,\theta^A) &\to& \kappa_\ms(u,\theta^A) -
        b_\ms(\theta^A) \delta (u - \varepsilon ),
        \eea
        \ees
        where we have used the convention explained in footnote \ref{uvdef},
and by taking the limit
$\varepsilon \to 0$.
The effect of making
this replacement and limit
in Eq.\ (\ref{pol1})
is to add to the corner term (\ref{finalcorner}) in the
        presymplectic form the terms
        \be
        \label{hsb}
         \omega_{\rm corner}'= \frac{1}{8 \pi} \int d^2 \theta \delta \sqrt{q} \wedge
          \delta b_{\rm r} +
                    \frac{\sqrt{2}}{16 \pi} \int d^2 \theta  \sqrt{q} \left(\delta {\bar \zeta}_{\rm e} - \frac{1}{2} \delta \Delta \zeta_{\rm e}\right) 
          \wedge \delta b_{\rm r},
                    \ee
where $b_{\rm r} = b_\ps - b_\ms$ is the relative boost parameter, the
difference between the boost parameters on the two surfaces $\Sp$ and
$\Sm$ \footnote{In the extended \phase space
the functions $\alpha_\ps^{\rm here}$ and $\alpha_\ms^{\rm here}$ used
in this section are related to those defined in Sec.\ \ref{sec:pse} by
$\alpha_\ps^{\rm here} = \alpha_\ps - \left. \alpha_\ps \right|_{S_0}
= \alpha_\ps - b_\ps$,
$\alpha_\ms^{\rm here} = \alpha_\ms - \left. \alpha_\ms \right|_{S_0}
= \alpha_\ps + b_\ms$.}.
Here we have used Eqs.\ (\ref{con66}), (\ref{alphadef}), (\ref{zetadef0}), (\ref{zetaeo}) and (\ref{zetabardef}).

There are three symplectically orthogonal sectors in the prephase space
with the presymplectic form (\ref{Omegafinal}) supplemented with
(\ref{hsb}): a spin two sector, an even parity
spin zero sector, and an odd parity spin zero sector.
The bulk spin two and spin zero pieces have
appeared in several previous analyses of single null surfaces, for
example Refs.\ \cite{Hawking:2016sgy,Hopfmuller2017a,Ciambelli:2023mir}. 
However, the corner terms terms
differ from previous analyses since we
are considering a pair of intersecting null surfaces, and because
we include the effect of half-sided boosts.
We now discuss these
different sectors one by one,
identify and eliminate edge modes that correspond to degeneracy
directions to obtain the physical phase space $\slp_\mfp$,
and derive the corresponding Poisson
brackets.

\subsubsection{Spin two sector}

The tensor sector is parameterized by the metric variables
${\tilde {\hat q}}_{{\rm e}\,AB}(U,\theta^A)$,
${\tilde {\hat q}}_{{\rm o}\,AB}(U,\theta^A)$, 
$\Delta {\hat q}_{{\rm e}\,AB}(\theta^A)$,
${\bar {\hat q}}_{{\rm e}\,AB}(\theta^A)$ and 
$\Delta {\hat q}_{{\rm o}\,AB}(\theta^A)$.
The corresponding presymplectic form is given by the first line of the
bulk expression (\ref{finalbulk}) and the first two lines of the corner
expression (\ref{finalcorner}).

        If we set to zero the variables ${\tilde {\hat q}}_{{\rm
            e}\,AB}$ and
${\tilde {\hat q}}_{{\rm o}\,AB}$ that depend on $U$, then
        the presymplectic form in the spin two sector is degenerate, with degeneracy
        direction $\partial / \partial \Delta {\hat
	  q}_{{\rm o}\,AB}$, and in this sector of \phase space we
        could fix the degeneracy by gauge fixing
\be
\Delta {\hat	  q}_{{\rm o}\,AB}=0.
\label{degeneracy}
\ee
More generally when we allow ${\tilde {\hat q}}_{{\rm     e}\,AB}$ and
${\tilde {\hat q}}_{{\rm o}\,AB}$ to be non zero, the degeneracy is
lifted and replaced by the presymplectic form being weakly but not strongly nondegenerate.
Nevertheless, we argue in
Appendix \ref{app:zero} that the correct prescription for dealing with
the weak non-degeneracy is to enforce the condition (\ref{degeneracy}), the same
prescription that would apply to a degeneracy.  After doing so it is
possible to invert the symplectic form to obtain the nonzero Poisson brackets
\bes
\bea
\left\{
{\tilde {\hat q}}_{{\rm     e}\,AB}(U,\theta)
,
{\tilde {\hat q}}_{{\rm     e}\,A'B'}(U',\theta')
\right\} &=& 8 \pi P_{ABA'B'}(\theta,\theta') \left[ H(U-U') -
  \frac{1}{2} \right] \delta^2(\theta,\theta'),\\
\left\{
{\tilde {\hat q}}_{{\rm     o}\,AB}(U,\theta)
,
{\tilde {\hat q}}_{{\rm     o}\,A'B'}(U',\theta')
\right\} &=& 8 \pi P_{ABA'B'}(\theta,\theta') \left[ H(U-U') -
  \frac{1}{2} \right] \delta^2(\theta,\theta'),\\
\label{spin2corner}
\left\{
{\Delta {\hat q}}_{{\rm     e}\,AB}(\theta)
,
{\check q}_{{\rm     e}\,A'B'}(\theta')
\right\} &=& 16 \pi P_{ABA'B'}(\theta,\theta')  \delta^2(\theta,\theta'),
\eea
\ees
where $H$ is the Heaviside step function, $\delta^2(\theta,\theta')$
is the covariant delta function on the sphere,
\be
P_{ABC'D'} = {\hat q}_{0\, AB} {\hat q}_{0\,C'D'} - {\hat q}_{0\,AC'}
{\hat q}_{0\,BD'} - {\hat q}_{0\,AD'} {\hat q}_{0\,BC'},
\ee
and ${\hat q}_{0\,BC'}$ is the parallel transport tensor from $\theta$
to $\theta'$ associated with the metric ${\hat q}_{0\,AB}$.
We have also defined 
\be
{\check q}_{{\rm e}\,CD} = 	 {\bar {\hat q}}_{{\rm e}\,CD} + 2 \int dU g'
	 {\tilde {\hat q}}_{{\rm e}\,CD} .
\ee

\subsubsection{Odd spin zero sector}
\label{sec:spin0odd}

The odd spin zero sector is parameterized by the bulk variables
${\tilde \zeta}_{\rm o}(U,\theta)$, ${\tilde \alpha}_{\rm
  o}(U,\theta)$ and the corner variables ${\bar \zeta}_{\rm o}(\theta)$ and ${\bar \alpha}_{\rm
  o}(\theta)$.
The corresponding presymplectic form is given by the second and fourth
terms of the second line of the
bulk expression (\ref{finalbulk}), and the odd terms in lines three
through five of the corner expression (\ref{finalcorner}).
In Appendix \ref{app:zero} we show that the variables
${\bar \zeta}_{\rm o}(\theta)$ and ${\bar \alpha}_{\rm
  o}(\theta)$ are essentially gauge degrees of freedom and can be set
to zero,
\be
\label{zzz}
{\bar \zeta}_{\rm o}(\theta) = {\bar \alpha}_{\rm
  o}(\theta) =0,
\ee
and that the Poisson brackets for the remaining variables are
\bes
\label{pb44}
\bea
\left\{
{\tilde \zeta}_{{\rm     o}}(U,\theta)
,
{\tilde \zeta}_{{\rm     o}}(U',\theta')
\right\} &=& \frac{8 \pi}{c_7} \left[ H(U-U') -
  \frac{1}{2} \right] \delta^2(\theta,\theta'),\\
\left\{
{\tilde \alpha}_{{\rm     o}}(U,\theta)
,
{\tilde \alpha}_{{\rm     o}}(U',\theta')
\right\} &=& 8 \pi c_7 \left[ H(U-U') -
  \frac{1}{2} \right] \delta^2(\theta,\theta'),\\
\label{unusual}
\left\{
{\tilde \zeta}_{{\rm     o}}(U,\theta)
,
{\tilde \alpha}_{{\rm     o}}(U',\theta')
\right\} &=& 4 \pi \delta^2(\theta,\theta').
\eea
\ees
This is a kind of horizon algebra of the sort discussed by Wall
\cite{Wall:2011hj}.  It is an unusual algebra due to the nonzero right hand side
of Eq.\ (\ref{unusual}), which is independent of $U$ and $U'$.

\subsubsection{Even spin zero sector}
\label{sec:spin0even}

The even spin zero sector is parameterized by the bulk variables
${\tilde \zeta}_{\rm e}(U,\theta)$, ${\tilde \alpha}_{\rm
  e}(U,\theta)$ and the corner variables $\Delta \zeta_{\rm
  e}(\theta)$, ${\bar \zeta}_{\rm e}(\theta)$, ${\bar \alpha}_{\rm
  e}(\theta)$, $\sqrt{q}(\theta)$ and $b_{\rm r}(\theta)$.
The corresponding presymplectic form is given by the first and third
terms of the second line of the
bulk expression (\ref{finalbulk}), the even terms in lines three
through six of the corner expression (\ref{finalcorner}), and the
additional corner contribution (\ref{hsb}).

We can simplify the presymplectic form by defining the following new
variables, an adjusted area element
\be
{\cal A} = \ln \sqrt{q} + \frac{1}{\sqrt{2}} {\bar \zeta}_{\rm e} -
\frac{1}{2 \sqrt{2}} \Delta \zeta_{\rm e},
\ee
and an adjusted boost parameter
\be
   {\cal B} = 2 b_{\rm r} + \sqrt{2} c_7 \Delta \zeta_{\rm e}
   + 2 \sqrt{2} c_7 {\bar \alpha}_{\rm e} - \sqrt{2} (1 - c_7) ({\bar
     \zeta}_{\rm e} - \Delta \zeta_{\rm e}/2).
\ee
The corner terms in the symplectic potential then become
	\bea
	\label{finalcorner1}
	\omega_{\rm corner,\,spin\ 0,\,even} &=&
	- 4 c_7  \delta \Delta \zeta_{\rm e} \wedge
	  \int dU g'
	\delta {\tilde \zeta}_{\rm e} 
	+ 8 c_7^{-1}  \delta {\bar \alpha}_{\rm e} \wedge
	   \int dU g'
	\delta {\tilde \alpha}_{\rm e} 
	+ 4   \delta \Delta \zeta_{\rm e} \wedge
	\delta {\bar \alpha}_{\rm e} 
        \nonumber \\
	&&
	+ 2 \delta {\cal A} \wedge \delta {\cal B}.
	\eea
The symplectic potential now no longer depends on ${\bar \zeta}_{\rm e}$,
which is therefore a degeneracy direction and can be set to zero.
Also the first line of Eq.\ (\ref{finalcorner1}) together with the even bulk terms
in Eq.\ (\ref{finalbulk}) have the same form as the odd sector treated
Sec.\ \ref{sec:spin0odd} above, and can be treated as described
there.  In particular we obtain that $\Delta \zeta_{\rm e}$ and ${\bar
  \alpha}_{\rm e}$ can be set to zero:
\be
\label{zzzz}
{\bar \zeta}_{\rm e} = \Delta \zeta_{\rm e} = {\bar \alpha}_{\rm e}
=0.
\ee
This then simplifies the definitions of ${\cal A}$ and ${\cal B}$ to
\be
{\cal A} = \ln \sqrt{q}, \ \ \ \ {\cal B} = 2 b_{\rm r}.
\ee
The final results for the Poisson brackets in the even sector are then
\bes
\label{pb45}
\bea
\label{pb45a}
\left\{
{\tilde \zeta}_{{\rm     e}}(U,\theta)
,
{\tilde \zeta}_{{\rm     e}}(U',\theta')
\right\} &=& \frac{8 \pi}{c_7} \left[ H(U-U') -
  \frac{1}{2} \right] \delta^2(\theta,\theta'),\\
\left\{
{\tilde \alpha}_{{\rm     e}}(U,\theta)
,
{\tilde \alpha}_{{\rm     e}}(U',\theta')
\right\} &=& 8 \pi c_7 \left[ H(U-U') -
  \frac{1}{2} \right] \delta^2(\theta,\theta'),\\
\label{pb45c}
\left\{
{\tilde \zeta}_{{\rm     e}}(U,\theta)
,
{\tilde \alpha}_{{\rm     e}}(U',\theta')
\right\} &=& 4 \pi \delta^2(\theta,\theta'),\\
\label{pb45d}
\left\{ {\cal A}(\theta) , {\cal B}(\theta') \right\} &=& - 16 \pi \delta^2(\theta,\theta').
\eea
\ees
Note that the Poisson brackets (\ref{pb44}) and (\ref{pb45a}) -- (\ref{pb45c}) can be rewritten in terms of
the fields ${\tilde \zeta}_\ps$, ${\tilde \alpha}_\ps$, ${\tilde
  \zeta}_\ms$ and ${\tilde \alpha}_\ms$, giving results of the same form.

\subsection{Area operator}
\label{sec:area}

The Poisson brackets (\ref{pb45}) imply that the weighted area
functional on phase space
\be
\int d^2 \theta \sqrt{q} w(\theta),
\ee
where $w(\theta)$ is some weighting function, generates a
transformation on phase space given by increasing the boost parameter
$b_{\rm r}(\theta) \to b_{\rm r}(\theta) - 8 \pi w(\theta)$.
This corresponds to changing the inaffinity in the convention we use
for how to fix the rescaling freedom.  However one could instead
adopt the convention (\ref{altconv}) for fixing the rescaling freedom.
In this case the transformation would have the
effect of rescaling both the covariant and contravariant normals on
both $\Sp$ and $\Sm$ by the same boost factor, as discussed in
Sec.\ \ref{sec:pse} above.
The fact that the area element and boost parameter are canonical
conjugates is well known from various special cases
\cite{1993PhRvD..47.3275H,Carlip:1993sa,Lehner2016,Hopfmuller2017a},
however we believe that the derivation given here is the first derivation within the context of a complete phase space for two intersecting null surfaces. 

Our approach is directly amenable to Lorentzian canonical
quantization, wherein one could directly study the crossed product
degree of freedom controlling the area term in the generalized entropy
\cite{Chandrasekaran:2022eqq}.
In particular, our results should
be straightforwardly generalizable to boost surfaces more general than
the bifurcation surface, and to non-stationary backgrounds.

\section{Interpretation of charge non-uniqueness}
\label{sec:nonunique}

	In Sec.\ \ref{sec:decomp_charge} above we found that the
        gravitational charges depend  
	on two arbitrary parameters $c_2$ and $c_5$.  As discussed in a
	general context in Sec.\ \ref{sec:gp}, this lack of uniqueness
	reflects a dependence on a choice of polarization for the
        theory \cite{Odak:2021axr,Odak:2022ndm,Rignon-Bret:2023fjq,Freidel:2020xyx,Freidel:2020svx,Freidel:2020ayo}. 
        In this section we will make the connection explicit by deriving the
	required polarization choice (\ref{polarizationanswer}) for arbitrary values of the parameters,
        as well as the corresponding subspace $\fs_\mfs$ of the field configuration space.

	As discussed in Sec.\ \ref{sec:actionprinciple} above,
	in order to find a good action principle
        for the pair of
	intersecting null surfaces, we need to find a subspace
	$\fs_\mfs$ of the field configuration space $\fs$ which make the
	corner and boundary flux terms vanish in the polarization
	(\ref{polarizationanswer}).
	We cannot simply fix the induced metrics on $\Sp$ and $\Sm$, 
	since $\Sp$ and $\Sm$ form an initial data surface and this restriction
	would be too strong a constraint.  Instead, the idea is that we want to fix half of the
	physical degrees of freedom on $\Sp$ and $\Sm$, so that imposing an additional similar
	constraint on a final data surface determines a unique
	solution\footnote{This is analogous
		to imposing vanishing variations
		of generalized coordinates at initial and final times but allowing
		generalized momenta to vary freely in classical mechanics.}.
	
	We can define such a field configuration subspace $\fs_\mfs$ as
	follows.  The polarization $\Xi$ corresponding to
        the parameters $c_2$ and $c_5$ can be obtained by adding the expression (\ref{pol2}) from $\Sp$ to a corresponding expression from $\Sm$, and rewriting in terms of the variables defined in Sec.\ \ref{sec:sf}.  Linearizing about the stationary background and dropping a total variation, the result is
	\be
	\label{Xifinal}
	\Xi = \frac{1}{32 \pi} \int_{-\infty}^\infty dU \int d^2 \theta \sqrt{q}
	\, \Xi_{\rm bulk}(U,\theta^A)  +\frac{1}{32 \pi} \int d^2 \theta \sqrt{q}
	\, \Xi_{\rm corner}(\theta^A),
	\ee
	where the bulk term is
	\bea
	\label{finalbulkXi}
	\Xi_{\rm bulk} &=&
	{\hat q}_0^{AC} {\hat q}_0^{BD} \left[
	{\tilde {\hat q}}_{{\rm e}\,AB,U}
	\left(   \delta {\tilde {\hat q}}_{{\rm   e}\,CD} + c_0 \delta {\tilde {\hat q}}_{{\rm   e}\,CD,U}
	\right)
	+     {\tilde {\hat q}}_{{\rm o}\,AB,U}
	\left(   \delta {\tilde {\hat q}}_{{\rm   o}\,CD} + c_0 \delta {\tilde {\hat q}}_{{\rm   o}\,CD,U}
	\right)
	\right]
\nonumber \\ &&
+2 c_7 {\tilde  \zeta}_{\rm e} \delta {\tilde \zeta}_{{\rm e},U}
	+2 c_7 {\tilde  \zeta}_{\rm o} \delta {\tilde \zeta}_{{\rm o},U} 
	-2 c_7^{-1}  {\tilde  \alpha}_{\rm e} \delta {\tilde \alpha}_{{\rm e},U}
	-2 c_7^{-1}  {\tilde  \alpha}_{\rm o} \delta {\tilde \alpha}_{{\rm o},U}         
        \eea
	We will not write out the corner term here, but we note
        that it can be made to vanish by fixing on $S_0$ the
        quantities $\Delta {\hat q}_{{\rm e}\,AB}$ and $\sqrt{q}$,
        from Eqs.\ (\ref{degeneracy}), (\ref{zzz}) and (\ref{zzzz}).

	We now decompose the metric variables as
	\be
	{\tilde {\hat q}}_{\iota\,AB}(U,\theta^A) = \sum_\Gamma \int_{-\infty}^\infty d\omega
	e^\Gamma_{ij}(\theta^A) q_{\iota\Gamma\omega} e^{- i \omega U},
	\ee
	where $e^\Gamma_{ij}$ are polarization tensors for $\Gamma = 1,2$,
	and $\iota$ can be ${\rm e}$ or ${\rm o}$.
	We write $q_{\iota\Gamma\omega}$ in
	terms of an amplitude and phase as
	\be
	q_{\iota\Gamma\omega} = a_{\iota\Gamma\omega} e^{i \varphi_{\iota\Gamma\omega}}.
	\ee
	Then for $c_0=0$ we define $\fs_\mfs$ by fixing values of the phases
	$\varphi_{\iota\Gamma\omega}$ but allowing the amplitudes $a_{\iota\Gamma\omega}$ to vary
	freely.  This forces the large square brackets in Eq.\ (\ref{finalbulkXi}) to vanish.
	We similarly decompose ${\tilde \zeta}_{\rm e}, {\tilde \zeta}_{\rm o},
        {\tilde \alpha}_{\rm e}$ and ${\tilde \alpha}_{\rm o}$ as
	\be
	{\tilde \zeta}_\iota(U,\theta^A) = \int_{-\infty}^\infty d
        \omega {\tilde \zeta}_{\iota\omega} e^{- i \omega U},
        \ \ \ \ \ 
	{\tilde \alpha}_\iota(U,\theta^A) = \int_{-\infty}^\infty d
        \omega {\tilde \alpha}_{\iota\omega} e^{- i \omega U},
	\ee
	and defining ${\tilde \zeta}_{\iota\omega} = b_{\iota\omega} e^{i
	    \vartheta_{\iota\omega}}$, ${\tilde \alpha}_{\iota\omega}
          = c_{\iota\omega} e^{i \chi_{\iota\omega}}$
          we require that the phases
	$\vartheta_{\iota\omega}$ and $\chi_{\iota\omega}$ be fixed. 
	This forces the spin 0 terms in Eq.\ (\ref{finalbulkXi}) to vanish,
	just like for the spin 2 terms, even though the spin 0 terms have a
	Neumann-like form instead of the Dirichlet-like form of the spin 2 terms.
	Together these conditions together with the corner conditions
	mentioned above are sufficient to make the presymplectic potential (\ref{Xifinal}) vanish 
	and give a good variational principle.
	
	For the more general case $c_0\ne0$ the definition of $\fs_\mfs$
	must be modified to fixing the values of 
	\be
	\varphi_{\iota\Gamma\omega} - c_0 \ln a_{\iota\Gamma\omega}
	\ee
	and allowing $a_{\iota\Gamma\omega}$ to vary freely, while the
	requirement on $\vartheta_{\iota\omega}$ and $\chi_{\iota\omega}$ is unchanged.


	
	\section*{Acknowledgments}
	
	We thank Laurent Freidel, Gloria Odak, 
 	Kartik Prabhu, Pranav Pulakkat, Antoine Rignon-Bret and Simone Speziale
	for helpful discussions and/or correspondence,
        Luca Ciambelli for helpful discussions and comments on the manuscript,
        and Gabriel Sánchez-Pérez and Marc Mars for helpful comments
	and for pointing out an error in Sec.\ \ref{sec:initialvalue} in the first version of
	this paper.  E.F. is supported in part by NSF grant PHY-2110463 and by a
	Simons Foundation Fellowship. VC is supported by a grant from the
	Simons Foundation (816048, VC).

	\appendix

	\section{Choice of phase space polarization determined by boundary and
		corner fluxes}
	\label{app:pol}

	In this appendix, we show that the presymplectic potential 
	(\ref{polarizationanswer}) on \phase space
	obtained from boundary and corner fluxes
	is consistent with the
	presymplectic form $\Omega$ defined in Eq.\ (CFSS,2.18).
	As in Sec.\ \ref{sec:gp}, we use the notation (CFSS,XX) to mean
	Eq.\ (XX) of Chandrasekaran, Flanagan, Shehzad and Speranza (CFSS)
	\cite{Chandrasekaran:2021vyu}.

	Starting from the definition
	(CFSS,2.18), we can deform the spatial slice $\Sigma$ to 
	become the portion of the boundary $\partial M$ that is to the past of
	$\Sigma$, since $ d \omega' =0$.
	The first term in
	Eq.\ (CFSS,2.18) then becomes an integral over a
	subset of the boundary components $\ns_j$ and over $\Delta \ns_{\bar
		j}$.  There is an overall sign flip due to the fact that the orientation
	of $\Sigma$ is opposite to that of the components of $\partial M$ to
	the past of $\Sigma$, from Sec.\ \ref{covr}.
	Using Eqs.\ (CFSS,2.11), (CFSS,2.12) and (CFSS,2.43) now gives
	$\Omega = \delta \Xi$, where
	\bea
	\label{polarizationanswer0}
	\Xi &=& -\sum^\prime_j \int_{\ns_j}
	\left[ - \delta \ell_j' - \delta d c_j' + d \lambda' + \beom_j +
	d \cflx_j
	\right]
	- \int_{\Delta
		\ns_{\bar j}}
	\left[ - \delta \ell'_{\bar j} - \delta d c'_{\bar j} + d \lambda' +
	\beom_{\bar j} +
	d \cflx_{\bar j}
	\right] \nonumber \\
	&&+ \int_{\partial \Sigma} \left[ \lambda' - \delta c'_{\bar j} +
	d\gamma'_{\bar j} +
	\cflx_{\bar j} \right].
	\eea
	Here the primed sum is over the boundary components $\nb_j$ to the
	past of $\Sigma$.
	The total variation terms involving $\ell'$ and $c'$ in
	Eq.\ (\ref{polarizationanswer0}) drop out of the expression for $\Omega$ since
	$\delta^2=0$, and so we can drop them from the expression for $\Xi$. 
	The exact term $d \gamma'$ vanishes when integrated
	over $\partial \Sigma$.  The contributions from $\lambda'$ vanish
	because $\lambda'$ is continuous across the corners, as discussed after
	Eq.\ (CFSS,5.4), and because of the cancellation between the second and
	third terms of the contributions
	from $\lambda'$ at $\partial \Sigma$.  With these simplifications
	the expression (\ref{polarizationanswer0}) reduces to
	\be
	\label{polarizationanswer1}
	\Xi = -\sum^\prime_j \int_{\ns_j}
	( \beom_j +
	d \cflx_j
	)
	- \int_{\Delta
		\ns_{\bar j}}
	( \beom_{\bar j} +
	d \cflx_{\bar j}
	)
	+ \int_{\partial \Sigma} \cflx_{\bar j}.
	\ee
	Making use of the orientation convention (\ref{orien}),
	using the relation (\ref{rels}) and noting that there is a cancellation
	between the second and third terms of the integral over $\delta
	\Sigma$, the expression (\ref{polarizationanswer1}) reduces to 
	the simpler form (\ref{polarizationanswer}).

	\section{Review of geometric quantities defined on a null surface}
	\label{app:nullreview}
	
	In this appendix we review various geometric quantities that can be defined on
	null surfaces.  For more details see
	Refs.\ \cite{Ashtekar:2001jb,Gourgoulhon:2005ng,Chandrasekaran:2020wwn,CFP}.
	
	Consider a spacetime $(M, g_{ab})$ containing a null surface ${\cal N}$.
	We denote by $\Pi^a_i$ the pullback operator that maps spacetime
	tensors to intrinsic tensors, and we use $\hateq$ to mean equality
	when evaluated on ${\cal N}$.
	We pick a smooth normal covector $\ell_a$ on ${\cal N}$, and define the
	nonaffinity $\nonaffinity$, a function on ${\cal N}$, by
	\be
	\ell^a \nabla_a \ell^b \hateq \nonaffinity \ell^b.
	\label{kappadef}
	\ee
	The contravariant normal $\ell^a = g^{ab} \ell_b$, when evaluated on $\scri$, can be
	viewed as an intrinsic vector $\ell^i$, since $\ell^a \ell_a =0$.  We denote by
	$\inducedmetric_{ij}$ the degenerate induced metric, and by
	$\volume_{ijk}$ the 3-volume form on ${\cal N}$ given by taking the pullback
	of $\volume_{abc}$ where $\volume_{abc}$ is any three form with $4 \volume_{[abc} \ell_{d]} = \varepsilon_{abcd}$.
	Finally we define a 2-volume form by
	\be
	\label{volumesmall}
	\volumesmall_{ij} =  \volume_{ijk} \ell^k.
	\ee
	
	Next, we take the pullback on the index $a$ of $\nabla_a \ell^b$, which
	is then orthogonal to $\ell_b$ on the index $b$.  This quantity therefore
	defines an intrinsic tensor ${\shape}_i^{\ j}$ called the Weingarten
	map \cite{Gourgoulhon:2005ng}.  The second fundamental form or shape
	tensor is $K_{ij} = {\shape}_i^{\ k} \inducedmetric_{kj}$, which can be decomposed
	as
	\be
	\label{Kdecompos}
	K_{ij} = \frac{1}{2} \expansion \inducedmetric_{ij} + \sigma_{ij}
	\ee
	in terms of an expansion $\expansion$ and a symmetric traceless shear tensor $\sigma_{ij}$.
	These fields on a null surface obey the relations \cite{Gourgoulhon:2005ng,CFP}
	\begin{subequations}
		\label{grelations1}
		\begin{eqnarray}
		\label{grelations1a}
		\inducedmetric_{ij} \ell^j &=& 0, \\
		\label{grelations1a1}
		\sigma_{ij} \ell^j &=& 0, \\
		\label{grelations1a2}
		{\shape}_i^{\ j} \ell^i &=& \nonaffinity \ell^j,\\
		\label{grelations1b}
		( \lie_\ell - \expansion) \inducedmetric_{ij} &=& 2 \sigma_{ij}, \\
		\label{grelations1c}
		( \lie_\ell - \expansion) \volume_{ijk} &=& 0, \\
		\label{grelations1d}
		( \lie_\ell - \expansion) \volumesmall_{ij} &=& 0, \\
		\label{grelations1e}
		( \lie_\ell - \nonaffinity) \expansion &=& - \frac{1}{2} \expansion^2 -  \sigma_{ij} \sigma_{kl} q^{ik} q^{jl} - R_{ab} n^a n^b,\\
		\label{addi2}
		d (f \volumesmall) &=& \volume (\lie_\ell + \Theta) f, 
		\end{eqnarray}
	\end{subequations}
	where $f$ is any function on ${\cal N}$, $\lie_{\ell}$ is the Lie derivative, and
	$q^{ij}$ is any tensor that satisfies
	\be
	q_{ij} q^{jk} q_{kl} = q_{il}.
	\label{partialinverse}
	\ee
	Note that the shear squared source term in the Raychaudhuri equation (\ref{grelations1e}) is independent of the choice of tensor $q^{ij}$.
	
	Consider now rescaling the normal according to
	\be
	\ell^i \to e^\sigma \ell^i,
	\label{rescale00}
	\ee
	where $\sigma$ is a smooth function on ${\cal N}$.
	Under this transformation the various fields
	transform as
	\begin{subequations}
		\label{contransformc}
		\begin{eqnarray}
		\label{contransformc2}  
		\ell_a &\to& e^{\sigma} \ell_a, \\
		\label{contransformc3}
		\inducedmetric_{ij} &\to&  \inducedmetric_{ij}, \\
		\label{contransformc4a}
		\volumesmall_{ij} &\to&  \volumesmall_{ij}, \\
		\label{contransformc4}
		\volume_{ijk} &\to& e^{ - \sigma} \volume_{ijk}, \\
		\label{contransformc5}
		\nonaffinity &\to& e^{\sigma} (\nonaffinity + \lie_\ell \sigma),\\
		\label{contransformc6}
		\expansion &\to& e^{\sigma} \expansion,\\
		\label{contransformc7a}
		K_{ij} &\to& e^{\sigma} K_{ij} ,\\
		\label{contransformc7}
		{\shape}_i^{\ j} &\to& e^{\sigma} \left[{\shape}_i^{\ j} +
		D_i\sigma \ell^j  \right],
		\end{eqnarray}
	\end{subequations}
	where $D_i$ is any derivative operator on ${\cal N}$.
	These transformation laws preserve the relations (\ref{grelations1}).

	\subsection{Structures that depend on choice of auxiliary normal}
	
	An auxiliary normal on the null surface ${\cal N}$ is a vector field $n^a$
	which satisfies
	\be
	\label{auxnormal}
	n_a n^a = 0, \ \ \ \ \ \ n_a \ell^a = -1.
	\ee
	Such vectors are not unique, since given any vector $Z^a$ orthogonal
	to $\ell^a$ and $n^a$, the mapping
	\be
	\label{Deltandef}
	n^a  \goesto n^a + Z^a + \frac{1}{2}Z^2 \ell^a
	\ee
	preserves the conditions (\ref{auxnormal}).  Although the choice of
	$n^a$ is arbitrary, we will find it useful to define several
	quantities that 
	depend on this choice.
	
	The covector $n_a$ can be pulled back to the surface to yield
	$n_i = \pullback_i^a n_a$, which satisfies $n_i \ell^i =-1$ from
	Eqs.\ (\ref{auxnormal}) and (CFP,3.7).  However there is no natural
	definition of a contravariant intrinsic vector $n^i$.
	Next, the tensor $\delta^a {}{}_b + n^a \ell_b$ maps spacetime vectors
	$v^a$ to vectors orthogonal to $\ell^a$, that is, intrinsic
	vectors. We write this mapping as 
	$
	v^a \goesto v^i = \pushback^i_a v^a.
	$
	The quantities $\pushback^i_a$ and $\pullback_i^a$ satisfy
	\be
	\delta^i_j = \pushback^i_a \, \pullback^a_j,
	\ \ \ \ \ \ \delta^a_{\ \,b}+n^a \ell_b = \pullback^a_i \,
	\pushback^i_b.
	\ee
	We define spacetime tensors that correspond to the induced
	metric
	\be
	q_{ab} = \pushback^i_a \pushback^j_b q_{ij} = g_{ab} + 2 \ell_{(a}
	n_{b)}, \label{qabdef} 
	\ee
	and shear tensor
	\be
	\sigma_{ab} = \pushback^i_a \pushback^j_b \sigma_{ij}.
	\ee
	It follows that the quantities $q^{ab}$ and $\sigma^{ab}$ are
	orthogonal to $\ell_a$, and so define intrinsic tensors $q^{ij}$ and
	$\sigma^{ij}$.  The mixed index quantity $q^a_{\ b}$ is orthogonal to
	$\ell_b$ on its $a$ index so is intrinsic on that index.  If we
	perform a pullback on the $b$ index we then obtain the quantity
	\be
	q^i_{\ j} = \delta^i_{\ j} + \ell^i n_j.
	\label{qmixed}
	\ee
	The various intrinsic tensors $q$ are related by
	$q^{ij} q_{jk} = q^i_{\ k}$ and $q^i_{\ j} q_{ik} = q_{jk}$, which
	follow from (CFP,3.6) and $q^a_{\ b} q^b_{\ c} = q^a_{\ c}$.
	
	Finally we define the rotation one form
	\be
	\omega_i = - n_j {\cal K}_i^{\ j}
	\label{twist}
	\ee
	which satisfies $\omega_i \ell^i=\kappa$, and its projected version
	\be
	\varpi_i = q_i^{\ j} \omega_j = \omega_i + \kappa n_i
	\label{ptwist}
	\ee
	which is orthogonal to $\ell^i$.
	The Weingarten map can be written in terms of these quantities and the
	second fundamental form as 
	\be
	\label{Kident}
	{\cal K}_i^{\ j} = q^{jk} K_{ik} + \ell^j( \varpi_i - \kappa n_i).
	\ee

	\section{Covariant double null initial value formulation}
	\label{app:ivf}
	
	In this appendix we derive the existence and uniqueness theorem for
	double null initial data stated in Sec.\ \ref{sec:theorem}.
	The derivation is based on the harmonic gauge result of Rendall
	on the existence of a solution of the vacuum Einstein equation given
	suitable initial data on two intersecting null surfaces
	\cite{70785209-fc5c-312c-9bd2-7c68ce777f34}.
	
	Rendall uses a coordinate system $(x^1, x^2, x^3, x^4)$, which we will
	denote by $(x^+, x^-, \theta^A)$, for which the two surface $S_0$ is
	given by $x^+ = x^- =0$, the surface $\Sp$ is $x^-=0$, and $\Sm$ is $x^+=0$.
	The gauge is chosen to be harmonic and in addition to satisfy on $S_0$
	\be
	\label{init1}
	g_{\ps\ms} = -1, \ \ \ \ g_{\ps\ps} = 0, \ \ \ \ g_{\ms\ms} = 0,
	\ \ \ \ \ g_{\ps A}=0, \ \ \ \ g_{\ms A}=0.
	\ee
	On $\Sp$ the coordinates satisfy
	\be
	g_{\ps\ps}=0, \ \ \ \ g_{\ps A} =0,
	\ee
	while on $\Sm$ the conditions are
	\be
	g_{\ms\ms} = 0, \ \ \ \ g_{\ms A}=0.
	\ee
	
	We now show an initial data structure
	\be
	[q_{AB}, {\bar \omega}_A, m,
	\Theta_\ps, \Theta_\ms, \ell^i_\ps, \kappa_\ps, {\bar q}^\ps_{ij},
	\ell^{i'}_\ms, \kappa_\ms, {\bar q}^\ms_{i'j'}]
	\label{ec000}
	\ee
	of the type discussed in Sec.\ \ref{sec:initialvalue} determines a set of initial
	data that satisfy all the conditions required by Rendall.  
	Here we have parameterized the conformal equivalence classes of
	metrics in terms
	of the rescaled metrics defined in Sec. \ref{sec:idsalternative}.  We
	follow the following series of steps:
	
	\begin{enumerate}
		\item By using the rescaling freedom (\ref{2rescalings}), we specialize to a
		representative of the equivalence class with $\kappa_\ps = 0$ on
		$\Sp$, $  \kappa_\ms =0$ on $\Sm$, and $m=0$ on $S_0$.
		
		\item We choose coordinates $\theta^A$ on $S_0$ and extend them along
		$\Sp$ and $\Sm$ by demanding that they be constants along integral
		curves of ${\vec \ell}_\ps$ and ${\vec \ell}_\ms$.  We define a coordinate
		$x^+$ on $\Sp$ such that ${\vec \ell}_\ps = \partial / \partial x^+$
		and $x^+ =0$ on $S_0$, and similarly for $x^-$.
		
		\item The metric components on the surface $S_0$ are given as
		$g_{AB} = q_{AB}$, $g_{\ps\ms} = -1$, with all other components vanishing.
		
		\item From the rescaled metrics ${\bar q}^\ps_{ij}$ and ${\bar
			q}^\ms_{i'j'}$ and the initial convergences $\Theta_\ps$,
		$\Theta_\ms$, we can recover the metric components $g_{AB}$ on  
		$\Sp$ and $\Sm$ using the procedure described in 
		Sec.\ \ref{sec:idsalternative}, the same procedure used by Rendall.
		
		\item The condition $\kappa_\ps=0$ on $\Sp$ yields the constraint
		\be
		\label{kappavanish}
		g_{\ps\ps,\ms}     = 2 g_{\ps\ms,\ps}.
		\ee
		Rendall shows that this combined with the harmonic gauge condition yields a
		first order differential equation for $g_{\ps\ms}$ along $\Sp$,
		\be
		\label{fode}
		\partial_+ \ln g_{\ps\ms} = \Theta_\ps/2,
		\ee
		which can be solved
		using the initial condition (\ref{init1}).  We similarly obtain
		$g_{\ps\ms}$ along $\Sm$.
		
		\item Rendall derives from the vacuum harmonic gauge Einstein equations a second order differential equation for 
		$g_{\ps\ps}$ along $\Sm$, of the form $\partial_\ms \partial_\ms
		g_{\ps\ps} = \ldots$.  The solution is determined by the initial
		data on $S_0$, with  $g_{\ps\ps}$ on $S_0$ given by Eq.\ (\ref{init1}) and
		$g_{\ps\ps,\ms}$ on $S_0$ given by Eqs.\ (\ref{kappavanish}) and (\ref{fode}).  Similarly
		$g_{\ms\ms}$ along $\Sp$ is obtained.
		
		\item Finally, Rendall similarly derives second order
		differential equations for $g_{\ms A}$ along $\Sp$ and $g_{\ps
			A}$ along $\Sm$ \cite{70785209-fc5c-312c-9bd2-7c68ce777f34}:
		\be
		\label{eez}
		\partial_\ps \partial_\ps g_{\ms A} = \ldots, \ \ \ \ \        \partial_\ms \partial_\ms g_{\ps A} = \ldots.
		\ee
		These metric components vanish on $S_0$, and their initial
		derivatives are given by combining the expression 
		\be
		{\bar \omega}_A =   \frac{1}{2} ( g_{A\ms,\ps} - g_{A\ps,\ms})
		\ee
		for the H\'a\'ji\^{c}ek one-form ${{\bar \omega}_A}$ from the definition
		(\ref{rotoneform}), with the constraint from harmonic gauge
		\be
		g^{BD} {}^{(2)}\Gamma^A_{BD} = g^{AC} ( g_{C\ps,\ms} +
		g_{C\ms,\ps}).
		\ee
		Here ${}^{(2)}\Gamma^A_{BC}$ are the connection coefficients of the
		two dimensional metric $g_{AB}$ on $S_0$.  Thus Eqs.\ (\ref{eez})
		determine the metric coefficients $g_{\ps A}$ and $g_{\ms A}$ along $\Sp$ and $\Sm$.
		
	\end{enumerate}
	
	To summarize, we have obtained all the metric components on $\Sp$ and
	$\Sm$, and these components satisfy the conditions needed by Rendall.
	The remaining harmonic gauge conditions along $\Sp$ can be satisfied by adjusting
	the derivatives $g_{A\ps,\ms}$ and $\partial_\ms {\rm det}(g_{AB})$ of
	the metric components off the surface, and similarly along $\Sm$.    
	The construction uses
	all the information in the initial data structure.  It follows from
	Rendalls result that there exists a solution to the vacuum Einstein
	equation in a neighborhood of the two null surfaces which is compatible with this initial data, and thus also
	compatible with our initial data structure.
	
	If we choose a different representative of the initial data structure, then by 
	step 1 above the two representatives are related by a rescaling of
	the form (\ref{2rescalings}) where $\sigma_\ps = - \sigma_\ms$ and both depend
	only on the angles $\theta^A$.  The resulting solutions
	$g_{\alpha\beta}(x^\gamma)$ are then
	related by the harmonic diffeomorphism whose restriction to $\Sp$ and
	$\Sm$ takes the form $x^+ \to e^{\alpha(\theta^A)}
	x^+$, $x^- \to e^{-\alpha(\theta^A)} x^-$ for some $\alpha$.
	Thus the mapping from initial data structures to spacetime geometries is well
	defined.
	We could also define a unique mapping from initial data structures to
	harmonic gauge metrics, which we denote by
	\be
	\mfi \to g_{ab}^{\rm har}[\mfi],
	\ee
	by imposing in step 1 that ${\bar \omega}_A$ be parity odd, from
	Eq.\ (\ref{rescaleomega}).

	The prescription described in this Appendix generates generic harmonic gauge solutions of the
	vacuum equations \cite{70785209-fc5c-312c-9bd2-7c68ce777f34}.  If we
	assume that acting with diffeomorphisms yields fully generic
	solutions, then it follows that solutions are unique up to gauge.
	To see this, suppose that $g_{ab}$ is a generic solution with
	$\mfi[g_{ab}]  = \mfi_1$.  By assumption there exists
	a bulk diffeomorphism $\psi$ and an initial data structure
	$\mfi_2$ for which
	\be
	\label{fd}
	g_{ab} = \psi_* g_{ab}^{\rm har}[\mfi_2].
	\ee
	Also without loss of generality we can take the restriction of $\psi$
	to the boundary to be the identity map.  Now acting on Eq.\ (\ref{fd})
	with $\mfi[..]$ yields that $\mfi_1 = \mfi_2$,
	so that $g_{ab} = \psi_* g_{ab}^{\rm har}[\mfi_1]$.  It
	follows that any two such metrics are related by a diffeomorphism.
	A more general proof of uniqueness up to gauge of the double null initial value
	formalism can be found in Refs.\ \cite{Mars:2022gsa,Mars:2023hty}.

        \section{Examples of half-sided boosts}
        \label{app:boosts}

        In this appendix we give the explicit details of how half-sided
        boosts act on the maximally extended Schwarzschild
        spacetime.  We show that their action on the initial data is to
        insert a delta function in the affinity, in agreement with our
        definition of Sec.\ \ref{halfsided} specialized to vanishing expansion.

	We write the Schwarzschild metric in Kruskal coordinates
	\be
	ds^2 = - e^{\alpha(UV)} dU dV + {\cal R}(UV)^2 d\Omega^2
	\ee
        for some functions $\alpha$ and ${\cal R}$.
	Consider now the boost mapping
	\be
	U = e^{-\beta({\bar U}, {\bar V})} {\bar U}, \ \ \ \ V = e^{\beta({\bar U}, {\bar V})} {\bar V}.
	\ee
	When the parameter $\beta$ is a constant, this is an exact symmetry of the metric.  Now suppose
	that we take $\beta$ to take different constant values in each of the four quadrants:
	\be
	\beta({\bar U}, {\bar V}) = \sum_{\gamma,\sigma = \pm} \beta_{\gamma\sigma}  H(\gamma {\bar U}) H(\sigma {\bar V}),
	\ee
        where $H$ is the Heaviside step function.
	The metric becomes
	\be
	ds^2 = - e^{\alpha({\bar U}{\bar V})} d{\bar U} d{\bar V} + {\cal R}({\bar U}{\bar V})^2 d\Omega^2
	+ g_{{\bar U}{\bar U}} d {\bar U}^2
	+ g_{{\bar V}{\bar V}} d {\bar
		V}^2,
	\label{newmetric}
	\ee
        where
        \bes
        \label{metricc}
        \bea
        g_{{\bar U}{\bar U}}  &=& 	 - e^\alpha {\bar V} \delta({\bar U}) \left[ (\beta_{\ps\ps} - \beta_{\ms\ps}) H({\bar V}) +(\beta_{\ps\ms} - \beta_{\ms\ms}) H(-{\bar V}) \right]       , \\
        g_{{\bar V}{\bar V}} &=& 	 e^\alpha {\bar U} \delta({\bar V}) \left[ (\beta_{\ps\ps} - \beta_{\ps\ms}) H({\bar U}) +(\beta_{\ms\ps} - \beta_{\ms\ms}) H(-{\bar U}) \right].
        \eea
        \ees
        The transformed metric contains a leftward propagating shock
        proportional to $\delta({\bar U})$ and a rightward propagating shock
        proportional to $\delta({\bar V})$.  Here we have not made any
        linearized approximation.

       The transformation to the metric, specialized to the initial
       data surface ${\bar U} = - {\bar V}$, matches the kink
       transform defined in Ref.\ \cite{Bousso:2020yxi}.
 The perturbation to the induced metric on this surface vanishes, while the
 perturbation to the extrinsic curvature consists of a delta
 function\footnote{The calculation involves the products of delta
 functions and step functions, which {\it a priori} are ill defined.
 However,
 here the delta functions arise as derivatives of the step functions, and
 if we replace the step functions at the start of the
 calculation with regulated versions, and take the regularization
 parameter to zero at the end of the calculation, we obtain the rule
 $\delta(x) H(x) \to \delta(x)/2$ which gives a unique result.}
 in the radial-radial components, localized on the bifurcation twosphere.

 Consider now how the transformation affects initial data specified on
 the intersecting null surfaces ${\bar U} =0, {\bar V} \ge 0$ and
 ${\bar V}=0, {\bar U} \ge 0$.  We regulate the location of the null
 surfaces to place them just before the shocks, by defining
 \be
    {\hat U} = {\bar U} + \ve, \ \ \ \ {\hat V} = {\bar V} + \ve,
      \ee
      where $\ve > 0$ is infinitesimal, and considering the initial
      null surfaces
      ${\hat U} =0, {\hat V} \ge 0$ and
 ${\hat V}=0, {\hat U} \ge 0$.  Consider first the rightward null
      surface ${\hat V}=0$.  On this surface, in the limit
      $\varepsilon \to 0$, the metric components (\ref{metricc}) vanish but their
      derivatives do not.  We choose the normal covector to be
      $\ell_a = - e^{\alpha_0} (d {\hat V})_a/2$, where $\alpha_0 =
      \alpha(0)$, yielding from the transformed metric (\ref{newmetric})
 the contravariant normal ${\vec \ell} = \partial / \partial {\hat U}$.
      We now compute from the metric (\ref{newmetric}) the various
      components of the initial data 
      structure (\ref{ids2}), obtaining 
      that the only quantity
      which is affected by the transformation is the inaffinity.  The
      transformed inaffinity is
      \be
      \kappa_\ps = - (\beta_{\ps\ms} - \beta_{\ms\ms}) \delta({\hat U}
      - \varepsilon).
      \ee
      This matches the general result (\ref{transformed1}) derived in
      Sec.\ \ref{sec:pse}, specialized to the case of vanishing expansion and
      shear of the background.  A similar result applies for
      $\kappa_\ms$.

	\section{Shocks and phase space extension in 1+1 dimensions}
        \label{app:enlarge}

        In Sec.\ \ref{sec:pse} in the body of the paper we defined an enlargement of the gravitational phase space
        associated with allowing shocks to be present in the initial data.  Here we
        describe an analogous definition of
a phase space enlargement involving shocks in the simpler context of a
free scalar field in 1+1 dimensions, to give some intuition for the
more complicated gravitational case.

In 2D Minkowski spacetime with metric $ds^2 = - du dv$, consider the
space of smooth solutions $\phi(u,v)$ for a free scalar field within
the future light cone of the origin.
We can parameterize the phase space in terms of the initial data on
the right and left portions of the light cone
\be
\phi_R(u) = \phi(u,0), \ \ \ \ u \ge 0
\ee
and
\be
\phi_L(v) = \phi(0,v), \ \ \ \ v \ge 0.
\ee
These functions are constrained by the continuity condition
\be
\phi_R(0) = \phi_L(0).
\label{cont}
\ee
We will also assume that the limit $u\to\infty$ of $\phi_R(u)$ exists
and is finite, that $|\phi_R'(u)|$ falls off faster than $1/u$ as $u
\to \infty$, and similarly for $\phi_L(v)$.
The symplectic form on this phase space is then
\be
\label{Omega00}
\Omega = \int_0^\infty du \, \phi_R \wedge \phi_{R,u} + \int_0^\infty dv \, \phi_L \wedge \phi_{L,v}.
\ee

We now consider an enlarged solution space obtained by adding 
solutions of the form
\be
\Delta \phi(u,v) = \beta H(u-\varepsilon) H(-v + \varepsilon) - \beta
H(-u + \varepsilon) H(v - \varepsilon)
\ee
to the smooth solutions.
Here $H$ is the Heaviside step function, $\beta$ is a constant and $\varepsilon>0$ is an
infinitesimal parameter.  This solution contains a leftward
propagating discontinuity and a rightward propagating discontinuity
just to the future of the initial data surface.  If we now compute the symplectic form
for the solution $\phi + \Delta \phi$ and then take the limit
$\varepsilon \to 0$, the effect is to add to 
the expression (\ref{Omega00}) the correction term
\be
\left[ \phi_L(\infty) - \phi_R(\infty) \right] \wedge \beta.
\ee
The same correction to the symplectic form can be obtained by simply
making the replacements $\phi_R(u) \to \phi_R(u) + \beta$, $\phi_L(v)
\to \phi_L(v) - \beta$ in the expression (\ref{Omega00}).
The enlargement of the phase space can therefore be described in terms of
relaxing the continuity requirement (\ref{cont}), while keeping the same
expression (\ref{Omega00}) for the symplectic form.

We can isolate the symplectically orthogonal degrees of freedom for
both the original and enlarged phase spaces using
the techniques described in Sec.\ \ref{sec:sf}.
We define the even and odd combinations of fields
\be
	\phi_{\rm e} = \frac{1}{\sqrt{2}} ( \phi_R + \phi_L), \ \ \ \
	\phi_{\rm o} = \frac{1}{\sqrt{2}} ( \phi_R - \phi_L).
\ee
We introduce the notations for the limiting values
\be
\phi_{\rm e}(\pm \infty) = {\bar \phi}_{\rm e} \pm \Delta \phi_{\rm
  e}, \ \ \ \ \phi_{\rm o}(\pm \infty) = {\bar \phi}_{\rm o} \pm \Delta \phi_{\rm
  o}.
\ee
Finally we fix a monotonic smooth function $g(u)$ with $g(\pm\infty) =
\pm 1/2$ and define ${\tilde \phi}_{\rm e}(u) = \phi_{\rm e}(u) - {\bar
  \phi}_{\rm e} - g(u) \Delta \phi_{\rm e}$, and similarly for ${\tilde
  \phi}_{\rm o}$.
The symplectic form (\ref{Omega00}) can then be written in terms of these notations as
	\bea
	\label{ss11}
	\Omega &=& \int \left[ {\tilde \phi}_{\rm e} \wedge {\tilde
            \phi}_{\rm e}^\prime
	+  {\tilde \phi}_{\rm o} \wedge {\tilde \phi}_{\rm o}^\prime
	\right]
	+ \left({\bar \phi}_{\rm e} + 2 \int g' {\tilde \phi}_{\rm e}\right) \wedge \Delta \phi_{\rm e}
	+ \left({\bar \phi}_{\rm o} + 2 \int g' {\tilde \phi}_{\rm
          o}\right) \wedge \Delta \phi_{\rm o}. \ \ \ 
	\eea
In the original phase space, the continuity condition (\ref{cont})
imposes the relation ${\bar \phi}_{\rm o} =  \Delta \phi_{\rm o}/2$,
which eliminates the first term in the bracket of the third term of Eq.\ (\ref{ss11}).  Then the
analysis of Appendix \ref{app:zero} shows that $\partial/\partial \Delta \phi_{\rm o}$
is effectively a degeneracy or gauge direction, and so $\Delta \phi_{\rm o}$ can be
set to zero, eliminating the entire third term.  Thus, in
the original phase space, one is left with the first two terms in the
expression (\ref{ss11}).  By contrast, in the enlarged phase space, all
three terms are present.

	\section{Symmetry transformations of initial data}
	
	In this appendix we derive how the initial data elements (\ref{ids2}) on two intersecting null surfaces
	transform under a bulk diffeomorphism when using the fixing (\ref{conv1}) of the perturbative rescaling freedom.
	We show that the transformation depends not just on the boundary
	diffeomorphisms (\ref{bds}) but also on the additional Weyl rescaling
	pieces (\ref{gammadef}) of the bulk diffeomorphism.  
	
	We start by picking a representative of the extended initial data
	structure (\ref{ids2}):
	\be
	\label{ids2a}
	\left(q_{AB}, {\bar \omega}_A, m,
	\Theta_\ps, \Theta_\ms, \ell_{\ps\,a}, \ell^i_\ps, \kappa_\ps, [q^\ps_{ij}],
	\ell_{\ms\,a}, \ell^{i'}_\ms, \kappa_\ms, [q^\ms_{i'j'}] \right).
	\ee
	If we act on this with a diffeomorphism $\psi$,
	all the fields get transformed by the pullback $\psi_*$ or by the
	pullbacks of the boundary diffeomorphisms (\ref{bds}):
	\bea
	\label{ids2b}
	&&\bigg( \varphi_{0\,*}
	q_{AB}, \, \varphi_{0\,*} {\bar \omega}_A, \, \varphi_{0\,*} m,\,
	\varphi_{0\,*} \Theta_\ps,\, \varphi_{0\,*} \Theta_\ms,\, e^{\gamma_\ps}
	\ell_{\ps\,a}, \, \varphi_{\ps\,*} \ell^i_\ps, \, \varphi_{\ps\,*}
	\kappa_\ps, \, [\varphi_{\ps\,*} q^\ps_{ij}], \nonumber \\
	&&
	e^{\gamma_\ms} \ell_{\ms\,a}, \varphi_{\ms\,*} \ell^{i'}_\ms,
	\varphi_{\ms\,*} \kappa_\ms, [\varphi_{\ms\,*} q^\ms_{i'j'}] \bigg).
	\eea
	Here we have used the definitions (\ref{gammadef}) to replace the
	pullbacks of $\ell_{\ps\,a}$ and $\ell_{\ms\,a}$ by the rescaled one forms.
	We now note that the convention given by Eq.\ (\ref{conv1}) or Eq.\ (\ref{fixell}) requires the transformed
	normal covectors to coincide with the original ones.  We can implement
	this with a rescaling transformation of the form (\ref{2rescalings}), with
	$\sigma_\ps = -\gamma_\ps$ and $\sigma_\ms = - \gamma_\ms$.  This
	yields
	\bea
	\label{ids2c}
	&&\bigg( \varphi_{0\,*}
	q_{AB}, \, \varphi_{0\,*} {\bar \omega}_A - D_A(\gamma_\ps -
	\gamma_\ms)/2, \, \varphi_{0\,*} m - \gamma_\ps - \gamma_\ms,\,
	e^{-\gamma_\ps} \varphi_{0\,*} \Theta_\ps,\, e^{-\gamma_\ms}
	\varphi_{0\,*} \Theta_\ms,
	\nonumber \\
	&& \,  \ell_{\ps\,a}, 
	\, e^{-\gamma_\ps} \varphi_{\ps\,*} \ell^i_\ps, \, e^{-\gamma_\ps}
	( \varphi_{\ps\,*} \kappa_\ps - \lie_{\ell_\ps} \gamma_\ps), \, [\varphi_{\ps\,*} q^\ps_{ij}], \nonumber \\
	&&\,  \ell_{\ms\,a}, 
	\, e^{-\gamma_\ms} \varphi_{\ms\,*} \ell^{i'}_\ms, \, e^{-\gamma_\ms}
	( \varphi_{\ms\,*} \kappa_\ms - \lie_{\ell_\ms} \gamma_\ms), \, [\varphi_{\ms\,*} q^\ms_{ij}]
	\bigg).
	\eea
	Specializing now to linearized transformations and dropping the $+$ and $-$
	subscripts for simplicity we obtain the transformation laws
	\bes
	\label{symt}
	\bea
	\label{symt1}
	\delta q_{ij} &=& \lie_\chi q_{ij}, \\
	\label{symt2}
	\delta \ell^i &=& (\lie_\chi \ell)^i - \gamma \ell^i, \\
	\label{symt3}
	\delta \Theta &=& \lie_\chi \Theta - \gamma \Theta, \\
	\label{symt4}
	\delta \kappa &=& \lie_\chi \kappa - \gamma \kappa - \lie_\ell \gamma.
	\eea
	\ees
	If we specialize to the restricted \phase space $\sls_\mfp$, we obtain
	from Eqs.\ (\ref{iidd1}) and (\ref{symt2}) that
	\be
	\delta \ell^i = \Psi \ell^i = (\beta - \gamma) \ell^i,
	\label{symt2a}
	\ee
	cf.\ Eq.\ (\ref{Psidef}).
	These results are used in Secs.\ \ref{sec:chargeder} and \ref{sec:fluxder}.

        \section{Fluxes of boost charges}
        \label{app:boostflux}
        
In this appendix we derive the charges and fluxes associated with the 
boosts and half-sided boosts discussed in Sec.\ \ref{halfsidedboosts}.
We then generalize slightly a result of
Ciambelli, Freidel and Leigh (CFL) \cite{Ciambelli:2023mir} and show that
the charges increase monotonically to the future if one chooses the
value of the parameter $\alpha_0$ entering the definition (\ref{boostdef2})
to have the value (\ref{vvv}).

We first consider the charges associated with the normal boost (\ref{boostdef1}),
working in the maximal phase space $\slp$.
We use the general expression \eqref{chargenew3} for the charge for a symmetry of this supertranslation
form, and simplify using Eqs.\ (\ref{volumesmall}), (\ref{iidd23}) and
(\ref{boost3}) to obtain
\be
	{\tilde H}_{\xi} = \int_{\partial \Sigma} {\tilde h}_\xi = 
	\frac{1}{8\pi} \int_{\partial \Sigma} \mu_{ij} ( \lie_\ell +
        \kappa - c_2 \Theta) f.
	\label{boost33}
        \ee
Next we specialize to the coordinate system $(u,\theta^A)$ discussed in
Sec.\ \ref{halfsidedboosts}, where ${\vec \ell} = \partial_u$, the
boost has the explicit form (\ref{prop1}), the condition (\ref{CFLcondt}) is
satisfied,
and the boost surface $B$ is $u = u_0(\theta^A)$.
This gives
\be
\tilde{H}_{\xi} = \frac{1}{8\pi}\int_{\partial \Sigma} d^2\theta
  \sqrt{q} \, b(\theta) \left\{ 1 - (\alpha_0 +c_2)
     \left[ u - u_0(\theta)\right] \Theta(u,\theta) \right\}.
\label{bc99}
     \ee
In particular evaluated on the boost surface $B$ the charge is just
$\int_{\partial \Sigma} d^2 \theta \sqrt{q} b / (8 \pi)$, the average
of the boost parameter $b$ over the surface, as shown by CFL.

If we consider instead the half-sided boost (\ref{hsb33}), the effect
on the charge is to multiply the integrand in Eq.\ (\ref{bc99}) with the
Heaviside function
\be
H[u - u_0(\theta) ].
\ee
Thus the charge is unaltered if evaluated on a surface $\partial \Sigma$ to the future of the boost
surface $B$, and is zero to the past of $B$. It jumps discontinuously
at $B$.

We can easily compute the fluxes of these charges to the future of $B$
using the identity
(\ref{addi2}), which gives
\begin{align}
(d\tilde{h}_{\xi})_{ijk} = \eta_{ijk} F_{\xi}/(8 \pi), 
\end{align}
with $F_\xi = (\lie_\ell + \Theta) (\lie_\ell + \kappa - c_2\Theta) f$.
Simplifying this expression using the explicit form (\ref{prop1}) of the
boost, the condition (\ref{boostdef2}) on the parameter $u$ along the geodesics,
and the Raychaudhuri equation (\ref{grelations1e}) we get after some algebra
\begin{align}
F_{\xi} = b\left[(1- c_2 - \alpha_0)\Theta + (\alpha_0 +
  c_2)(\alpha_0 - \frac{1}{2}) (u - u_0) \Theta^2 + (\alpha_0 +
  c_2) (u-u_0)\sigma^2 \right],
\end{align}
where $\sigma^2 = \sigma_{ij} \sigma_{kl} q^{ik} q^{jl}$.
We want this quantity to be nonnegative for all $u \ge u_0$ and for all
$\Theta$ and $ \sigma^2$, in order for the boost charges to increase
monotonically along the null surface to the future. Assuming $b \ge 0$
this yields three conditions:
\bes
\bea
(1 - c_2 - \alpha_0)\Theta \geq 0,\\
(\alpha_0 + c_2)(\alpha_0 - \frac{1}{2}) \geq 0,\\
\alpha_0 + c_2 \geq 0. 
\eea
\ees
For a generic null surface $\Theta$ has no fixed sign, so we must
choose the value 
\be
\alpha_0 = 1 - c_2
\ee
of $\alpha_0$ in order
to satisfy the first inequality.
The remaining constraints are then satisfied if $c_2 \le 1/2$ (or
equivalently if $\alpha_0 \ge 1/2$)\footnote{If instead we specialize
to future event horizons, where $\Theta
\ge 0$ is guaranteed, then the boost charges are monotonic if the parameters satisfy
$0 \le \alpha_0 + c_2 \le 1$ and $\alpha_0 \ge 1/2$.}.  This generalizes the result of 
CFL which was specialized to $c_2 = 1/2$.

\section{Construction of physical phase space from \phase space}
	\label{app:zero}
	
	In this appendix we justify setting to zero the variable $\Delta {\hat
	  q}_{{\rm o}\,AB}$ in the presymplectic form (\ref{Omegafinal}), and derive the form of the Poisson brackets (\ref{pb44}).
        
	The general procedure for passing from the \phase space $\sls$ to
	the phase space $\slp$ is to mod out by degeneracy directions of the
	presymplectic form $\Omega$, as described in detail by Harlow and Wu
	\cite{Harlow:2019yfa}.  Here degeneracy directions are defined in
	terms of vectors $X$ on $\sls$ whose contractions with $\Omega$
	vanish. Such degeneracy directions correspond to true gauge degrees of
	freedom (and in practice one can often simply gauge fix these degrees of
	freedom to construct $\slp$).  In finite dimensions one can then obtain a nondegenerate
	symplectic form on $\slp$ which can be inverted to define Poisson brackets.
	
	However a complication that can arise in infinite dimensions is that,
	even in the absence of degeneracy directions, the
	symplectic form can be weakly but not strongly nondegenerate
	\cite{Ashtekar:1981bq,Prabhu:2022zcr}.  In other words, the map
	$X \to [ Y \to \Omega(X,Y) ]$ can be injective but not surjective
	(which depends on the choice of topology).  In this case it is not possible to
	invert the symplectic form to define Poisson brackets.
	
	A general method to circumvent this difficulty in defining Poisson
	brackets has been suggested by Prabhu, Satishchandran and Wald \cite{Prabhu:2022zcr}.
	Given a phase space $\slp$ with weakly nondegenerate symplectic form
	$\Omega$, we define a class of functions $F : \slp \to {\bf R}$ 
	by the requirement that there exists a
	vector field ${\vec v}$ for which
	\be
	\delta F = - I_{v} \Omega,
	\label{specialfunctions}
	\ee
	where $I_{v}$ is the inner product operation defined in
	Sec.\ \ref{sec:dsc}.
	Given any two such functions $F$ and $G$
	corresponding to vector fields ${\vec v}$ and ${\vec w}$,
	the Poisson bracket is 
	defined as\footnote{If $\Omega$ is degenerate then ${\vec v}$ and ${\vec w}$ are not uniquely defined, nevertheless the right hand side of Eq.\ (\ref{pb}) is well defined from the definition (\ref{specialfunctions}).}
	\be
	\label{pb}
	\left\{ F , G \right\} = - \Omega( v, w).
	\ee
        This function belongs to the class (\ref{specialfunctions}) with
	the vector field being $L_w v$ \footnote{This follows from
        Cartan's magic formula, $\lie_v \Omega= i_v d \Omega + d
        i_v \Omega = d i_v \Omega = -d^2 F = 0$, and $d \{ F,G\} = - d
        \Omega(v,w) = - d i_w i_v \Omega = - \lie_w i_v \Omega + i_w d
        i_v \Omega = - \lie_w i_v \Omega = -i_{\lie_w v} \Omega - i_v
        \lie_w \Omega = -i_{\lie_w v} \Omega$.}. The bracket also
        obeys the Jacobi identity\footnote{To see this consider three
        functions $F$, $G$ and $H$ with corresponding vector fields
        ${\vec v}$, ${\vec w}$ and ${\vec x}$.  We have
        $\{ F, \{ G, H\} \}
        +\{ G, \{ H, F\} \}
        +\{ H, \{ F, G\} \}= - \Omega(v,\lie_xw) - \Omega(w, \lie_v x)
        - \Omega(x, \lie_w v) = - \lie_x \Omega(v,w) -
        \Omega(x,\lie_wv)$.
        However we also have $\lie_x \Omega(v,w) = \lie_x i_w i_v
        \Omega = i_x d i_w i_v \Omega = i_x \lie_w i_v \Omega = i_x
        i_{\lie_w v} \Omega$.
               }. We thus obtain a definition of Poisson  
	brackets on a preferred class of functions on the phase space,
	which correspond to the physically allowed observables.
	
        \subsection{Spin two sector}
        \label{app:zero2}
        
	We now apply this general theory to the symplectic form (\ref{Omegafinal}).
	To simplify the presentation we drop the irrelevant variables, keeping
	only the fourth term in Eq.\ (\ref{finalbulk}) and the fourth term in
	Eq.\ (\ref{finalcorner}).  We also neglect the irrelevant dependence
	on the angular variables $\theta^A$, and simplify the notation by
	writing $\varphi(U)$ for ${\tilde {\hat q}}_{{\rm o}\,AB}$ and $\psi$
	for $\Delta {\hat q}_{{\rm o}\,AB}$.  Thus
	we consider a phase space
	$\slp$ parameterized by $(\varphi(U), \psi)$ with $\varphi(U) \to 0$
	as $U \to \pm \infty$, with symplectic form
	\be
	\label{pb0}
	\Omega = \int dU \delta \varphi \wedge \delta \varphi' + 2 \int dU
	g' \delta \varphi \wedge \delta \psi.
	\ee
	We can also parameterize the phase space in terms of the function
	\be
	\label{zetadef}
	\zeta(U) = \varphi(U) + g(U) \psi,
	\ee
	which is subject to the boundary condition $\zeta(\infty) = - \zeta(-\infty)$.

	We want to justify setting $\psi$ to zero in the phase space $\slp$.
	We define the phase space $\slp_0$ by setting $\psi$ to
	zero, with symplectic form $\Omega_0$ given by setting $\psi=0$ in
	Eq.\ (\ref{pb0}).  The corresponding Poisson bracket $\left\{
	... \right\}_0$ on $\slp_0$ on functionals $F_0 = F_0[ \varphi(U)]$
	is given by the standard expression
	\be
	\left\{ F_0,G_0 \right\}_0 = -\frac{1}{4} \int_{-\infty}^\infty dU
	\int_{-\infty}^U d {\bar U} \left[
	\frac{\delta F_0}{\delta
		\varphi({\bar U}) } \frac{\delta G_0}{\delta
		\varphi(U) }-
	\frac{\delta G_0}{\delta
		\varphi({\bar U}) } \frac{\delta F_0}{\delta
		\varphi(U) } \right].
	\label{pb4a}
	\ee
	Our task now is to analyze the full phase space $\slp$ with symplectic
	form (\ref{pb0}), and to show that the Poisson algebra obtained from the
	Prabhu-Satishchandran-Wald procedure is isomorphic to the $\psi=0$
	algebra (\ref{pb4a})

	We first show that there are no degeneracy directions in the phase
	space $\slp$.  We write a
	general vector field as
	\be
	{\vec v} = \int dU v^\varphi(U) \frac{\delta} {\delta \varphi(U)} +
	v^\psi \frac{\partial}{\partial \psi},
	\label{vf0}
	\ee
	where the coefficients are functions of $\psi$ and functionals of
	$\varphi$.  Contracting with the symplectic form (\ref{pb0}) yields
	\be
	I_v \Omega = \int dU \left[ - 2 v^\psi g'(U) - 2 v^{\varphi}_{,U}(U) \right]
	\delta \varphi(U) + 2 \int dU g' v^\varphi \, \delta \psi.
	\label{ivw}
	\ee
	Setting this to zero yields $v^\psi g'(U) + 2 v^{\varphi}_{,U}(U)=0$,
	which is incompatible with the boundary condition
	\be
	v^\varphi \to 0, \ \ \ U \to \pm \infty
	\label{bc1}
	\ee
	combined with Eq.\ (\ref{gcondt}) unless $v^\psi = v^\varphi =0$.
	
	Next we determine the allowed class (\ref{specialfunctions}) of functionals
	$F[\varphi(U),\psi]$ on phase space.  Taking a variation of $F$
	and using Eq.\ (\ref{ivw}) gives the
	conditions
	\bes
	\label{Fsoln}
	\bea
	\frac{ \partial F}{\partial \psi} &=& 2 \int dU g' v^\varphi, \\
	\frac{ \delta F}{\delta \varphi(U)} &=& - 2  v^\psi g'(U) - 2 v^\varphi_{,U}(U).
	\eea
	\ees
	Given the functional $F$, the general solution of these equations for a vector field
	${\vec v}$ satisfying the boundary condition (\ref{bc1}) is
	\bes
	\label{vf1}
	\bea
	v^\psi &=& - \frac{1}{2} \int dU \frac{ \delta F} {\delta \varphi(U)},\\
	v^\varphi(U)  &=& \frac{1}{2} \left[ \int d{\bar U} \frac{\delta
		F}{\delta \varphi({\bar U})} \right] \left[ g(U) + \frac{1}{2}
	\right] - \frac{1}{2} \int_{-\infty}^U d {\bar U} \frac{\delta
		F}{\delta \varphi({\bar U})}.
	\eea
	\ees
	Combining Eqs.\ (\ref{Fsoln}) and (\ref{vf1})  yields the equation for $F$
	\be
	\label{Feqn}
	\frac{ \partial F}{\partial \psi} = \int d U g(U) \frac{ \delta
		F}{\delta \varphi(U)}.
	\ee

	The general solution of Eq.\ (\ref{Feqn}) can be expressed in terms of a functional
	$F_0[\varphi(U)]$ as
	\be
	\label{iso}
	F[\varphi(U),\psi] = F_0\left[ \varphi(U) + \psi g(U) \right].
	\ee
	This result is not surprising, since it says that the functional must
	be independent of the choice of function $g(U)$ when expressed in
	terms of the variable (\ref{zetadef}), a property it inherits from the
	symplectic form (\ref{pb0}).  In addition, the functional $F_0$
        is constrained by the differential equation (\ref{Feqn}): it
        must be
	determined by continuity\footnote{This is just a heuristic
		argument since we have not specified the topology we are using on
		$\slp$.  It would be useful to make this argument precise by using the
		description by Ashtekar of $\slp$ as a Fr\'echet space\cite{Ashtekar:1981bq}.}
	by its restriction to $\slp_0$.  For example, the functional
	\be
	F[\varphi(U),\psi] = F_0[ \zeta(U) ] = -4 \lim_{U \to \infty} \zeta(U) =
	-2 \psi
	\ee
	is not an allowed functional since it violates Eq.\ 
        (\ref{Feqn}), and it 
        vanishes identically on
	$\slp_0$.  It formally corresponds to the vector field $\int dU
	\partial / \partial \varphi(U)$ whose associated one parameter family
	of diffeomorphisms maps $\varphi(U) \to \varphi(U) + \varepsilon$,
	which violates the boundary conditions on $\varphi(U)$.  Thus there is
	no vector field for this functional that satisfies the required
	condition (\ref{specialfunctions}).

	Finally we compute the Poisson bracket (\ref{pb}) of two allowed
	functionals $F$ and $G$ corresponding to vector fields ${\vec v}$ and
	${\vec w}$.  Using Eqs.\ (\ref{pb0}), (\ref{vf0}) and (\ref{vf1}) we obtain
	\be
	\left\{ F,G \right\} = -\frac{1}{4} \int_{-\infty}^\infty dU
	\int_{-\infty}^U d {\bar U} \left[
	\frac{\delta F}{\delta
		\varphi({\bar U}) } \frac{\delta G}{\delta
		\varphi(U) }-
	\frac{\delta G}{\delta
		\varphi({\bar U}) } \frac{\delta F}{\delta
		\varphi(U) } \right];
	\label{pb4}
	\ee
	note that the derivative terms with respect to $\psi$ have canceled
	out.
	Now as argued above there is a isomorphism between functionals $F_0$ on $\slp_0$ and
	allowed functionals $F$ on $\slp$, given by Eq.\ (\ref{iso}).  This isomorphism
	preserves the Poisson bracket structure:
	\be
	\left\{ F , G \right\} = H  \ \ \  \iff \ \ \ \left\{ F_0, G_0 \right\}_0 = H_0,
	\ee
	as can be seen by evaluating the bracket (\ref{pb4}) at $\psi=0$ and
	comparing
	with Eqs.\ (\ref{pb4a}) and (\ref{iso}).
	Hence setting $\psi=0$ at the start of the analysis reproduces the
	correct Poisson bracket algebra.  Even though $\psi$ does not
	correspond to a degeneracy direction, it effectively acts like one.

	\subsection{Odd spin zero sector}

        A similar analysis applies to the odd spin zero sector of the symplectic form discussed in Sec.\ \ref{sec:spin0odd}.
        From Eq.\ (\ref{Omegafinal}) the symplectic form of this sector can be written as
        \be
        \label{Omeganormalize}
        \Omega_{{\rm spin\ 0, odd}} = \frac{c_7}{16 \pi} \int d^2 \theta \sqrt{q} {\tilde \Omega}
        \ee
        with
        \be
        \label{tildeOmega}
        {\tilde \Omega} = \int dU \left[ \psi' \wedge \psi - \varphi'
          \wedge \varphi \right] - 4 {\bar \varphi} \wedge \int dU g' \varphi
        + 4 {\bar \psi} \wedge \int dU g' \psi + 4 {\bar \varphi}
        \wedge {\bar \psi}.
        \ee
        Here we have simplified the notation by defining $\varphi(U) =
        \delta {\tilde \zeta}_{\rm o}(U)$, $\psi(U) = c_7^{-1} \delta
               {\tilde \alpha}_{\rm o}(U)$, ${\bar \varphi} = \delta
               {\bar \zeta}_{\rm o}$ and ${\bar \psi} = c_7^{-1}
               \delta {\bar \alpha}_0$, and we have suppressed the dependence
               on $\theta$.  The fields $\varphi$ and $\psi$ obey the
               boundary conditions $\varphi(U) \to 0, \psi(U) \to 0$
               as $U \to \pm \infty$.
        An analysis similar to that of Sec.\ \ref{app:zero2} starting
        from the symplectic form (\ref{tildeOmega}) now shows that
        allowed functionals $F[\varphi(U), \psi(U), {\bar
            \varphi}, {\bar \psi}]$ are given in terms of functionals
        $F_0$ by
	\be
	\label{iso1}
 F[\varphi(U), \psi(U), {\bar
            \varphi}, {\bar \psi}] 
 = F_0\left[ \varphi(U) + 2 {\bar \varphi} g(U) - {\bar \psi},
   \psi(U) + 2 {\bar \psi} g(U) - {\bar \varphi}
   \right].
	\ee
        As before the Poisson algebra (\ref{pb}) of these functionals defined on the full
        phase space is isomorphic to a Poisson algebra defined on the
        space with ${\bar \varphi} = {\bar \psi} = 0$, with the
        Poisson bracket on functionals $F[\varphi(U), \psi(U)]$ being
        given by\footnote{Note that the result (\ref{pb11}) differs
        from that obtained by substituting ${\bar \varphi} = {\bar
          \psi} = 0$ in Eq.\ (\ref{tildeOmega}) and inverting to obtain Poisson
        brackets, which gives only the first two lines of
        Eq.\ (\ref{pb11}).  Instead one must use the prescription based on
        Eq.\ (\ref{pb}) described in Sec.\ \ref{app:zero2}.}
        \bea
        \label{pb11}
	\left\{ F,G \right\} &=&
        -\frac{1}{4} \int_{-\infty}^\infty dU
	\int_{-\infty}^U d {\bar U} \left[
	\frac{\delta F}{\delta
		\varphi({\bar U}) } \frac{\delta G}{\delta
		\varphi(U) }-
	\frac{\delta G}{\delta
		\varphi({\bar U}) } \frac{\delta F}{\delta
		\varphi(U) } \right] \nonumber \\
        &&+\frac{1}{4} \int_{-\infty}^\infty dU
	\int_{-\infty}^U d {\bar U} \left[
	\frac{\delta F}{\delta
		\psi({\bar U}) } \frac{\delta G}{\delta
		\psi(U) }-
	\frac{\delta G}{\delta
		\psi({\bar U}) } \frac{\delta F}{\delta
	  \psi(U) } \right] \nonumber \\
                &&+\frac{1}{4} \int_{-\infty}^\infty dU
	\int_{-\infty}^\infty d {\bar U} \left[
	\frac{\delta F}{\delta
		\varphi({\bar U}) } \frac{\delta G}{\delta
		\psi(U) }-
	\frac{\delta G}{\delta
		\varphi({\bar U}) } \frac{\delta F}{\delta
		\psi(U) } \right].
	\label{pb41}
	\eea
This Poisson bracket is equivalent to those given in
Eqs.\ (\ref{pb44}), when we restore the angular dependence and adjust the
normalization according to Eq.\ (\ref{Omeganormalize}).

	\bibliographystyle{JHEP}
	\bibliography{hamiltonians-null-surfaces,asycps}
	
\end{document}